\documentclass[12pt]{article}
\pdfoutput=1
\usepackage{amsmath,amssymb}
\usepackage{graphicx}

\makeatletter  % section number
 \renewcommand{\theequation}{%
   \thesection.\arabic{equation}}
  \@addtoreset{equation}{section}
\makeatother

\makeatletter
\def\appendix{
\def\theequation{\Alph{section}.\arabic{equation}}
\setcounter{equation}{0}
\def\thesection{\Alph{section}}
\@addtoreset{equation}{section}
\setcounter{section}{0}
\def\@seccntformat##1{
\@nameuse{prefix@##1}
\@nameuse{the##1}
\@nameuse{postfix@##1}\quad}
\def\prefix@section{Appendix~}
} 
\makeatother

\def\slashchar#1{\setbox0=\hbox{$#1$} % set a box for #1
\dimen0=\wd0 % and get its size
\setbox1=\hbox{/} \dimen1=\wd1 % get size of /
\ifdim\dimen0>\dimen1 % #1 is bigger
\rlap{\hbox to \dimen0{\hfil/\hfil}} % so center / in box
#1 % and print #1
\else % / is bigger
\rlap{\hbox to \dimen1{\hfil$#1$\hfil}} % so center #1
/ % and print /
\fi}

\renewcommand{\thefootnote}{\fnsymbol{footnote}}

\DeclareMathOperator*{\wa}{=}
\DeclareMathOperator*{\nyoro}{\approx}

\begin{document}

\title{ Dynamical view of pair creation \\in uniform electric and magnetic fields } 
\author{Naoto Tanji\hspace{0.5pt}\footnotemark[2]\footnotemark[0] }
\date{ \normalsize{\textit{Institute of physics, University of Tokyo, Komaba, Tokyo 153-8902, Japan} } }

\maketitle

\footnotetext[2]{\textit{predoctoral fellow of the Japan Society for the Promotion of Science} }
\footnotetext{\textit{E-mail address}: tanji@nt1.c.u-tokyo.ac.jp }

\renewcommand{\thefootnote}{$*$\arabic{footnote}}

\begin{abstract}
Pair creation in a uniform classical electromagnetic field (Schwinger mechanism) is studied
focusing on the time evolution of the distribution of created particles. 
The time evolution of the distribution in time-dependent fields is also presented
as well as effects of back reaction. 
Motivated by the Glasma flux tube, which may be formed at the initial stage of heavy-ion collisions,
we investigate effects of a magnetic field parallel to an electric field, 
and find that the magnetic field makes the evolution of a fermion system faster. 
\end{abstract}

\vspace{-340pt}
\begin{flushright}
UT-Komaba/08-15
\end{flushright}
\vspace{300pt}

\tableofcontents

\section{Introduction} \label{sec:intro}
Pair creation in an external field has a long history pioneered by the discovery of the Klein paradox 
\cite{Klein}, which was further investigated by Sauter \cite{Sauter}. 
They found that tunneling from a positive frequency state to a negative frequency state can occur 
in relativistic quantum mechanics. 
This is a paradox in view of single-particle theory. 
If one treats it as a problem in quantum field theory, the paradox is resolved and 
particle pair creation from vacuum is concluded. 

Another approach to pair creation in an external field is based on Heisenberg--Euler effective Lagrangian 
\cite{Heisenberg-Euler,Dunne}. 
Heisenberg and Euler found that the 1-loop effective action of QED in a classical electric field
has an imaginary part, which means a vacuum in an electric field is unstable against creation of 
particle--antiparticle pairs. 
In 1951, Schwinger fully formalized it using the proper time method \cite{Schwinger},
and thus particle creation in a classical electric field is called Schwinger mechanism. 
He derived the famous formula for pair creation probability per unit volume and time:
\begin{equation}
w = \frac{2(eE)^2 }{(2\pi )^3 } \sum _{n=1} ^\infty \frac{1}{n^2 } \exp \left( -\frac{n\pi m^2 }{eE} \right) , 
\label{pcp}
\end{equation} 
from a vacuum persistence probability
\begin{equation}
P_0 = |\langle 0,\mathrm{out} |0,\mathrm{in} \rangle |^2 = \exp \left( -\int \! d^4 x \ w\right) , \label{P0_0}
\end{equation}
where $E$ is an electric field strength and $m$ is mass of electron. 

This non-perturbative particle creation mechanism in strong external fields has a wide range of applications; 
not only original QED problems but also pair creation in non-Abelian electromagnetic fields
\cite{Yildiz-Cox,Ambjorn-Hughes,Gyulassy-Iwazaki,Nayak2005a,Nayak2005b} 
and gravitational backgrounds \cite{Brout,Damour,Birrell-Davies,Fulling-textbook}. 
Furthermore, the Mott transition (metal-insulator transition) can be discussed in analogue with 
the Schwinger mechanism \cite{Aoki,Green}. 

For $e^+ e^-$ pair creation, an electric field strength of the order or above the critical value 
$E_c = m^2 /e \sim 10^{16} $ V/cm is needed, which is beyond current technological capabilities. 
The prospects to observe the Schwinger mechanism at future X-ray laser facilities are investigated 
in Ref.\cite{Ringwald}. 
Furthermore, the Coulomb field of superheavy nuclei may cause non-perturbative $e^+ e^-$ creation 
\cite{Soffel,Greiner-textbook}. 
Also the possibility that a supercritical electric field is generated at the surface of bare strange stars
is proposed\cite{Usov}.

In contrast, sufficiently strong field strengths are expected to be obtained in color electric fields
which are formed just after collisions between high energy particles, such as $e^+ e^-$ and heavy ions.
For example, in the framework of Color Glass Condensate (CGC), an initial field strength 
$gE \sim Q_s ^2 \sim 1 \ \text{GeV}^2 $ 
($Q_s $ is a saturation scale of colliding nuclei and $g$ is a strong coupling)
is realized in heavy-ion collisions at RHIC (e.g. \cite{Iancu}). 
Thus, the Schwinger mechanism is supposed to be a relevant mechanism for hadronic production. 
Pair creation by the Schwinger mechanism in $e^+ e^-$ collisions was first studied by Casher et al.\cite{Casher}
with the color flux tube model, in which a color electric field between $q\bar{q} $ is treated as
a constant and classical electric field. 
A considerable number of studies on pair creation in flux tubes have followed 
shifting its stage from $e^+ e^-$ collisions to heavy-ion collisions;
the Lund model \cite{Andersson}, finite-time corrections \cite{Neuberger}, 
finite-size corrections for longitudinal direction (the direction of the electric field) \cite{Wang-Wong,Kim2007a}
and for transverse direction \cite{Gatoff-Wong,Wong-Wang-Wu,Schonfeld}, 
perturbation around a background field \cite{Gitman}
and effects of back reaction \cite{Glendenning,Kluger1991-1993,Kluger1993,Kluger1998}. 
Furthermore, kinetic equations incorporating pair creation from backgrounds have been investigated
based on semi-classical theory \cite{Kajantie,Gatoff,Asakawa-Matsui} and quantum field theory
\cite{Rau,Rau-Muller,Smolyansky,Bloch,Vinnik,Tarakanov}. 

The pair creation probability has been applied to a number of studies on 
particle production in high energy collisions.
From Eq.\eqref{P0_0} it is recognized that the pair creation probability expresses that for collapse of vacuum.
Therefore, the pair creation probability has no direct information about number of created particles\footnote{$w$ and
$dN/dtdx^3  $ are confused in some papers, however, they are different quantity.
$w$ can be equated with $dN/dtdx^3 $ only if $w\ll 1$ (see \cite{Nikishov1970b,Gavrilov2007}). } and
becomes invalid after vacuum collapses, or particles are created.
Furthermore, there is a possibility that pair annihilation occurs. 
In view of these circumstances, it is more straightforward to treat a particle mean number 
$\langle 0,\mathrm{in} |a_\mathbf{p} ^\dagger a_\mathbf{p} |0,\mathrm{in} \rangle $, which has direct information about particle number,
than to treat the pair creation probability. 
Thus, we deal with mostly a particle mean number, or a distribution function in this paper. 

To get a particle mean number, we must quantize charged particle fields
and define a particle picture specifically. 
In other words, we need to define creation and annihilation operators of a particle and an antiparticle.
For this purpose, canonical quantization is more suitable than path-integral quantization. 
Formulating the Schwinger mechanism in terms of canonical quantization has been done by Nikishov 
\cite{Nikishov1970a,Narozhny-Nikishov,Nikishov1970b} as the resolution of the Klein paradox. 
However, how the distribution evolves in time has been unknown in that method 
because only asymptotic states, in and out, are defined. 
Also longitudinal kinetic momentum distributions of created particles have not been obtained 
because kinetic momentum is inseparably connected with time in an electric field.

This obscurity of a longitudinal momentum distribution is closely related to the following paradox
\cite{Kajantie}. 
Suppose that observer \textit{A} watches pair creation in a homogeneous and steady electric field, 
and gets a phase space distribution function of created particles $n(\mathbf{p} )$, which is
independent of the space--time coordinates because the electric field is constant in space and time. 
Furthermore, let observer \textit{B} move relatively to \textit{A} along the direction of the electric field. 
Then, observer \textit{B} obtains the same distribution with \textit{A}, namely $n(\mathbf{p} )$, 
because a phase space distribution is scalar with respect to Lorentz boost. 
If rewritten in terms of momentum in \textit{B}'s coordinates $\mathbf{p} ^\prime $, 
this distribution takes a different functional form
from \textit{A}'s ; $n(\mathbf{p} ) = n^\prime (\mathbf{p} ^\prime )$,
except the case that $n(\mathbf{p} )$ is independent of the momentum in the direction of the electric field. 
For example, if \textit{A} observes that a particle is created with 0 momentum, 
then \textit{B} observes this particle is created with non-zero momentum. 
This inequivalence of two observers contradicts the fact that the electric field is boost-invariant along its
direction. 
That is to say, the uniform electric field takes the same strength and the same configuration 
in both \textit{A}'s and \textit{B}'s frame, 
and thus \textit{A} and \textit{B} are on an equal footing. 
Therefore, the distribution which depends on the longitudinal momentum conflicts 
with the boost invariance of the system. 
This paradox has made a longitudinal momentum distribution in a homogeneous and constant electric field 
unidentified. 

Another obstacle to revealing a momentum distribution in an electric field is ambiguity of particle picture. 
In an external field which breaks Poincar$\acute{\text{e}} $ invariance, we lose the criterion to define
{\lq \lq}what is particle", while in free Minkowski space we have the criterion of positive and negative frequency 
with regard to Poincar$\acute{\text{e}} $ invariance. 
Because of this difficulty, longitudinal momentum distributions have been obtained only after an electric field 
is switched off \cite{Gavrilov1996,Gavrilov2007,Hallin}. 

In this paper, we reveal a longitudinal momentum distribution and its time evolution
by introducing instantaneous positive and negative frequency solutions. 
We decompose the field operator into positive and negative frequency parts instantaneously. 
This procedure extracts the information of a particle picture from the field operator. 
Creation and annihilation operators of a particle and an antiparticle are obtained by this decomposition.
Because the field operator evolves non-trivially under the influence of an electric field, 
the creation and annihilation operators have time-dependence; $a_\mathbf{p} (t), b_\mathbf{p} (t) $. 
This time-dependent particle picture enables us to see the time evolution of the momentum distribution
$\langle 0,\mathrm{in} |a_\mathbf{p} ^\dagger (t) a_\mathbf{p} (t) |0,\mathrm{in} \rangle $. 

To get a dynamical view of how an intense electric field decays into particles is important 
to gain an understanding of the initial stage of heavy-ion collisions. 
How quarks and gluons thermalize after heavy-ion collisions, 
especially short thermalization time $\lesssim 1$ fm$/c$ expected from hydrodynamic calculations (e.g.\cite{Heinz}), 
is not fully understood, 
although there has been numerous attempts to understand the thermalization process 
(e.g. \cite{Mrowczynski} and references therein).
The CGC predicts that there are strong longitudinal color electric and magnetic fields 
just after collisions of heavy nuclei, the state of which is called Glasma \cite{Lappi}. 
Thus, the color flux tube model is supported by the CGC and attracts renewed interest. %emerges from the CGC. 
In this circumstances, decay of a color electric field due to the Schwinger mechanism \cite{Kajantie,Gatoff} 
plays a pivotal role at the initial stage of heavy-ion collisions. 
Recently, a thermalization scenario based on the analogy between the Schwinger mechanism 
and the Hawing-Unruh effect has been proposed \cite{Kharzeev2005,Kharzeev2007}. 
However, as already noted, there remains a dearth of understanding of the Schwinger mechanism in spite of its long history.
In particular, we need to study effects of a longitudinal magnetic field\footnote{The Schwinger mechanism 
in a magnetic field has been 
already studied in some papers \cite{Tarakanov,Nikishov1970a,Gavrilov1996,Gavrilov2007,Popov,Kim2007b}. 
However, its effects on a system with back reaction have not been studied. }, 
which is a new point in the Glasma flux tube. 
Thus, in this paper we re-investigate the Schwinger mechanism as well as effects of a longitudinal magnetic field
strictly based on quantum field theory, not on the semi-classical approximation \cite{Casher,Kharzeev2005,Popov}
or the kinetic approach \cite{Kajantie,Gatoff,Asakawa-Matsui}.

\vspace{11pt}
The remainder of this paper is organized as follows. 
In the next section, we review the formulation of the Schwinger mechanism in terms of canonical quantization
for both bosons (scalar QED) and fermions (spinor QED) based on mainly Ref.\cite{Nikishov1970a}. 
A kinetic momentum distribution in the direction of the electric field cannot be obtained by this method because 
only asymptotic states at $t\to \pm \infty $, which are labeled by not kinetic momentum but canonical momentum, 
are defined. 

In Sec.\ref{sec:PL}, we fully reveal the momentum distribution of created particles in a homogeneous and steady
electric field by defining a particle picture instantaneously.  
The resultant distributions are static and do not give us dynamical view of pair creation,
because the background electric field is static. 
We explain how the longitudinal momentum dependence of the distribution is compatible with the electric field
which is boost-invariant to the longitudinal direction. 
We also show that a magnetic field parallel to the electric field suppresses pair creation of bosons
and enhances that of fermions. 

In Sec.\ref{sec:Nonste}, we deal with non-steady electric fields, strength of which is 0 at first time and
non-zero later. 
This set-up of electric fields enables us to investigate the time evolution of distributions. 
The particle picture associated with instantaneous positive and negative frequency solutions offers us
physically reasonable behavior of the distributions, although it has been criticized in cosmological particle creation. 
Also the time evolution of total particle number density and charge current is shown. 

In Sec.\ref{sec:BR}, we consider back reaction, which is an electric field generated by particles created
from an initial electric field.  
The distributions show plasma oscillation, which is a classical dynamics. 
They are also influenced by quantum effect, namely Bose enhancement or Pauli blocking. 
We estimate the time when the initial electric field decays into particles. 
 
Sec.\ref{sec:summary} is devoted to summary and discussion. 

In appendix, 2-component formalism of the Klein--Gordon field and a method to solve the Dirac equation 
in an electromagnetic field are reviewed. 

Throughout this paper, we use the units in which $\hbar =1$, $c=1$ and $k_B =1 $.

%%%%%%%%%%%%%%%%%%%%%%%%%%%%%%%%%%%%%%%%%%%%%%%%%%%%%%%%%%%%%%%%%%%%%%%%%%%%%%%%%%%%%%%%%
\section{Canonical quantization in an external electromagnetic field} \label{sec:quanti}

In this section, we review canonical quantization in a constant electromagnetic field,
and derive a particle mean number \cite{Nikishov1970a} (see also \cite{Ambjorn-Hughes,Brout}),
however a longitudinal momentum distribution of created particles cannot be obtained in this section
(longitudinal means the direction of the electric field).

%%%%%%%%%%%%%%%%%%%%%%%%%%%%%%%%%%%%%%%%%%
\subsection{Bosons in a constant electromagnetic field} \label{subsec:boson}
\subsubsection{Pure electric field} \label{subsubsec:boson-pure}
We treat charged scalar fields interacting with a classical background electric field:
\begin{equation}
\mathcal{L} = (\partial _{\mu} -ieA_{\mu} )\phi ^{\dagger} 
              \! \cdot \! (\partial ^{\mu} +ieA^{\mu} )\phi -m^2 \phi ^{\dagger} \phi . \label{KGlag}
\end{equation}
At first, we take a constant and homogeneous electric field directing along $z$-axis, and choose the gauge potential as
\begin{equation}
A^\mu = (0,0,0,-Et) .
\end{equation}
This gauge greatly simplifies our calculation because it is independent of the space coordinates\footnote{Other choice
of gauge, e.g. $A^\mu =(-Ez, 0,0,0)$, brings some difficulties (see \cite{Fulling1976,Ambjorn-Wolfram}).}
and thus we use the gauge $A^0 =0$ throughout this paper. 
The equation of motion for $\phi $ is
\begin{equation}
[\partial _0 ^2 -\partial _{\mathrm{T} } ^2 -(\partial _3 +ieEt)^2 +m^2] \phi=0 , \label{KG}
\end{equation} 
where $\partial _\text{T} ^2 \equiv \partial _x ^2 +\partial _y ^2 $. 
The $\phi $ field may be expanded as
\begin{equation}
\phi (t,\mathbf{x} ) = \int \frac{d^3 \! p}{(2\pi )^{3/2} } \phi _{\mathbf{p} } (t) e^{i\mathbf{p} \cdot \mathbf{x} } ,
\end{equation}
because the equation of motion \eqref{KG} does not depend on the space coordinates explicitly. 
The mode function $\phi _{\mathbf{p} } (t) $ obeys the equation
\begin{equation}
\biggl[ \frac{d^2 }{dt^2 } +(eEt+p_z )^2 +m_{\mathrm{T} } ^2 \biggr] \phi _{\mathbf{p} } (t) =0 , \label{Schr}
\end{equation}
in which transverse mass $m_{\mathrm{T} } $ is introduced as
\begin{equation}
m_{\mathrm{T} } ^2 \equiv m^2 +p_{\mathrm{T} } ^2 = m^2 +p_x ^2 +p_y ^2 .
\end{equation}
Note that the role of transverse degrees of freedom is only shifting the mass.

Changing variable as
\begin{equation}
\xi \equiv \sqrt{\frac{2}{eE} } (eEt+p_z )
\end{equation}
converts Eq.\eqref{KG} into a Schr\"{o}dinger-like equation
\begin{equation}
\biggl[ \frac{d^2 }{d\xi ^2 } +\frac{1}{4} \xi ^2 +a \biggr] \phi _\mathbf{p} (t) =0 , \label{rKG}
\end{equation}
in which $a$ is dimension-less parameter defined by
\begin{equation}
a = \frac{m_\mathrm{T} ^2}{2eE} .
\end{equation}

If we choose two solutions of Eq.\eqref{rKG} ${}_\pm \! \phi _\mathbf{p} (t) $ satisfying conditions
\begin{gather}
i {}_+ \! \phi ^* _\mathbf{p} (t) \overleftrightarrow{\frac{d}{dt} } {}_+ \! \phi _\mathbf{p} (t) = 1 \notag \\
i {}_- \! \phi ^* _\mathbf{p} (t) \overleftrightarrow{\frac{d}{dt} } {}_- \! \phi _\mathbf{p} (t) = -1 \label{Bnorm1} \\
i {}_+ \! \phi ^* _\mathbf{p} (t) \overleftrightarrow{\frac{d}{dt} } {}_- \! \phi _\mathbf{p} (t) = 0 , \notag
\end{gather}
then the mode functions defined by 
${}_{\pm } \! f_\mathbf{p} (x) \equiv {}_{\pm } \! \phi _\mathbf{p} (t) e^{i\mathbf{p} \cdot \mathbf{x} } /(2\pi )^{3/2} $
fulfill the orthonormal conditions 
\begin{gather}
({}_+ \! f_\mathbf{p} ,{}_+ \! f_\mathbf{q} ) = \delta ^3 (\mathbf{p} -\mathbf{q} ) \notag \\
({}_- \! f_\mathbf{p} ,{}_- \! f_\mathbf{q} ) = -\delta ^3 (\mathbf{p} -\mathbf{q} ) \label{Bnorm2} \\
({}_+ \! f_\mathbf{p} ,{}_- \! f_\mathbf{q} ) = 0 , \notag
\end{gather}
where the inner product $(f,g)$ is defined by
\begin{equation}
(f,g) = i\int \! d^3 x f^* \overleftrightarrow{\partial _0 } g . \label{Binner}
\end{equation}
Because of the orthonormal conditions \eqref{Bnorm2}, the field operator may be decomposed as
\begin{equation}
\begin{split}
\phi (x) &= \int \! d^3 p [{}_+ \! \phi _\mathbf{p} (t) \frac{e^{i\mathbf{p} \cdot \mathbf{x} } }{(2\pi )^{3/2} } a_\mathbf{p} 
             + {}_- \! \phi _\mathbf{p} (t) \frac{e^{i\mathbf{p} \cdot \mathbf{x} } }{(2\pi )^{3/2} }  b ^{\dagger } _{-\mathbf{p}} ] \\
         &= \int \! d^3 p [{}_+ \! f_\mathbf{p} (x) a_\mathbf{p} + {}_- \! f_\mathbf{p} (x) b ^{\dagger } _{-\mathbf{p}} ] , \label{Bfield}
\end{split}
\end{equation}
and relations
\begin{equation}
\begin{matrix}
a_\mathbf{p} = ({}_+ \! f_\mathbf{p} ,\phi ) \\
b ^{\dagger } _{-\mathbf{p} } = -({}_- \! f_\mathbf{p} ,\phi )
\end{matrix} \label{ab}
\end{equation}
hold in the same way as the free field.
Imposing the canonical commutation relation
$[\phi (t,\mathbf{x} ) , \pi (t,\mathbf{x} ^{\prime } )]=i\delta ^3 (\mathbf{x} -\mathbf{x} ^{\prime } ) $,
where $\pi (x) $ is canonical conjugate momentum $\pi (x) = \dot{\phi } ^{\dagger } (x) $
and using Eq.\eqref{ab}, one yields commutation relations 
\begin{equation}
[a_\mathbf{p} ,a^{\dagger } _\mathbf{q} ] = [b_\mathbf{p} ,b ^{\dagger } _\mathbf{q} ] 
 = \delta ^3 (\mathbf{p} -\mathbf{q} ) , \ \text{others} = 0 . \label{B_CM}
\end{equation}
The charge operator $\hat{Q} $ and the canonical momentum operator $\hat{P} ^i $ are
expressed as
\begin{equation}
\begin{split}
\hat{Q} &= ie\int \! d^3 x \phi ^{\dagger } \overleftrightarrow{\frac{d}{dt} } \phi \\
        &= e(\phi,\phi) \\
        &= e\int \! d^3 p (a^{\dagger } _\mathbf{p} a_\mathbf{p} -b^{\dagger } _\mathbf{p} b_\mathbf{p} )
\end{split} \label{B_Q}
\end{equation}
\begin{equation}
\begin{split}
\hat{P} ^i &= -\int \! d^3 x (\partial _0 \phi ^{\dagger } \partial _i \phi 
              + \partial _i \phi ^{\dagger } \partial _0 \phi ) \\
           &= \int \! d^3 p \ p^i (a^{\dagger } _\mathbf{p} a_\mathbf{p} + b^{\dagger } _\mathbf{p} b_\mathbf{p} ) \hspace{10pt} (i=1,2,3) .
\end{split} \label{B_Pj}
\end{equation}
In the derivation of these equations, 
we have used Eq.\eqref{Bnorm2} and subtracted the infinite vacuum contribution by normal ordering.
Because of Eqs.\eqref{B_CM},\eqref{B_Q} and \eqref{B_Pj},
$a^{\dagger } _\mathbf{p} $ and $a_\mathbf{p} $ can be interpreted as the creation and annihilation operator of
a particle with charge $+e$ and canonical momentum $\mathbf{p} $ respectively, and $b^{\dagger } _\mathbf{p} $ and $b_\mathbf{p} $
as the creation and annihilation operator of an antiparticle with charge $-e$ and canonical momentum $\mathbf{p} $ respectively.
Notice that $\hat{Q} $ can be diagonalized by the creation and annihilation operators 
owing to the orthonormal condition \eqref{Bnorm2}, and $\hat{P} ^i $ is diagonalized owing to the orthonormal condition 
\eqref{Bnorm2} and spatial translational symmetry as well. 
These are independent of an actual choice of the solutions of Eq.\eqref{rKG} ${}_\pm \! \phi _\mathbf{p} (t) $.

One set of solutions of Eq.\eqref{rKG} satisfying the orthonormal conditions \eqref{Bnorm1} can be expressed
in terms of a parabolic cylinder function $D_\nu (x) $ \cite{Whittaker,Abramowitz} as
\begin{equation}
\begin{matrix}
{}_+ \! \phi _\mathbf{p} (t) = \frac{e^{-\frac{\pi }{4} a} }{(2eE)^{\frac{1}{4} } }
                      D_{-ia -\frac{1}{2} } (e^{\frac{\pi }{4} i} \xi ) \\
{}_- \! \phi _\mathbf{p} (t) = \frac{e^{-\frac{\pi }{4} a} }{(2eE)^{\frac{1}{4} } }
                      D^{*} _{-ia -\frac{1}{2} } (e^{\frac{\pi }{4} i} \xi ) .
\end{matrix} \label{phi^out}
\end{equation}
Choice of these solutions is not unique. 
There remains the freedom of the Bogoliubov transformation:
\begin{equation}
\begin{matrix}
{}_+ \! \tilde{\phi } _\mathbf{p} (t) = \alpha _\mathbf{p} {}_+ \! \phi _\mathbf{p} (t) +\beta _\mathbf{p} ^{*} {}_- \! \phi _\mathbf{p} (t) \\[+3pt]
{}_- \! \tilde{\phi } _\mathbf{p} (t) = \alpha _\mathbf{p} ^{*} {}_- \! \phi _\mathbf{p} (t) +\beta _\mathbf{p} {}_+ \! \phi _\mathbf{p} (t)
\end{matrix}
\end{equation}
with the constraint $|\alpha _\mathbf{p} |^2 -|\beta _\mathbf{p} |^2 =1$.
Since these Bogoliubov transformed solutions ${}_\pm \! \tilde{\phi } _\mathbf{p} $ 
also satisfy the conditions \eqref{Bnorm1},
quantization procedure can be done choosing any ${}_\pm \! \tilde{\phi } _\mathbf{p} $
from these infinite sets of solutions, and thus there are infinite degrees of freedom how to expand the field operator
\eqref{Bfield}, from which the creation and annihilation operators are defined. 
Accordingly vacua defined by $a_\mathbf{p} |0\rangle =b_\mathbf{p} |0\rangle =0 $ exist in infinite way and any two of them are not 
equivalent \cite{Fulling1973,Umezawa}.

The identical arbitrariness also exists in quantization of the free field. 
But in that case, it is natural to choose mode functions
as eigenfunctions of $i\partial _{\mu } $, that are positive and negative frequency solutions with regard to
Poincar$\acute{\mathrm{e} }$ invariance.

In an external field, decomposition of mode functions into positive and negative frequency solutions is not straightforward
because the Lagrangian depends on space--time coordinates explicitly. 
In other words, particle concept becomes ambiguous due to interaction with the external field.

Therefore, we need a criterion to define {\lq\lq}particle".
Now, we employ {\lq \lq}asymptotic WKB criterion" \cite{Nikishov1970a,Ambjorn-Hughes},
which leads to the same result as obtained in the proper time method by Schwinger. 
In this criterion, 
we determine a particle picture with the same idea as in a scattering problem in field theory.
Namely, decide a particle picture from the behavior of mode functions in asymptotic regions, 
infinite past and future.
But now, there is no asymptotic free region and the mode functions do not approach the plane wave solution
$\exp (\mp ipx) $,
because we treat the steady electric field which lies from infinite past to infinite future\footnote{Instead of using 
the asymptotic WKB criterion, one can employ an electric field which asymptotically vanishes at infinite past and future,
and obtain the same result with us \cite{Narozhny-Nikishov,Nikishov1970b}. }.
Thus, we define positive and negative frequency mode functions in such a way that
the positive frequency mode function approaches $\exp [iS_{cl} (t) ] $ instead of $\exp (-ipx) $ 
and the negative frequency one $\exp [-iS_{cl} (t) ] $ instead of $\exp (ipx) $ in asymptotic 
region\footnote{Notice that a classical action of a free particle is $-px=-\sqrt{\mathbf{p} ^2 +m^2 } t+\mathbf{p} \cdot \mathbf{x} $. },
where $S_{cl} (t) $ is a classical action of a charged particle under the electric field $A^\mu = (0,0,0,-Et) $:
\begin{equation}
\begin{split}
S_{cl} (t) &= -\frac{1}{2} eEt\sqrt{t^2 +\left( \frac{m}{eE} \right) ^2 }
         -\frac{m^2 }{2eE} \ln \left( \sqrt{eE} t+\sqrt{eEt^2 +\frac{m^2 }{eE} } \right) \\
 &\approx  \begin{cases}
            -\frac{1}{2} eEt^2 -a\ln \left( \sqrt{eE}  t\right) & \ ( t\rightarrow +\infty ) \\
            \frac{1}{2} eEt^2 +a\ln \left( \sqrt{eE} |t|\right) & \ ( t\rightarrow -\infty ).
           \end{cases} \label{kotensayou}
\end{split}
\end{equation}
One can confirm that the mode functions \eqref{phi^out} behave like as
\begin{equation}
\begin{matrix}
{}_+ \! \phi _\mathbf{p} (t) \approx (\sqrt{eE} |t|)^{-\frac{1}{2} }
                             \exp \{ -\frac{i}{2} eEt^2 -ia\ln (\sqrt{eE} |t|) \} \\[+7pt]
{}_- \! \phi _\mathbf{p} (t) \approx (\sqrt{eE} |t|)^{-\frac{1}{2} }
                             \exp \{ \frac{i}{2} eEt^2 +ia\ln (\sqrt{eE} |t|) \}
\end{matrix} \label{out_behave}
\end{equation}
at $t\to +\infty $ modulo a constant factor, using the formula \cite{Abramowitz}
\begin{equation}
D_\nu (z) \underset{|z|\to \infty }{\approx } \left\{
 \begin{array}{ll}
 z^\nu e^{-\frac{1}{4} z^2 } & \left( |\arg z|<\frac{3}{4} \pi \right) \\
 z^\nu e^{-\frac{1}{4} z^2 } -\frac{\sqrt{2\pi } }{\Gamma (-\nu ) } e^{\nu \pi i} z^{-\nu -1} e^{\frac{1}{4} z^2 }
                         & \left( \frac{1}{4} \pi <|\arg z|<\frac{5}{4} \pi \right) \\
 z^\nu e^{-\frac{1}{4} z^2 } -\frac{\sqrt{2\pi } }{\Gamma (-\nu ) } e^{-\nu \pi i} z^{-\nu -1} e^{\frac{1}{4} z^2 }
                         & \left( -\frac{5}{4} \pi <|\arg z|<-\frac{1}{4} \pi \right).
 \end{array}
 \right. \label{asymD}
\end{equation}
Therefore ${}_+ \! \phi _\mathbf{p} (t) $ and ${}_- \! \phi _\mathbf{p} (t) $
given by Eq.\eqref{phi^out} are settled as the positive and negative frequency solution, respectively.

At $t\to -\infty $, however, these functions do not behave like as
${}_\pm \! \phi _\mathbf{p} (t) \approx \exp [\pm iS_{cl} (t)] $ and another mode functions do.
This is the very consequence of pair creation;
particle picture in infinite past and that in infinite future are not equivalent.
Thus, let ${}_{\pm } \phi _\mathbf{p} (t) $ defined by Eq.\eqref{phi^out} be ${}_{\pm } \phi ^{\mathrm{out} } _\mathbf{p} (t) $.
On the other hand, one can find that
\begin{equation}
\begin{matrix}
{}_+ \! \phi ^{\mathrm{in} } _\mathbf{p} (t) = \frac{e^{-\frac{\pi }{4} a} }{(2eE)^{\frac{1}{4} } }
                                       D^{*} _{-ia-\frac{1}{2} } (-e^{\frac{\pi }{4} i} \xi ) \\[+7pt]
{}_- \! \phi ^{\mathrm{in} } _\mathbf{p} (t) = \frac{e^{-\frac{\pi }{4} a} }{(2eE)^{\frac{1}{4} } }
                                       D _{-ia-\frac{1}{2} } (-e^{\frac{\pi }{4} i} \xi )
\end{matrix} \label{phi^in}
\end{equation}
behave like as ${}_\pm \! \phi _\mathbf{p} (t) \approx \exp [\pm iS_{cl} (t)] $, respectively, at $t\to -\infty $.

${}_\pm \phi _\mathbf{p} ^\mathrm{in} $ and ${}_\pm \phi _\mathbf{p} ^\mathrm{out} $ are related by the Bogoliubov translation as
\begin{equation}
\begin{matrix}
{}_+ \phi ^{\mathrm{in} } _\mathbf{p} (t) = \alpha _\mathbf{p} {}_+ \! \phi ^{\mathrm{out} } _\mathbf{p} (t)
                                   + \beta _\mathbf{p} ^{*} {}_- \! \phi ^{\mathrm{out} } _\mathbf{p} (t) \\[+7pt]
{}_- \phi ^{\mathrm{in} } _\mathbf{p} (t) = \alpha _\mathbf{p} ^{*} {}_- \! \phi ^{\mathrm{out} } _\mathbf{p} (t)
                                   + \beta _\mathbf{p} {}_+ \! \phi ^{\mathrm{out} } _\mathbf{p} (t)
\end{matrix} \label{phiBogo}
\end{equation}
with coefficients
\begin{equation}
\alpha _\mathbf{p} = -e^{-\frac{\pi }{2} a+\frac{\pi }{4} i } \frac{\sqrt{2\pi } }{\Gamma (\frac{1}{2} -ia) } , \
\beta _\mathbf{p} = ie^{-\pi a} . \label{alpha_beta}
\end{equation}
One can confirm the equation
\begin{equation}
|\alpha _\mathbf{p} |^2 -|\beta _\mathbf{p} |^2 =1 \label{boson-albe}
\end{equation}
which guarantees the unitarity, holds.

We have chosen two sets of mode functions ${}_\pm \phi _\mathbf{p} ^\mathrm{in} $ and ${}_\pm \phi _\mathbf{p} ^\mathrm{out} $
from infinite candidates using the asymptotic WKB criterion. 
Then, we can define two distinct particle pictures
by expanding the field operator using ${}_{\pm } \! f^{\mathrm{in/out} } _\mathbf{p} (x)
\equiv {}_{\pm } \! \phi ^{\mathrm{in/out} } _\mathbf{p} (t) e^{i\mathbf{p} \cdot \mathbf{x} } /(2\pi )^{3/2} $:
\begin{equation}
\begin{split}
\phi (x) &= \int \! d^3 p [{}_+ \! f^{\mathrm{in} } _\mathbf{p} (x) a^{\mathrm{in} } _\mathbf{p} 
             + {}_- \! f^{\mathrm{in} } _\mathbf{p} (x) b ^{\mathrm{in} \dagger } _{-\mathbf{p}} ] \\
           &= \int \! d^3 p [{}_+ \! f^{\mathrm{out} } _\mathbf{p} (x) a^{\mathrm{out} } _\mathbf{p} 
             + {}_- \! f^{\mathrm{out} } _\mathbf{p} (x) b ^{\mathrm{out} \dagger } _{-\mathbf{p}} ] , \label{Bfield2}
\end{split}
\end{equation}
and two distinct vacua are introduced as
\begin{gather}
a^{\mathrm{in} } _\mathbf{p} |0,\mathrm{in} \rangle = b ^{\mathrm{in} } _\mathbf{p} |0,\mathrm{in} \rangle =0 \label{in-vac} \\
a^{\mathrm{out} } _\mathbf{p} |0,\mathrm{out} \rangle = b ^{\mathrm{out} } _\mathbf{p} |0,\mathrm{out} \rangle =0 .
\end{gather}
Using \eqref{Bnorm2}, \eqref{ab}, \eqref{phiBogo} and \eqref{Bfield2}, the relations between creation or
annihilation operators of in and out are derived:
\begin{equation}
\begin{matrix}
\begin{split}
a^{\mathrm{out} } _\mathbf{p} &= ({}_+ \! f^{\mathrm{out} } _\mathbf{p} ,\phi ) \\
              &= \int \! d^3 q [({}_+ \! f^{\mathrm{out} } _\mathbf{p} ,{}_+ \! f^{\mathrm{in} } _\mathbf{q} )a^{\mathrm{in} } _\mathbf{q}
                 +({}_+ \! f^{\mathrm{out} } _\mathbf{p} ,{}_- \! f^{\mathrm{in} } _\mathbf{q} )b^{\mathrm{in} \dagger } _{-\mathbf{q}} ] \\
              &= \alpha _\mathbf{p} a^{\mathrm{in} } _\mathbf{p} + \beta _\mathbf{p} b^{\mathrm{in} \dagger } _{-\mathbf{p}}
\end{split} \\[+7pt]
\begin{split}
b^{\mathrm{out} \dagger } _{-\mathbf{p}} &= -({}_- \! f^{\mathrm{out} } _\mathbf{p} ,\phi ) \\
              &= -\int \! d^3 q [({}_- \! f^{\mathrm{out} } _\mathbf{p} ,{}_+ \! f^{\mathrm{in} } _\mathbf{q} )a^{\mathrm{in} } _\mathbf{q}
                 +({}_- \! f^{\mathrm{out} } _\mathbf{p} ,{}_- \! f^{\mathrm{in} } _\mathbf{q} )b^{\mathrm{in} \dagger } _{-\mathbf{q}} ] \\
              &= \alpha ^{*} _\mathbf{p} b^{\mathrm{in} \dagger } _{-\mathbf{p}} + \beta ^{*} _\mathbf{p} a^{\mathrm{in} } _\mathbf{p} .
\end{split}
\end{matrix} \label{abBogo}
\end{equation}

Now we can calculate a mean number of particle pair produced by the electric field using the relations \eqref{abBogo}.
Assume now that the state is $|0,\mathrm{in} \rangle $, which is the vacuum of the particle picture defined at $t\to -\infty $
and which is invariable because we use the Heisenberg picture.
By definition \eqref{in-vac}, the mean number of in-particles is 0: 
$\langle 0,\mathrm{in} |a^{\mathrm{in} \dagger } _\mathbf{p} a^{\mathrm{in} } _\mathbf{p} |0,\mathrm{in} \rangle 
= \langle 0,\mathrm{in} |b^{\mathrm{in} \dagger } _\mathbf{p} b^{\mathrm{in} } _\mathbf{p} |0,\mathrm{in} \rangle = 0$.
However, we must consider that particles observed really are out-particles, and they are condensed
in the in-vacuum.

We define the distribution function of particles and that of antiparticles by
\begin{equation}
n_\mathbf{p} \equiv \langle 0,\mathrm{in} |a^{\mathrm{out} \dagger } _\mathbf{p} a^\mathrm{out} _\mathbf{p} |0,\mathrm{in} \rangle \frac{(2\pi )^3 }{V}
 \label{distriP} 
\end{equation}
and
\begin{equation}
\bar{n} _\mathbf{p} 
 \equiv \langle 0,\mathrm{in} |b^{\mathrm{out} \dagger } _\mathbf{p} b^\mathrm{out} _\mathbf{p} |0,\mathrm{in} \rangle \frac{(2\pi )^3 }{V}
 \label{distriAP}
\end{equation}
respectively.
Because $a_\mathbf{p} ^\dagger a_\mathbf{p} $ ($b_\mathbf{p} ^\dagger b_\mathbf{p} $) is the operator which represents the total (anti)particle
number in whole space, their expectation values diverge proportionally to space volume. 
Therefore, we have divided them by
infinite space volume $V=\int \! d^3 x =(2\pi )^3 \delta ^3 (0) $ and converted them into the (anti)particle number
per unit volume, which is finite. 
Furthermore $(2\pi )^3 $, which is really $(2\pi \hbar )^3 $ and represents the unit of
phase space volume, has been multiplied to get the (anti)particle number per the unit of phase space volume
$(2\pi \hbar )^3 $, that is a phase space distribution function.

Using Eq.\eqref{alpha_beta} and \eqref{abBogo}, the distribution functions \eqref{distriP} and \eqref{distriAP} become
\begin{gather}
n_\mathbf{p} = |\beta _\mathbf{p} |^2 
       = e^{-2\pi a}
       = \exp \biggl( -\frac{\pi m_{\mathrm{T} } ^2 }{eE} \biggr)  \label{Bn_p1} \\
\bar{n} _\mathbf{p} = |\beta _{-\mathbf{p} } |^2 , \notag
\end{gather}
where $n_\mathbf{p} = \bar{n} _{-\mathbf{p} } $ holds for general $\alpha _\mathbf{p} $ and $\beta _\mathbf{p} $
because of the charge and the momentum conservation.
Thus, here and henceforth we call $n_\mathbf{p} $ particle pair distribution function, which means
the number of particles with momentum $\mathbf{p} $ and that of antiparticles with momentum $-\mathbf{p} $ simultaneously.

Note that $n_\mathbf{p} $ is independent of the longitudinal momentum $p_z $.
In view of Lorentz invariance, this fact is reasonable since the system with an electric field directed along $z$-axis
is invariant under Lorentz boost to $z$-direction.
However, it is strange that there equally exist particles with every momentum $p_z $ including opposite direction to
the electric field.

This paradox comes from confusion of canonical momentum with kinetic momentum.
Now the subscript $\mathbf{p} $ of $a_\mathbf{p} $ and $b_\mathbf{p} $ represents canonical momentum
because $a_\mathbf{p} $ and $b_\mathbf{p} $ diagonalize the canonical momentum operator \eqref{B_Pj}.
Because $a_\mathbf{p} ^\mathrm{out} $ and $b_\mathbf{p} ^\mathrm{out} $ have been introduced by \eqref{Bfield2}
in which all time-dependence was taken by ${}_{\pm } \! f^\mathrm{out} _\mathbf{p} (x) $
and they were determined to approach $\exp [\pm iS_{cl} (t)] $ at $t\to +\infty $,
$a_\mathbf{p} ^\mathrm{out} $ and $b_\mathbf{p} ^\mathrm{out} $ are the operator which
annihilate a particle or an antiparticle with \textit{canonical} momentum $\mathbf{p} $ at $t\to +\infty $, respectively.
With the gauge $A^\mu =(0,0,0,-Et) $, canonical momentum $\mathbf{p} $, which is a gauge-dependent quantity, has relation to 
kinetic momentum $\mathbf{\Pi } $ as
\begin{equation}
p_z = \Pi _z -eEt , \label{pp}
\end{equation}
which can be explained that kinetic momentum increases linearly with time while canonical momentum is
a constant of motion with regard to translational symmetry. 
(For transverse direction, canonical and kinetic momentum are identical because there is no external field in 
that direction.)
If we take a limit $t\to +\infty $ for fixed $p_z $, kinetic momentum $\Pi _z $ approaches infinity.
Thus, for any value of $p_z $, 
$n_\mathbf{p} $ represents the number of particles at $t\to +\infty $, when kinetic momentum of particles
becomes infinity due to acceleration by the electric field.
Of course in reality, there are particles with finite kinetic momentum at any time 
since pair creation happens constantly, 
however, they are invisible by this method because we took the limit $t\to +\infty $ for fixed $p_z $.
If one want to see a kinetic momentum distribution of created particles, a particle picture must be defined
at each time not only asymptotic region. This will be done in Sec.\ref{sec:PL}.

Although the $z$-component of canonical momentum is not a physical value, we can interpret it as an indicator of
the time when pair creation occurs. Letting the time of pair creation be $t_0 $ and assuming 
kinetic momentum of the created particle is 0 at this time, Eq.\eqref{pp} indicates $t_0 =-p_z/eE $.
Because the system is static and pair creation happens constantly, there is no special value for $t_0 $.
This is the reason why $n_\mathbf{p} $ is independent of $p_z $.

\vspace{11pt}
Also the pair creation probability can be derived by our method.
Although we have little interest in the pair creation probability as a physical value,
we show its derivation in order to clarify meanings of each calculation.

From Eq.\eqref{abBogo} one can find the relation between $|0,\mathrm{in} \rangle $ and $|0,\mathrm{out} \rangle $:
\begin{gather}
|0,\mathrm{in} \rangle = \prod _\mathbf{p} \mathcal{N} _\mathbf{p} ^{-1/2} \exp \left[ \frac{(2\pi )^3 }{V}
  \frac{\beta _\mathbf{p} }{\alpha ^{*} _\mathbf{p} } a^{\mathrm{out} \dagger } _\mathbf{p} b^{\mathrm{out} \dagger } _{-\mathbf{p}} \right] 
 |0,\mathrm{out} \rangle 
 \label{2MSS} \\
|\mathcal{N} _\mathbf{p} | = |\alpha _\mathbf{p} |^2 = 1+e^{-2\pi a} . \notag
\end{gather}
The in-vacuum is a two-mode squeezed state of the out-particle pairs.
Using this relation, we can calculate a vacuum persistence probability as follows:
\begin{equation}
P_0 = |\langle 0,\mathrm{out} |0,\mathrm{in} \rangle |^2
    = \prod _\mathbf{p} |\mathcal{N} _\mathbf{p} |^{-1}
    = \exp \left[ -\frac{V}{(2\pi )^3 } \int \! d^3 p \ln \left( 1+e^{-2\pi a}\right) \right] . \label{P0_1}
\end{equation}
The last expression's index diverges in two ways.
Because of this divergence, $\langle 0,\mathrm{out} |0,\mathrm{in} \rangle $ is zero, which means that the Fock space built on
$|0,\mathrm{in} \rangle $ and that on $|0,\mathrm{out} \rangle $ are inequivalent \cite{Umezawa}.

The first divergence occurs because of infinite volume: $V=(2\pi )^3 \delta ^3 (0) $.
That is to say, because the pair creation probability per unit volume is not 0, 
pair creation happens without fail in infinite volume system.

The second one comes from $\int \! dp_z $, which corresponds to the fact that the time interval when pair creation can
occur is $-\infty $ to $+\infty $, because $p_z $ is the indicator of the time when pair creation happens.
The relation $\int _{-\infty } ^{\infty } \! dp_z = eE\int _{-\infty } ^{\infty } \! dt_0 =eET $
follows from $t_0  = -p_z /eE $, where $T$ expresses whole time interval, which is infinite.

We can now rewrite the vacuum persistence probability \eqref{P0_1} as
\begin{equation}
P_0 = \exp \left[ -\frac{VTeE}{(2\pi )^3 } \int \! d^2 p_{\mathrm{T}} \ln \left( 1+e^{-2\pi a}\right) \right] ,
\end{equation}
from which the pair creation probability per unit volume and time can be read as
\begin{equation}
\begin{split}
w &= \frac{eE}{(2\pi )^3 } \int \! d^2 p_\mathrm{T} \sum _{n=0} ^{\infty } \frac{(-1)^{n+1} }{n} e^{-2\pi na} \\
  &= \frac{(eE)^2 }{(2\pi )^3 } \sum _{n=0} ^{\infty } \frac{(-1)^{n+1} }{n^2 } e^{-\frac{\pi nm^2 }{eE} } .
\end{split}
\end{equation}

%%%%%%%%%%%%%%%%%%%%%%%%%%%%%%%%%%%%
\subsubsection{Collinear electric and magnetic field} \label{subsubsec:boson-coll}
We take a collinear electric and magnetic field $\mathbf{E} = (0,0,E), \mathbf{B} =(0,0,B) $ 
motivated by the Glasma flux tube \cite{Lappi}. 
We use the gauge potential $A^\mu = (0,-By,0,-Et) $.

The Klein--Gordon equation is
\begin{equation}
[\partial _0 ^2 -(\partial _1 +ieBy)^2 -\partial _2 ^2 -(\partial _3 +ieEt)^2 +m^2] \phi(t,\mathbf{x} )=0 . \label{KG-M}
\end{equation} 
This can be solved by separation of variable: $\phi (t,\mathbf{x} ) = \varphi (t) \chi (y) \frac{1}{2\pi } e^{i(p_x x+p_z z)} $.
Then, $\varphi (t) $ and $\chi (y) $ obey equations
\begin{gather}
\left[ \frac{d^2 }{dy^2 } -(eBy+p_x )^2 +\zeta \right] \varphi (y) = 0 \label{mag-sector} \\
\left[ \frac{d^2 }{dt^2 } +(eEt+p_z )^2 +m^2 +\zeta \right] \chi (t) = 0 . \label{ele-sector}
\end{gather}
These can be rewritten as
\begin{gather}
\left[ \frac{d^2 }{d\eta ^2 } -\frac{\eta ^2 }{4} +\frac{\zeta }{2eB} \right] \varphi (y) = 0 \label{mag-sector2} \\
\left[ \frac{d^2 }{d\xi ^2 } +\frac{\xi ^2 }{4} +\frac{m^2 +\zeta }{2eE} \right] \chi (t) = 0 \label{ele-sector2}
\end{gather}
with variables
\begin{equation}
\xi \equiv \sqrt{\frac{2}{eE} } (eEt +p_z ), \ \eta \equiv \sqrt{\frac{2}{eB} } (eBy +p_x ) .
\end{equation}

Eq.\eqref{mag-sector2} has bounded solutions only if $\zeta $ takes discrete value, $\zeta _n = (2n+1)eB $
$(n=0,1,\cdots ) $,
and a solution is
\begin{equation}
\phi _n (y) = \sqrt{\frac{L}{2\pi } } \left( \frac{eB}{\pi } \right) ^\frac{1}{4} \frac{1}{\sqrt{n!} } D_n (\eta ) ,
\end{equation}
which is labeled by the Landau level $n$ and normalized as
\begin{equation}
\int \! dy \phi _n (y) \phi _m (y) = \frac{L}{2\pi } \delta _{n,m} \ ,
\end{equation}
where $L$ is the linear size of the system: $V=L^3 $.

Eq.\eqref{ele-sector2} takes the same form as that in the pure electric field case \eqref{rKG}
with the replacement $m^2 +\zeta _n \to m^2 +p_\mathrm{T} ^2 $.
Thus, we can get the solution of \eqref{ele-sector2} from the solutions in the pure electric case,
\eqref{phi^in} and \eqref{phi^out},
by the replacement $a=\frac{m^2 +p_\mathrm{T} ^2 }{2eE} \to a_n \equiv \frac{m^2 +\zeta _n }{2eE} $:
\begin{equation}
\left \{
\begin{matrix}
{}_+ \! \phi ^{\mathrm{in} } _\mathbf{p} (t) = \frac{e^{-\frac{\pi }{4} a_n } }{(2eE)^{\frac{1}{4} } }
                                       D^{*} _{-ia_n -\frac{1}{2} } (-e^{\frac{\pi }{4} i} \xi) \\
{}_- \! \phi ^{\mathrm{in} } _\mathbf{p} (t) = \frac{e^{-\frac{\pi }{4} a_n } }{(2eE)^{\frac{1}{4} } }
                                       D _{-ia_n -\frac{1}{2} } (-e^{\frac{\pi }{4} i} \xi )
\end{matrix} 
\right.
\ \text{or} \
\left \{
\begin{matrix}
{}_+ \! \phi ^\mathrm{out} _\mathbf{p} (t) = \frac{e^{-\frac{\pi }{4} a_n } }{(2eE)^{\frac{1}{4} } }
                                     D_{-ia_n -\frac{1}{2} } (e^{\frac{\pi }{4} i} \xi ) \ \\
{}_- \! \phi ^\mathrm{out} _\mathbf{p} (t) = \frac{e^{-\frac{\pi }{4} a_n } }{(2eE)^{\frac{1}{4} } }
                                     D^{*} _{-ia_n -\frac{1}{2} } (e^{\frac{\pi }{4} i} \xi ) ,
\end{matrix} 
\right. 
\end{equation}
where $\mathbf{p} $ is the abbreviated expression for $(p_x ,p_z ,n) $.

Consequently, we can connect the pure electric field case and the collinear electromagnetic field case
by the replacements
\begin{equation}
\begin{matrix}
\frac{1}{\sqrt{2\pi } } e^{ip_y y} & \longleftrightarrow 
 & \sqrt{\frac{L}{2\pi } } \left( \frac{eB}{\pi } \right) ^\frac{1}{4} \frac{1}{\sqrt{n!} } D_n (\eta ) \\[+5pt]
\int \! dp_y &\longleftrightarrow & \frac{2\pi }{L} \sum _n \\[+5pt]
\delta (p_y -q_y ) & \longleftrightarrow & \frac{L}{2\pi } \delta _{n,m} \\[+5pt]
a=\frac{m^2 +p_\mathrm{T} ^2 }{2eE} & \longleftrightarrow & a_n = \frac{m^2 +(2n+1)eB }{2eE} .
\end{matrix} \label{EtoEB}
\end{equation}
The pair distribution function in the presence of the collinear electromagnetic field is
\begin{equation}
n_\mathbf{p} = e^{-2\pi a_n } = \exp \left \{ -\frac{\pi [m^2 +(2n+1)eB]}{eE} \right \} , \label{nB_EB}
\end{equation} 
from which one sees that a longitudinal magnetic field makes particles {\lq \lq}heavy" and suppresses pair creation. 

\vspace{11pt} 
We can also derive the pair creation probability by the replacements \eqref{EtoEB}. 
The vacuum persistence probability \eqref{P0_1} changes to
\begin{equation}
P_0 = |\langle 0,\mathrm{out} |0,\mathrm{in} \rangle |^2
    = \exp \left[ -\frac{L^2 }{(2\pi )^2 } 
      \sum _n \int \! dp_x \int \! dp_z \ln \left( 1+e^{-2\pi a_n }\right) \right] . \label{P0_EB1}
\end{equation}
The divergence from $\int \! dp_z $ is treated in the same way as that in the pure electric field case:
$\int \! dp_z = eET$.

New divergence comes from $\int \! dp_x $ because the integrand is independent of $p_x $ in this case. 
This is related with degeneracy of the Landau orbit. 
Comparing Eq.\eqref{mag-sector} and the Schr\"{o}dinger equation of one-dimensional harmonic oscillator
\begin{equation}
\left( \frac{1}{2m} \frac{d^2 }{dx^2 } -\frac{1}{2} m\omega ^2 x^2 +\epsilon \right) \psi = 0 , 
\end{equation}
we can notice that $y_0 \equiv -\frac{p_x }{eB} $ is something like a center of circular motion.
Because a center of circular motion does not influence physical result, the distribution function \eqref{nB_EB}
and the integrand of \eqref{P0_EB1} is independent of $p_x $.
Using the relation $y_0 = -\frac{p_x }{eB} $, the divergence from $\int \! dp_x $ can be treated as
$\int \! dp_x =\int \! d(-eBy_0 ) = eBL $.

Then, the vacuum persistence probability is rewritten as
\begin{equation}
P_0 = \exp \left[ -\frac{VTe^2 EB}{(2\pi )^2 } \sum _n \ln \left( 1+e^{-2\pi a_n }\right) \right] ,
\end{equation}
from which the pair creation probability per unit volume and time can be read as
\begin{equation}
\begin{split}
w &= \frac{e^2 EB}{(2\pi )^2 } \sum _{n=0} ^\infty \sum _{l=0} ^\infty  \frac{(-1)^{l+1} }{l} e^{-2\pi l a_n} \\
  &= \frac{e^2 EB}{(2\pi )^2 } \sum _{l=0} ^\infty \frac{(-1)^{l+1} }{l} e^{-\frac{\pi l m^2 }{eE} }
     \frac{1}{2\sinh \pi l \frac{B}{E} } , \label{wB_EM}
\end{split}
\end{equation}
which is also suppressed by the magnetic field. 
This result was first derived by Popov with the imaginary time method \cite{Popov}.

%%%%%%%%%%%%%%%%%%%%%%%%%%%%%%%%%%%%%%%%%%
\subsection{Fermions in a constant electromagnetic field} \label{subsec:fermion}
Next, we study pair creation of fermions which obey the Lagrangian
\begin{equation}
\mathcal{L} = \bar{\psi } \left[ i\gamma ^{\mu } (\partial _{\mu } +ieA_{\mu } ) -m \right] \psi , \label{DiracL}
\end{equation}
in the same way as that of bosons.

\subsubsection{Pure electric field} \label{subsubsec:fermion-pure} 
The equation of motion derived from the Lagrangian is
\begin{equation}
\left[ \gamma ^{\mu } (i\partial _{\mu } -eA_{\mu } ) -m\right] \psi (x) = 0 ,
\end{equation}
of which solutions under the gauge potential $A^\mu = (0,0,0,-Et) $ are
${}_\pm \! \psi _{\mathbf{p} s} ^\mathrm{in} (x) $ and ${}_\pm \! \psi _{\mathbf{p} s} ^\mathrm{out} (x) $
given by Eq.\eqref{inpsi} and \eqref{outpsi} respectively (see Appendix \ref{sec:2ndD}). 
The meaning of in and out is the same as that in Sec.\ref{subsec:boson}.
The subscripts $\uparrow ,\downarrow $ represent spin directions.

These solutions are related by the Bogoliubov transformation:
\begin{equation}
\begin{matrix}
{}_+ \! \psi _{\mathbf{p} s} ^\mathrm{in} (x) = \alpha _\mathbf{p} {}_+ \! \psi _{\mathbf{p} s} ^\mathrm{out} (x) 
 - \beta _\mathbf{p} ^* {}_- \! \psi _{\mathbf{p} s} ^\mathrm{out} (x) \\[+7pt]
{}_- \! \psi _{\mathbf{p} s} ^\mathrm{in} (x) = \alpha _\mathbf{p} ^* {}_- \! \psi _{\mathbf{p} s} ^\mathrm{out} (x)
 + \beta _\mathbf{p} {}_+ \! \psi _{\mathbf{p} s} ^\mathrm{out} (x)
\end{matrix} \ \ 
(s=\uparrow ,\downarrow ) ,\label{DiracBogo1}
\end{equation}
where the coefficients are
\begin{equation}
\alpha _\mathbf{p} = \frac{\sqrt{2\pi } a^{\frac{1}{2} } }{\Gamma (1-ia)} e^{-\frac{\pi }{2} a+\frac{\pi }{4} i} , \ 
\beta _\mathbf{p} = e^{-\pi a} .
\end{equation}
These coefficients satisfy the relation
\begin{equation}
|\alpha _\mathbf{p} |^2 +|\beta _\mathbf{p} |^2 =1 . \label{Dirac-albe}
\end{equation}
Notice that the sign on the left-hand side is opposite to that of boson [Eq.\eqref{boson-albe}] as a consequence of spin statistics.
Because a distribution function is given by $|\beta _\mathbf{p} |^2 $ not only in the boson case but also fermion
[this will be shown later], Eq.\eqref{Dirac-albe} guarantees Pauli's exclusion principal: 
$n_\mathbf{p} =|\beta _\mathbf{p} |^2 \leq 1 $. 

The solutions ${}_\pm \! \psi _{\mathbf{p} s} ^\mathrm{in} $ and ${}_\pm \! \psi _{\mathbf{p} s} ^\mathrm{out} $ satisfy orthonormal conditions
\begin{equation}
\begin{matrix}
({}_+ \! \psi _{\mathbf{p} s} ^{\mathrm{as} } ,{}_+ \! \psi _{\mathbf{q} s^{\prime } } ^{\mathrm{as} } ) 
  = ({}_- \! \psi _{\mathbf{p} s} ^{\mathrm{as} } ,{}_- \! \psi _{\mathbf{q} s^{\prime } } ^{\mathrm{as} } ) 
  = \delta ^3 (\mathbf{p} -\mathbf{q} ) \delta _{s s^{\prime } } \\[+7pt]
({}_+ \! \psi _{\mathbf{p} s} ^{\mathrm{as} } ,{}_- \! \psi _{\mathbf{q} s^{\prime } } ^{\mathrm{as} } )
 = ({}_- \! \psi _{\mathbf{p} s} ^{\mathrm{as} } ,{}_+ \! \psi _{\mathbf{q} s^{\prime } } ^{\mathrm{as} } ) =0
\end{matrix} \ \
\left( \begin{matrix}
\mathrm{as} =\mathrm{in} ,\mathrm{out} \\
s,s^{\prime } = \uparrow ,\downarrow 
\end{matrix} \right) , \label{Dnorm1}
\end{equation}
where the inner product is defined as 
\begin{equation}
(\psi _1 ,\psi _2 )\equiv \int \! d^3 x \psi _1 ^{\dagger } \psi _2 . \label{Dinner}
\end{equation}
If one expands the field operator in two ways:
\begin{equation}
\begin{split}
\psi (x) &= \sum _{s=\uparrow ,\downarrow } \int \! d^3 p \left[ {}_+ \! \psi _{\mathbf{p} s} ^\mathrm{in} (x) a_{\mathbf{p} s} ^\mathrm{in} 
            + {}_- \! \psi _{\mathbf{p} s} ^\mathrm{in} (x) b_{-\mathbf{p} s} ^{\mathrm{in} \dagger } \right] \\
         &= \sum _{s=\uparrow ,\downarrow } \int \! d^3 p \left[ {}_+ \! \psi _{\mathbf{p} s} ^\mathrm{out} (x) a_{\mathbf{p} s} ^\mathrm{out}
            + {}_- \! \psi _{\mathbf{p} s} ^\mathrm{out} (x) b_{-\mathbf{p} s} ^{\mathrm{out} \dagger } \right] ,
\end{split} \label{2psi1}
\end{equation}
and imposes canonical anti-commutation relation
$\{ \psi (t,\mathbf{x} ),\pi (t,\mathbf{x} ^{\prime } ) \} =i\delta ^3 (\mathbf{x} -\mathbf{x} ^{\prime } ) ,\ (\pi (x) = i\psi ^\dagger (x) ) $,
following equations 
\begin{gather}
\{ a_{\mathbf{p} s} ^\mathrm{as} ,a_{\mathbf{q} s^{\prime } } ^{\mathrm{as} \dagger } \} 
 = \{ b_{\mathbf{p} s} ^\mathrm{as} ,b_{\mathbf{q} s^{\prime } } ^{\mathrm{as} \dagger }  \} 
 = \delta ^3 (\mathbf{p} -\mathbf{q} ) \delta _{ss^{\prime } } \ , \ \text{others} = 0 \label{D_CM} \\
\hat{Q} = e\sum _{s=\uparrow ,\downarrow } \int \! d^3 p 
          (a_{\mathbf{p} s} ^{\mathrm{as} \dagger } a_{\mathbf{p} s} ^\mathrm{as} 
          -b_{\mathbf{p} s} ^{\mathrm{as} \dagger }  b_{\mathbf{p} s} ^\mathrm{as} ) \label{D_Q} \\
\hat{P} ^i = \sum _{s=\uparrow ,\downarrow } \int \! d^3 p \ p^i (a_{\mathbf{p} s} ^{\mathrm{as} \dagger } a_{\mathbf{p} s} ^\mathrm{as} 
          +b_{\mathbf{p} s} ^{\mathrm{as} \dagger }  b_{\mathbf{p} s} ^\mathrm{as} ) \ \ \ (i=1,2,3) \label{D_Pj}
\end{gather} 
$(\text{as} =\mathrm{in} ,\mathrm{out} )$ are obtained with the help of the orthonormal conditions \eqref{Dnorm1}. 
With these equations, we can interpret $a_{\mathbf{q} s} ^{\mathrm{in} \dagger } $ [$b_{\mathbf{q} s} ^{\mathrm{in} \dagger } $] 
as the operator which creates a particle [antiparticle] with canonical momentum $\mathbf{p} $ and spin $s$ at infinite past, 
and $a_{\mathbf{q} s} ^{\mathrm{out} \dagger } $ [$b_{\mathbf{q} s} ^{\mathrm{out} \dagger } $] as that at infinite future.

The relations
\begin{equation}
\begin{matrix}
\begin{split}
a_{\mathbf{p} s} ^\mathrm{out} &= ({}_+ \! \psi _{\mathbf{p} s} ^\mathrm{out} ,\psi ) \\
  &= \alpha _\mathbf{p} a_{\mathbf{p} s} ^\mathrm{in} +\beta _\mathbf{p} b_{-\mathbf{p} s} ^{\mathrm{in} \dagger }
\end{split} \\[+7pt]
\begin{split}
b_{-\mathbf{p} s} ^{\mathrm{out} \dagger } &= ({}_- \! \psi _{\mathbf{p} s} ^\mathrm{out} ,\psi ) \\
                             &= \alpha _\mathbf{p} ^* b_{-\mathbf{p} s} ^{\mathrm{in} \dagger } -\beta _\mathbf{p} ^* a_{\mathbf{p} s} ^\mathrm{in}
\end{split}
\end{matrix} \label{abBogoD}
\end{equation}
are followed from Eq.\eqref{DiracBogo1}, \eqref{Dinner} and \eqref{2psi1}.

Now, we can calculate a distribution function of created pairs observed at time of $+\infty $:
\begin{equation}
\begin{split}
n_{\mathbf{p} s} 
 &= \langle 0,\mathrm{in} | a_{\mathbf{p} s} ^{\mathrm{out} \dagger } a_{\mathbf{p} s} ^\mathrm{out} |0,\mathrm{in} \rangle \frac{(2\pi )^3 }{V}
  = \langle 0,\mathrm{in} | b_{-\mathbf{p} s} ^{\mathrm{out} \dagger } a_{-\mathbf{p} } ^\mathrm{out} |0,\mathrm{in} \rangle \frac{(2\pi )^3 }{V} \\
 &= |\beta _\mathbf{p} |^2 
  = e^{-2\pi a}
\end{split} \label{Dirn_p}
\end{equation} 
Of course, this is independent of spin states because there is no magnetic field.

Notice that the distribution \eqref{Dirn_p} is identical to that of boson \eqref{Bn_p}.
It seems strange that distribution functions do not reflect spin statistics.
This paradox can be understood as follows:
we now see particles with \textit{canonical} momentum $\mathbf{p} $ at infinite future, and as mentioned earlier,
\textit{kinetic} momentum of these particles are infinite due to acceleration by the electric field at that time.
Therefore, spin statistics vanishes because of infinite energy.

\vspace{11pt}
Next, we calculate the pair creation probability.
Using Eq.\eqref{abBogoD} one can derive a relation between $|0,\mathrm{in} \rangle $ and $|0,\mathrm{out} \rangle $: 
\begin{gather}
|0,\mathrm{in} \rangle = \prod _\mathbf{p} \mathcal{N} _{\mathbf{p} \uparrow } ^{-1/2} 
                   \left[ 1+ \frac{(2\pi )^3 }{V} \frac{\beta _\mathbf{p} }{\alpha _\mathbf{p} ^* }
                   a_{\mathbf{p} \uparrow } ^{\mathrm{out} \dagger } b_{-\mathbf{p} \uparrow } ^{\mathrm{out} \dagger } \right] \cdot
                   \prod _\mathbf{q} \mathcal{N} _{\mathbf{q} \downarrow } ^{-1/2} 
                   \left[ 1+ \frac{(2\pi )^3 }{V} \frac{\beta _\mathbf{q} }{\alpha _\mathbf{q} ^* }
                   a_{\mathbf{q} \downarrow } ^{\mathrm{out} \dagger } b_{-\mathbf{q} \downarrow } ^{\mathrm{out} \dagger } \right] |0,\mathrm{out} \rangle \\
|\mathcal{N} _{\mathbf{p} \uparrow } | = |\mathcal{N} _{\mathbf{p} \downarrow } |= |\alpha _\mathbf{p} |^{-2} = \frac{1}{1-e^{-2\pi a} } . \notag
\end{gather}
This relation yields the vacuum persistence probability:
\begin{equation}
P_0 = |\langle 0,\mathrm{out} |0,\mathrm{in} \rangle |^2 
    = \prod _{s=\uparrow ,\downarrow } \prod _\mathbf{p} |\alpha _\mathbf{p} |^2 
    = \exp \left[ 2 \frac{VT}{(2\pi )^3 } \int \! d^2 p_{\mathrm{T}} \ln \left( 1-e^{-2\pi a}\right) \right] ,
\end{equation}
from which the pair creation probability is obtained as 
\begin{equation}
\begin{split}
w &= \frac{2}{(2\pi )^3 } \int \! d^2 p_{\mathrm{T}} \sum _{n=0} ^\infty \frac{1}{n} e^{-2\pi a} \\
  &= \frac{2eE}{(2\pi )^3 } \sum _{n=0} ^{\infty } \frac{1}{n^2 } e^{-\frac{\pi nm^2 }{eE} } .
\end{split}
\end{equation}
Of course, this is identical to the Schwinger's result. 

%%%%%%%%%%%%%%%%%%%%%%%%%%%%%%%%%%%%%%%%%%%%%%%%%%%%%%%
\subsubsection{Collinear electric and magnetic field} \label{subsubsec:fermion-coll}
Similarly to the boson case, we can connect the pure electric field case and the collinear electromagnetic field case
by the replacements
\begin{equation}
\begin{matrix}
\frac{1}{\sqrt{2\pi } } e^{ip_y y} & \longleftrightarrow 
 & \sqrt{\frac{L}{2\pi } } \left( \frac{eB}{\pi } \right) ^\frac{1}{4} \frac{1}{\sqrt{n!} } D_n (\eta ) \\
\int \! dp_y &\longleftrightarrow & \frac{2\pi }{L} \sum _n \\
\delta (p_y -q_y ) & \longleftrightarrow & \frac{L}{2\pi } \delta _{n,m} \\
a=\frac{m^2 +p_\mathrm{T} ^2 }{2eE} & \longleftrightarrow & \left \{
 \begin{matrix}
 a_n ^\uparrow \equiv \frac{m^2 +2neB }{2eE} \\
 a_n ^\downarrow \equiv \frac{m^2 +(2n+2)eB }{2eE}
 \end{matrix} \
 \ (n=0,1,\cdots ) \right. ,
\end{matrix} \label{F:EtoEB}
\end{equation}
where $n$ and $m$ are Landau level. 
Compared with the boson case \eqref{EtoEB}, energy levels are split due to the spin--magnetic field interaction.

Thus, the distribution in the collinear electromagnetic field is
\begin{equation}
n_{\mathbf{p} s} = \exp (-2\pi a_n ^s ) \ \ (s=\uparrow ,\downarrow ) , \label{Dirn_pEM}
\end{equation} 
in which $\mathbf{p} $ is the abbreviated expression for $(p_x ,p_z ,n) $.

Notice that, also in this case, the magnetic field behaves like mass and suppresses pair creation.
On the other hand, 
spin--magnetic field interaction makes a particle of $\uparrow $-mode lighter.
Especially, $n=0$ and $\uparrow $-mode is not suppressed by the magnetic field 
because $a_0 ^\uparrow $ is independent of $B$.

\vspace{11pt}
The replacements \eqref{F:EtoEB} changes the calculation of the vacuum persistence probability as follows:
\begin{equation}
\begin{split}
P_0 &= \exp \left[ \frac{L^2 }{(2\pi )^2 } \sum _{s=\uparrow ,\downarrow } \sum _n
       \int \! dp_x \int \! dp_z \ln \left( 1-e^{-2\pi a_n ^s }\right) \right] \\
    &= \exp \left[ \frac{VTe^2 EB}{(2\pi )^2 } \sum _{s=\uparrow ,\downarrow } \sum _n
       \ln \left( 1-e^{-2\pi a_n ^s }\right) \right]
\end{split}
\end{equation}
in which the divergences from $\int \! dp_x $ and $\int \! dp_z $ are treated in the same way as 
that in Sec.\ref{subsubsec:boson-coll}.
Then, the pair creation probability is
\begin{equation}
\begin{split}
w &= -\frac{e^2 EB}{(2\pi )^2 } \sum _{s=\uparrow ,\downarrow } \sum _n \ln \left( 1-e^{-2\pi a_n ^s }\right) \\
  &= \frac{e^2 EB}{(2\pi )^2 } \sum _{s=\uparrow ,\downarrow } \sum _n \sum _{l=0} ^\infty \frac{1}{l} e^{-2\pi l a_n ^s } \\
  &= \frac{e^2 EB}{(2\pi )^2 } \sum _{l=0} ^\infty \frac{1}{l} e^{-\frac{\pi l m^2 }{eE} } \coth \pi l\frac{B}{E} . 
\label{wF_EM}
\end{split}
\end{equation}
This $w$ increases with increasing $B$\,\footnote{This behavior was first pointed out by Nikishov 
\cite{Nikishov1970a}. }
while that of boson [Eq.\eqref{wB_EM}] decreases.
This is because (i) pair creation of fermion in $n=0 $ and $\uparrow $-mode is not suppressed by the magnetic field,
while that of bosons in any Landau level is suppressed exponentially, and 
(ii) the number of modes degenerating in one Landau level per unit area is $eB/2\pi $, 
which is proportional to the magnetic field strength, as is known in quantum dynamics 
(see for example \cite{Landau}). 

This phenomenon is common to other bulk quantities: expectation values of the Hamiltonian, the number operator 
or the current operator, etc.
For example, the in-vacuum expectation value of the total number operator for fermion out-particles is
\begin{equation}
\begin{split}
\langle 0,\mathrm{in} | \hat{N} |0,\mathrm{in} \rangle 
 &= \sum _{s=\uparrow ,\downarrow } \frac{2\pi }{L} \sum _n \int \! dp_x \int \! dp_z 
     \langle 0,\mathrm{in} | a_{\mathbf{p} s} ^{\mathrm{out} \dagger } a_{\mathbf{p} s} ^\mathrm{out} |0,\mathrm{in} \rangle \\
 &= \sum _{s=\uparrow ,\downarrow } \frac{eB}{2\pi } V \sum _n \int \! \frac{dp_z }{2\pi } n_{\mathbf{p} s} \\
 &\nyoro _{B\sim E \gg m^2 } \frac{eB}{2\pi } V \int \! \frac{dp_z }{2\pi } n_{p_z ,n=0,\uparrow } ,
\end{split} \label{NinB}
\end{equation} 
which is approximately proportional to $B$\,\footnote{The integral $\int \! \frac{dp_z }{2\pi } n_{\mathbf{p} s} $ 
in Eq.\eqref{NinB} diverges because now $n_{\mathbf{p} s} $ is independent of $p_z $. 
Finite expectation values will be obtained in Sec.\ref{sec:Nonste} where an electric field is imposed during finite time. }.
From Eq.\eqref{NinB} we can confirm that the degeneracy of one Landau level per unit area is 
surely $eB/2\pi $.

%%%%%%%%%%%%%%%%%%%%%%%%%%%%%%%%%%%%%%%%%%%%%%%%%%%%%%%%%%%%%%%%%%%%%%%%%%%%%%%%%%%%%%%%%
\section{Longitudinal momentum distribution} \label{sec:PL}

In the previous section, we have calculated the distribution function of created particle pairs 
using the asymptotic WKB criterion.
However, a kinetic momentum distribution in the direction of the electric field (longitudinal direction) 
could not be obtained because we saw only the asymptotic state, in which longitudinal momentum
becomes infinite due to acceleration by the electric field.

To study a longitudinal momentum distribution in the presence of the constant electric field, 
we must define a particle picture at each instant not only at infinite past or future.
That is, to expand the field operator by instantaneous positive and negative frequency solutions at each time.

We define instantaneous positive and negative frequency solution at time $t=t_0 $ as the plane wave solutions
under the pure gauge $\mathbf{A} = \mathbf{A} (t_0 ) $, 
where $\mathbf{A} (t) $ is a gauge potential for an electric field $\mathbf{E} (t) $ (we use the gauge $A^0 =0 $). 
For a scalar field in the constant electric field $A^3 (t) = -Et $, 
the instantaneous positive and negative frequency solutions
at $t=t_0 $ are, respectively,
\begin{equation}
{}_\pm \! f_\mathbf{p} ^{(t_0 )} (x) = \frac{1}{\sqrt{(2\pi )^3 2\omega _p } } 
                                 e^{\mp i\omega _p t +i(\mathbf{p} -\mathbf{E} t_0 )\cdot \mathbf{x} } , \label{t_0plane}
\end{equation} 
in which $\omega _p \equiv \sqrt{p_z ^2 +m_\mathrm{T} ^2 } $\hspace{1pt}\footnote{In 
Ref.\cite{Smolyansky,Bloch,Vinnik,Tarakanov}, 
\begin{equation}
{}_\pm \! f_\mathbf{p} (x) = \frac{1}{\sqrt{(2\pi )^3 2\omega _p (t) } } 
 \exp \left[ \mp i\int ^t \! dt^\prime \omega _\mathbf{p} (t^\prime )  +i\mathbf{p} \cdot \mathbf{x} \right] \label{other}
\end{equation}
where $\omega _\mathbf{p} (t) = \sqrt{ (\mathbf{p} -e\mathbf{A} (t))^2 +m^2 } $ 
is used as instantaneous positive and negative frequency solutions instead of Eq.\eqref{t_0plane}. 
Equations \eqref{t_0plane} and \eqref{other} give the essentially same particle picture 
although the meaning of $\mathbf{p} $ in \eqref{t_0plane} and \eqref{other} are different;
$\mathbf{p} $ is kinetic in \eqref{t_0plane} and canonical in \eqref{other}
[One can confirm it by calculating the canonical momentum operator]. 
Differences between us and Ref.\cite{Smolyansky,Bloch,Vinnik,Tarakanov} are as follows. 
In Ref.\cite{Smolyansky,Bloch,Vinnik,Tarakanov}, a kinetic equation is investigated based on quantum field theory. 
However, it is practically easier to treat the field equation itself, which is a second-order differential equation, 
than to treat the kinetic equation with a non-Markovian source term, which is derived from the field equation and 
is an integro-differential equation. 
Furthermore, the employed regularization methods are distinct. 
In our method, the Hamiltonian is regularized only by normal ordering,
and the expectation value of the current operator is finite without regularization.  
} .
Then, the field expansion is
\begin{equation}
\phi (x) = \int \! d^3 p [ {}_+ \! f_\mathbf{p} ^{(t_0 )} (x) a_\mathbf{p} (t_0 ) +
                           {}_- \! f_\mathbf{p} ^{(t_0 )} (x) b_{-\mathbf{p} } ^\dagger (t_0 ) ] , \label{t_0exp0}
\end{equation}
where $a_\mathbf{p} (t_0 ) $ and $b_{-\mathbf{p} } (t_0 ) $ are annihilation operators defined at $t=t_0 $.
Because the mode functions \eqref{t_0plane} are not solutions of the field equation in the electric field,
the creation and annihilation operators are forced to have time-dependence,
which enables us to see the time evolution of the momentum distribution.

The idea of instantaneous positive and negative frequency was used to study particle creation 
in expanding universes.
However, it encounters some difficulties and is strongly criticized \cite{Fulling1979,Fulling-textbook}.
The main subject for the criticism is the behavior of a Bogoliubov coefficient $\beta _\mathbf{p} $ as a function of $\mathbf{p} $. 
For the Robertson-Walker model, $\beta _\mathbf{p} $ falls off with negative power of $|\mathbf{p} |$ for high momentum $\mathbf{p} $.
Especially for anisotropic case, $\beta _\mathbf{p} $ behaves like 
$\beta _\mathbf{p} \sim |\mathbf{p} |^{-1} $ for high momentum, and therefore the particle number per unit volume 
$\int \! \frac{d^3 p}{(2\pi )^3 } |\beta _\mathbf{p} |^2 $ diverges \cite{Fulling-textbook}. This is physically unacceptable.

Because of above-mentioned problem, other approach is believed to be a correct one to treat cosmological particle creation.
It is adiabatic approach \cite{Parker-Fulling,Birrell-Davies,Fulling-textbook}.
In the adiabatic approach, a physical Fock space is fixed by
expanding the field operator by the exact solutions whose asymptotic form for large mass or momentum 
is a WKB-type solution\footnote{This procedure is identical to the asymptotic WKB criterion 
we have used in Sec.\ref{sec:quanti}. }. 
Then, one ceases defining particle picture at intermediate time and treats mostly
field observables like as the energy--momentum tensor and the charge current, 
which are renormalized by adiabatic regularization.
For an electric background field, this approach has been used by Kluger \textit{et al.} 
\cite{Kluger1991-1993,Kluger1993}.

In spite of these circumstances, we apply the idea of instantaneous positive and negative frequency 
and investigate a particle number at each instant of time. 
That is because, for an electromagnetic background, $\beta _\mathbf{p} $ falls off rapidly for high momentum if one take an
appropriate electric field configuration (see Sec.\ref{sec:Nonste}). 
[In this section, $\beta _\mathbf{p} $ does not fall off for high longitudinal momentum. 
It is not a defect of our method but a natural result since we take a constant electric field.] 
Thus, the application of the instantaneous positive and negative frequency to particle creation 
in an electromagnetic field can avoid the criticism subjected to cosmological particle creation.

Of course, as noted earlier, a definition of particle in the presence of a pair-creating background is 
rather ambiguous. 
The use of the particle picture defined by the instantaneous positive and negative frequency solutions 
is merely an ansatz.
An ansatz must be justified by physically reasonable results.
Actually the ansatz of the instantaneous particle picture will provide us with convincing behavior 
of distribution functions.
Especially in Sec.\ref{sec:Nonste}, we will get distributions which yield a finite total particle number.

A merit of this particle picture is that the divergent part of the energy--momentum tensor can be removed only 
by normal ordering; no further regularization is needed unlike the adiabatic approach. 
This good point benefits even those who do not trust the concept of particle number at each instant 
and accept only field observables. 
One can interpret defining a particle picture at each time as a means to identify zero-point energy. 
Furthermore, energy conservation is automatically and manifestly guaranteed in this particle picture 
because the energy operator at each time takes the same form as that of the free field:
\begin{equation}
\hat{P} ^0 (t) = \int \! d^3 p \ \omega _\mathbf{p} \left[ a_\mathbf{p} ^\dagger (t) a_\mathbf{p} (t) +b_\mathbf{p} (t) b_\mathbf{p} ^\dagger (t) 
\right] ,
\end{equation}
and the Bogoliubov transformation, which relates creation and annihilation operators of different times,
conserves the commutation relation: 
$b_\mathbf{p} (t) b_\mathbf{p} ^\dagger (t) - {:\! b_\mathbf{p} (t) b_\mathbf{p} ^\dagger (t)\! :} =\text{const.} $
This is not the case generally in the adiabatic approach, in which subtracted part is time-dependent. 
In this and next section, however, total energy of the system is not conserved 
because there is an external electric field which provides the system with energy.
If one takes account of back reaction and treat an electric field as a dynamical variable,
total energy will be conserved (Sec.\ref{sec:BR}).

%%%%%%%%%%%%%%%%%%%%%%%%%%%%%%%%%%%%%%%%%
\subsection{Bosons in a constant electromagnetic field} \label{subsec:PLboson}
\subsubsection{Pure electric field} \label{subsubsec:PLboson-pure}
We re-investigate the distribution function of created particles in the constant electric field using
the particle picture defined by the instantaneous positive and negative frequency solutions.
That is to say, we calculate the pair distribution function
\begin{equation}
n_\mathbf{p} (t) \equiv \langle 0,\mathrm{in} |a^{\dagger } _\mathbf{p} (t) a _\mathbf{p} (t) |0,\mathrm{in} \rangle \frac{(2\pi )^3 }{V}
           = \langle 0,\mathrm{in} |b^\dagger _{-\mathbf{p} } (t) b _{-\mathbf{p} } (t) |0,\mathrm{in} \rangle \frac{(2\pi )^3 }{V} .
 \label{Bnp(t)def}
\end{equation}
The instantaneous particle picture is introduced and related to the in-particle picture by the expansion:
\begin{equation}
\begin{split}
\phi (x) &= \int \! d^3 p [ {}_+ \! f_\mathbf{p} ^\mathrm{in} (x) a_\mathbf{p} ^\mathrm{in} +
                           {}_- \! f_\mathbf{p} ^\mathrm{in} (x) b_{-\mathbf{p} } ^{\mathrm{in} \dagger} ] \bigg| _{t=t_0 } \\
         &= \int \! d^3 p [ {}_+ \! f_\mathbf{p} ^{(t_0 )} (x) a_\mathbf{p} (t_0 ) +
                           {}_- \! f_\mathbf{p} ^{(t_0 )} (x) b_{-\mathbf{p} } ^\dagger (t_0 ) ] . \label{in-t0expansion}
\end{split}
\end{equation}
The creation and annihilation operator of in-particle and that of time $t_0 $ are related by the Bogoliubov transformation
whose coefficients $\alpha _{\mathbf{p},\mathbf{q} } (t) ,\beta _{\mathbf{p},\mathbf{q} } (t) $ are given by
\begin{equation}
\begin{matrix}
{}_{+} \! f^{\mathrm{in} } _\mathbf{p} (x) = \int \! d^3 \mathbf{q}  \left[ \alpha _{\mathbf{p},\mathbf{q} } (t) 
                                       {}_{+} \! f_\mathbf{q} ^{(t)} (x) 
                                       + \beta _{\mathbf{p},\mathbf{q} } ^{*} (t) {}_{-} \! f_\mathbf{q} ^{(t)} (x) \right] \ \\
{}_{-} \! f^{\mathrm{in} } _\mathbf{p} (x) = \int \! d^3 \mathbf{q}  \left[ \alpha _{\mathbf{p},\mathbf{q} } ^{*} (t)
                                       {}_{-} \! f_\mathbf{q} ^{(t)} (x) 
                                       + \beta _{\mathbf{p},\mathbf{q} } (t) {}_{+} \! f_\mathbf{q} ^{(t)} (x) \right] .
\end{matrix} \label{tdBogo}
\end{equation}
However, this expansion is not definite because the coefficients has time-dependence.
This ambiguity of the expansion comes from the fact that the Klein--Gordon equation is the second-order differential equation.
Accordingly, to make the expansion determined, we put the additional conditions
\begin{equation}
\begin{matrix}
{}_{+} \! \dot{f} ^{\mathrm{in} } _\mathbf{p} (x) |_{t=t_0 } 
 = \int \! d^3 \mathbf{q}  \left[ \alpha _{\mathbf{p},\mathbf{q} } (t_0 ) {}_{+} \! \dot{f} _\mathbf{q} ^{(t_0 )} (t_0 ,\mathbf{x} ) 
   + \beta _{\mathbf{p},\mathbf{q} } ^{*} (t_0 ) {}_{-} \! \dot{f} _\mathbf{q} ^{(t_0 )} (t_0 ,\mathbf{x} ) \right] \ \\
{}_{-} \! \dot{f} ^{\mathrm{in} } _\mathbf{p} (x) |_{t=t_0 } 
 = \int \! d^3 \mathbf{q}  \left[ \alpha _{\mathbf{p},\mathbf{q} } ^{*} (t_0 ) {}_{-} \! \dot{f} _\mathbf{q} ^{(t_0 )} (t_0 ,\mathbf{x} ) 
   + \beta _{\mathbf{p},\mathbf{q} } (t_0 ) {}_{+} \! \dot{f} _\mathbf{q} ^{(t_0 )} (t_0 ,\mathbf{x} ) \right] ,
\end{matrix} \label{dttdBogo}
\end{equation}
where dot denotes time-derivative. 

Incidentally, these conditions, \eqref{tdBogo} and \eqref{dttdBogo}, tell us that the particle number
defined by the instantaneous particle picture at time $t$ is identical to 
which is obtained if one switches off the electric field at time $t$. 
This accordance is merely the matter of calculation; we do not switch off the field 
but only define a particle picture at each time.

In our calculation, it is convenient to use the 2-component form of the Klein--Gordon field (see Appendix \ref{sec:1stKG}). 
If one put the same condition as Eq.\eqref{tdBogo} on the 2-component mode function ${}_\pm \! F_\mathbf{p} ^{(\cdot)} $,
which is made from ${}_\pm \! f_\mathbf{p} ^{(\cdot)} $ by Eq.\eqref{2-component}, the condition \eqref{dttdBogo} is
automatically satisfied. 

Furthermore, we can get the Bogoliubov coefficients easily using the 2-component formalism.
Because the inner product expressed by the 2-component form \eqref{2-inner} does not contain time-derivative,
the Bogoliubov coefficients can be extracted from Eq.\eqref{tdBogo} by the inner product:
$\alpha _{\mathbf{p},\mathbf{q} } (t_0 ) = ({}_{+} \! f_\mathbf{q} ^{(t_0 )} ,{}_{+} \! f_\mathbf{p} ^{\mathrm{in} } ) $, 
while the inner product of the 1-component form \eqref{Binner} contains time-derivative and cannot extract the Bogoliubov
coefficients from Eq.\eqref{tdBogo}.

The 2-component forms corresponding to 
${}_\pm \! f_\mathbf{p} ^\mathrm{in} (x) = {}_\pm \! \phi _\mathbf{p} ^\mathrm{in} (t) e^{i\mathbf{p} \cdot \mathbf{x} } /(2\pi )^{3/2} $ are
\begin{equation}
{}_\pm \! F^{\mathrm{in} } _\mathbf{p} (x) = \frac{1}{2m} \left(
                                     \begin{array}{c}
                                      m{}_\pm \! \phi _\mathbf{p} ^\mathrm{in} +i{}_\pm \! \dot{\phi } _\mathbf{p} ^\mathrm{in} \\
                                      m{}_\pm \! \phi _\mathbf{p} ^\mathrm{in} -i{}_\pm \! \dot{\phi } _\mathbf{p} ^\mathrm{in}
                                     \end{array}
                                     \right) \frac{e^{i\mathbf{p} \cdot \mathbf{x}} }{\sqrt{(2\pi )^3 } } ,
\end{equation}
where ${}_\pm \! \phi _\mathbf{p} ^\mathrm{in} (t) $ are given by Eq.\eqref{phi^in},
and those to the instantaneous positive and negative frequency mode functions ${}_\pm \! f_\mathbf{p} ^{(t_0 )} (x) $
[Eq.\eqref{t_0plane}] are
\begin{gather}
{}_+ \! F_\mathbf{p} ^{(t_0 )} (x) = \frac{1}{2m} \left(
                                             \begin{array}{c}
                                              m+\omega _p \\
                                              m-\omega _p
                                             \end{array}
                                             \right) {}_+ \! f_\mathbf{p} ^{(t_0 )} (x) \ \\
{}_- \! F_\mathbf{p} ^{(t_0 )} (x) = \frac{1}{2m} \left(
                                             \begin{array}{c}
                                              m-\omega _p \\
                                              m+\omega _p
                                             \end{array}
                                             \right) {}_- \! f_\mathbf{p} ^{(t_0 )} (x) .
\end{gather}
Using these expressions, one can calculate the inner products as follows: 
\begin{allowdisplaybreaks}
\begin{gather}
\begin{split}
& \!
({}_+ \! f_\mathbf{p} ^{(t_0 )} ,{}_+ \! f_\mathbf{q} ^{\mathrm{in} } ) \\
 &= 2m\int \! d^3 x {}_+ \! F_\mathbf{p} ^{(t_0 ) \dagger } (t_0 ,\mathbf{x} ) \sigma _3 {}_+ \! F_\mathbf{q} ^\mathrm{in} (t_0 ,\mathbf{x} ) \\
 &= \frac{e^{i\omega _{\mathbf{p} } t_0 } }{\sqrt{2\omega _{\mathbf{p} } } }
    \left[ \omega _\mathbf{p} {}_+ \! \phi _{\mathbf{p} -e\mathbf{E} t_0 } ^\mathrm{in} (t_0 ) 
    +i{}_+ \! \dot{\phi } _{\mathbf{p} -e\mathbf{E} t_0 } ^\mathrm{in} (t_0 ) \right] \delta ^3 (\mathbf{p} -\mathbf{q} -e\mathbf{E} t_0 ) \\
 &= \frac{e^{-\frac{\pi }{4} a } }{(2eE)^{1/4} } \frac{e^{i\omega _{\mathbf{p} } t_0 } }{\sqrt{2\omega _{\mathbf{p} } } }
    \left[ \left( \omega _{\mathbf{p} } +p_z \right) 
    D^* _{-ia-\frac{1}{2} } (-e^{\frac{\pi }{4} i} \sqrt{\frac{2}{eE} } p_z  )
     +e^{\frac{\pi }{4} i} \sqrt{2eE} D^* _{-ia+\frac{1}{2} } 
     (-e^{\frac{\pi }{4} i} \sqrt{\frac{2}{eE} } p_z  ) \right] \\ 
 & \hspace{20pt} \times \delta ^3 (\mathbf{p} -\mathbf{q} -e\mathbf{E} t_0 ) \\
 &\equiv \alpha _\mathbf{p} (t_0 ) \delta ^3 (\mathbf{p} -\mathbf{q} -e\mathbf{E} t_0 )
\end{split} \notag \\[+8pt]
\begin{split}
& \!
({}_- \! f_\mathbf{p} ^{(t_0 )} ,{}_+ \! f^{\mathrm{in} } _\mathbf{q} ) \\
 &= 2m\int \! d^3 x {}_- \! F_\mathbf{p} ^{(t_0 ) \dagger } (t_0 ,\mathbf{x} ) \sigma _3 {}_+ \! F_\mathbf{q} ^\mathrm{in} (t_0 ,\mathbf{x} ) \\
 &= -\frac{e^{-i\omega _\mathbf{p} t_0 } }{\sqrt{2\omega _\mathbf{p} } }
    \left[ \omega _\mathbf{p} {}_+ \! \phi _{\mathbf{p} -e\mathbf{E} t_0 } ^\mathrm{in} (t_0 ) 
    -i{}_+ \! \dot{\phi } _{\mathbf{p} -e\mathbf{E} t_0 } ^\mathrm{in} (t_0 ) \right] \delta ^3 (\mathbf{p} -\mathbf{q} -e\mathbf{E} t_0 ) \\
 &= -\frac{e^{-\frac{\pi }{4} a } }{(2eE)^{1/4} } \frac{e^{-i\omega _{\mathbf{p} } t_0 } }{\sqrt{2\omega _{\mathbf{p} } } }
    \left[ \left( \omega _{\mathbf{p} } -p_z  \right) 
           D^* _{-ia-\frac{1}{2} } (-e^{\frac{\pi }{4} i} \sqrt{\frac{2}{eE} } p_z  )
           -e^{\frac{\pi }{4} i} \sqrt{2eE} D^* _{-ia+\frac{1}{2} } 
          (-e^{\frac{\pi }{4} i} \sqrt{\frac{2}{eE} } p_z  ) \right] \\
 & \hspace{20pt} \times \delta ^3 (\mathbf{p} -\mathbf{q} -e\mathbf{E} t_0 ) \\
 &\equiv -\beta ^* _\mathbf{p} (t_0 ) \delta ^3 (\mathbf{p} -\mathbf{q} -e\mathbf{E} t_0 )
\end{split} \label{Bab} \\
({}_+ \! f_\mathbf{p} ^{(t_0 )} ,{}_- \! f^{\mathrm{in} } _\mathbf{q} ) 
 = 2m\int \! d^3 x {}_+ \! F_\mathbf{p} ^{(t_0 ) \dagger } (t_0 ,\mathbf{x} ) \sigma _3 {}_- \! F_\mathbf{q} ^\mathrm{in} (t_0 ,\mathbf{x} )
 = \beta _\mathbf{p} (t_0 ) \delta ^3 (\mathbf{p} -\mathbf{q} -e\mathbf{E} t_0 ) \ \notag \\
({}_- \! f_\mathbf{p} ^{(t_0 )} ,{}_- \! f^{\mathrm{in} } _\mathbf{q} )
 = 2m\int \! d^3 x {}_- \! F_\mathbf{p} ^{(t_0 ) \dagger } (t_0 ,\mathbf{x} ) \sigma _3 {}_- \! F_\mathbf{q} ^\mathrm{in} (t_0 ,\mathbf{x} )
 = -\alpha ^* _\mathbf{p} (t_0 ) \delta ^3 (\mathbf{p} -\mathbf{q} -e\mathbf{E} t_0 ) , \notag 
\end{gather}
\end{allowdisplaybreaks}
from which the relations
\begin{equation}
\begin{matrix}
{}_+ \! f^{\mathrm{in} } _\mathbf{p} (x) = \alpha _{\mathbf{p} +e\mathbf{E} t_0 } (t_0 ) 
                                     {}_+ \! f_{\mathbf{p} +e\mathbf{E} t_0 } ^{(t_0 )} (x)
                                     +\beta ^* _{\mathbf{p} +e\mathbf{E} t_0 } (t_0 )
                                     {}_- \! f_{\mathbf{p} +e\mathbf{E} t_0 } ^{(t_0 )} (x) \\[+7pt]
{}_- \! f^{\mathrm{in} } _\mathbf{p} (x) = \alpha ^* _{\mathbf{p} +e\mathbf{E} t_0 } (t_0 )
                                     {}_- \! f_{\mathbf{p} +e\mathbf{E} t_0 } ^{(t_0 )} (x) 
                                     +\beta _{\mathbf{p} +e\mathbf{E} t_0 } (t_0 )
                                     {}_+ \! f_{\mathbf{p} +e\mathbf{E} t_0 } ^{(t_0)} (x)
\end{matrix} \label{tenkai1}
\end{equation}
are obtained.
One can check that the equation
\begin{equation}
|\alpha _\mathbf{p} (t_0 ) |^2 -|\beta _\mathbf{p} (t_0 ) |^2 = 1
\end{equation}
holds using the explicit form of $\alpha _\mathbf{p} (t_0 ) $ and $\beta _\mathbf{p} (t_0 ) $, or the orthonormal conditions
for each mode functions.

Inserting \eqref{tenkai1} into \eqref{in-t0expansion}, we get 
\begin{equation}
\begin{matrix}
a_\mathbf{p} (t_0) = \alpha _\mathbf{p} (t_0 )a^{\mathrm{in} } _{\mathbf{p} -e\mathbf{E} t_0 } 
                                 +\beta _\mathbf{p} (t_0 )b ^{\mathrm{in} \dagger } _{-\mathbf{p} +e\mathbf{E} t_0 } \ \\[+7pt]
b_{-\mathbf{p} } ^{\dagger } (t_0) = \alpha ^* _\mathbf{p} (t_0 )b ^{\mathrm{in} \dagger } _{-\mathbf{p} +e\mathbf{E} t_0 } 
                                            +\beta ^* _\mathbf{p} (t_0 )a^{\mathrm{in} } _{\mathbf{p} -e\mathbf{E} t_0 } .
\end{matrix} \label{BtdBogo2}
\end{equation}
If one calculate the $z$-component of the canonical momentum operator using the second expression in
Eq.\eqref{in-t0expansion}, 
\begin{equation}
\hat{P} ^3 = \int \! d^3 p \left[ (p_z -eEt_0 )a_\mathbf{p} ^{\dagger } (t_0 ) a_\mathbf{p} (t_0 )
                +(p_z +eEt_0 )b_\mathbf{p} ^{\dagger } (t_0 ) b_\mathbf{p} (t_0 ) \right] 
\end{equation}
is obtained, 
from which $a_\mathbf{p} ^{\dagger } (t_0 ) $ [$b_\mathbf{p} ^{\dagger } (t_0 ) $] is identified as
the operator which creates a particle [antiparticle] with \textit{kinetic} momentum $\mathbf{p} $ at time $t_0 $.

Now, we can calculate the pair distribution function at time $t$.
Substituting \eqref{BtdBogo2} into \eqref{Bnp(t)def} yields
\begin{equation}
\begin{split}
n_\mathbf{p} (t) &= |\beta _\mathbf{p} (t)|^2 \\
           &= \frac{e^{-\frac{\pi }{2} a } }{\sqrt{2eE} 2\omega _{\mathbf{p} } }
              \left| \left( \omega _{\mathbf{p} } -p_z \right) D_{-ia-\frac{1}{2} } ( -e^{\frac{\pi }{4} i} 
              \sqrt{\frac{2}{eE} } p_z )
              -e^{-\frac{\pi }{4} i} \sqrt{2eE} D _{-ia+\frac{1}{2} } ( -e^{\frac{\pi }{4} i} \sqrt{\frac{2}{eE} } p_z ) \right| ^2 ,
\end{split} \label{Bn_p}
\end{equation}
in which $p_z $ denotes, as noted above, kinetic momentum. 

Notice that Eq.\eqref{Bn_p} is independent of $t$. In other words, the distribution is static.
This is a natural result because we take the constant electric field which has existed from infinite past.
Of course, each particles are accelerated by the electric field. The distribution, however, has already reached static state
since the electric field has continued to create particles from infinite past.
How the distribution reaches static state will be clarified in Sec.\ref{subsec:sudden}.

Furthermore, one may notice that the distribution \eqref{Bn_p} depends on the longitudinal momentum $p_z $. 
As mentioned in the introduction, longitudinal momentum dependence of the distribution
conflicts with the boost-invariance of the electric field. 
In our calculation, the boost-invariance is broken by the quantization procedure. 
To define a particle picture, canonical quantization is needed, which is done on an equal-time line
(or at least a space-like line). 
That is to say, particle is global concept and its definition breaks the boost invariance.
To say more physically: before particles are created, the system with a constant electric field is boost-invariant
along its direction. However, once a particle is created, it brings a special frame, namely 
center of mass frame \cite{Kajantie}.

\begin{figure}[t]
 \begin{tabular}{cc}
  \begin{minipage}{0.5\textwidth}
   \begin{center}
    \includegraphics[scale=0.50,bb=0 0 353 268]{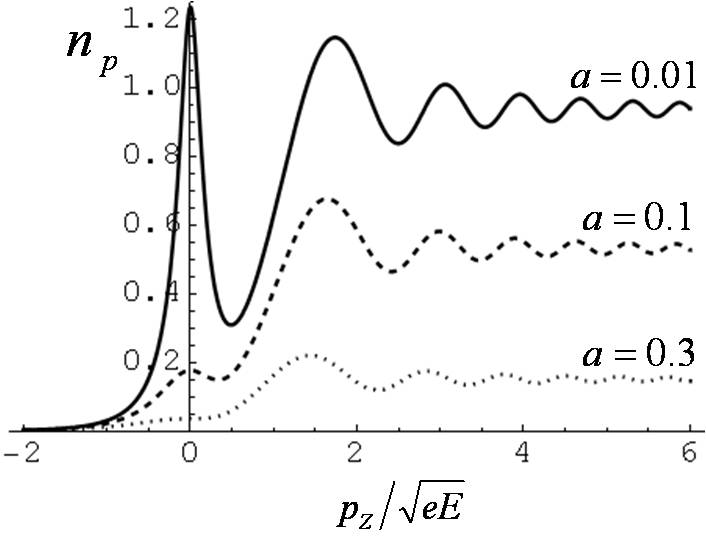}
   \end{center}
  \end{minipage} &
  \begin{minipage}{0.5\textwidth} 
   \begin{center}
    \includegraphics[scale=0.50,bb=0 0 353 272]{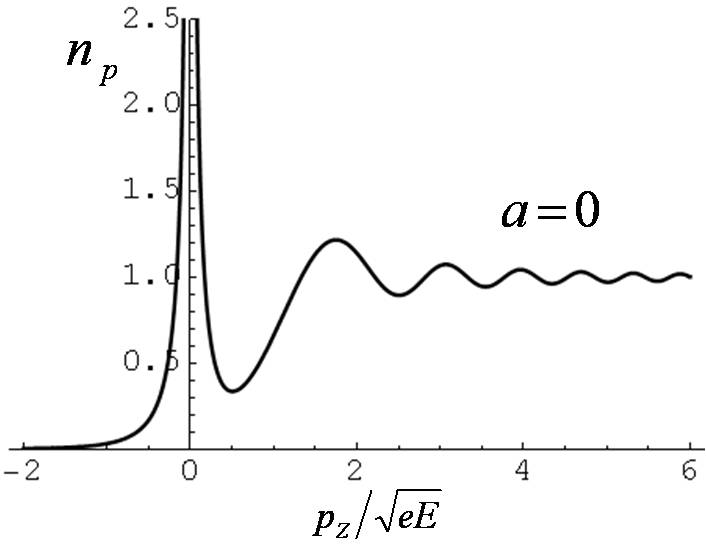}
   \end{center}
  \end{minipage}
 \end{tabular}
 \caption{Boson distributions}
 \label{fig:Bn_p}
 
\end{figure}

Now the following question may arise: 
\textit{Is there a special preferred frame to describe pair creation?}
Suppose that observer \textit{A} finds the distribution $n_\mathbf{p} $ and observer \textit{B} moves 
relatively to \textit{A} along the longitudinal direction. 
\textit{A} and \textit{B} are on an equal footing, because the electric field is boost-invariant along the longitudinal
direction and thus \textit{A} and \textit{B} feel the same electric field. 
Therefore, \textit{A} and \textit{B} must observe the same phenomena in each frame:
\textit{A} observes $n_\mathbf{p} $ and \textit{B} observes $n_{\mathbf{p} ^\prime } $,
where $\mathbf{p} ^\prime $ is the momentum in the \textit{B}'s frame and each $n_{(\cdot )} $ has 
the common functional form. 
However, if one supposes a phase space distribution is scalar with respect to Lorentz boost\footnote{This is
a matter of course by definition in classical theory.}, 
the distribution observed by \textit{B} takes the different functional form from that by \textit{A}:
\textit{A} observes $n_\mathbf{p} $, while \textit{B} observes $n_\mathbf{p} =n^\prime _{\mathbf{p} ^\prime } $. 
To say more specifically, if \textit{A} observes that a particle is created with longitudinal momentum of 0
(which is expected from the form of the distribution; Fig.\ref{fig:Bn_p}),
then \textit{B} observes this particle is created with non-zero momentum. 
In this situation, observer \textit{A} looks like a special observer.  

Of course, it is not the case. 
Although the functional form of 
the distribution $n_\mathbf{p} $ [Eq.\eqref{Bn_p}] is not boost-invariant along the longitudinal direction 
(i.e. $n_{(\cdot )} \neq n^\prime _{(\cdot )} $ ),
every observers observe the same distribution  
[modulo the ambiguity of particle picture in a background field]. 
Observer \textit{A} finds the distribution $n_\mathbf{p} $ based on the particle picture defined 
on an equal-time line of \textit{A}'s frame:
$n_\mathbf{p} = \langle 0,\mathrm{in} |a_\mathbf{p} ^\dagger (t) a_\mathbf{p} (t) |0,\mathrm{in} \rangle \frac{(2\pi )^3 }{V} $.
On the other hand, observer \textit{B} finds the distribution with the same functional form with \textit{A}'s: 
$n_{\mathbf{p} ^\prime } = \langle 0,\mathrm{in} ^\prime |a_{\mathbf{p} ^\prime } ^\dagger (t^\prime ) 
a_{\mathbf{p} ^\prime } (t^\prime ) |0,\mathrm{in} ^\prime \rangle \frac{(2\pi )^3 }{V^\prime } $.
This is because \textit{B} uses the particle picture defined on an equal-time line of \textit{B}'s frame, 
which is different from \textit{A}'s.
Thus, if one simply boosts the \textit{A}'s distribution to the \textit{B}'s frame, 
it is different from \textit{B}'s distribution 
because the boosted \textit{A}'s equal-time line is different from the \textit{B}'s. 
That is to say, our distribution \eqref{Bnp(t)def} is not scalar with respect to Lorentz boost unlike that in classical
theory because quantization procedure requires specifying a Lorentz frame.
If we take account of the fact that quantization procedure is not boost-invariant 
in the presence of a pair-creating background field, 
we can get the distribution, of which functional form itself is not boost-invariant, however,
which is compatible with the boost-invariance of the background field. 

\vspace{11pt}
In Fig.\ref{fig:Bn_p} the pair distribution function \eqref{Bn_p} is plotted as a function of 
the longitudinal momentum $p_z $ for several values of $a=\frac{m_\mathrm{T} ^2 }{2eE} $.
Hereafter we show all figures with dimension-less values which are scaled by $\sqrt{eE} $ 
[or $\sqrt{eE_0 } $ in a non-steady field case, where $E_0 $ is an initial field strength].

The shape of the distributions can be understood roughly as follows:
particles [antiparticles] are created with approximately 0-momentum, 
and accelerated by the electric field to get positive [negative] momentum. 
This process have been continued from infinite past and
makes, roughly speaking, a step function-like shape of the distribution.

The asymptotic forms of the distribution \eqref{Bn_p} at $p_z \to \pm \infty $ are
\begin{equation}
\begin{matrix}
n_\mathbf{p} \rightarrow e^{-2\pi a} & \ (p_z \to +\infty ) \ \\
n_\mathbf{p} \rightarrow 0 & \ (p_z \to -\infty ) .
\end{matrix}
\end{equation}
The value at $p_z \to +\infty $ agrees with which has been obtained in Sec.\ref{sec:quanti}, where the asymptotic WKB criterion
was used. This agreement is consistent with the opinion made in Sec.\ref{sec:quanti}, that we can see only particles
with infinite longitudinal kinetic momentum using the asymptotic WKB criterion.

\begin{figure}
\begin{center}
\includegraphics[scale=0.50,bb=0 0 353 272]{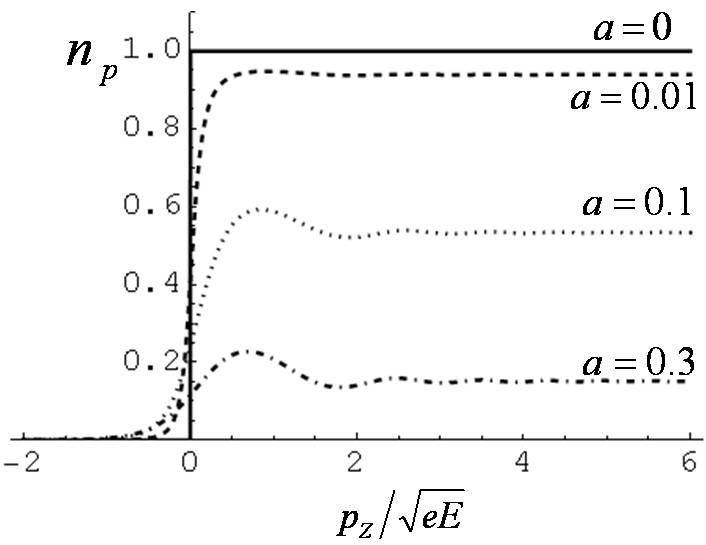} 
\end{center}
 \vskip -\lastskip \vskip -3pt
\caption{Fermion distributions}
\label{fig:Fn_p}
\end{figure}

One notable point seen in Fig.\ref{fig:Bn_p} is divergence of the distribution of massless particle 
at $p\equiv |\mathbf{p} | =0 $ :
\begin{equation}
n_\mathbf{p} \nyoro _{\begin{smallmatrix} m=0 \\ p\sim 0 \end{smallmatrix} } 
 \frac{\left \{ \Gamma \left( \frac{3}{4} \right) \right \} ^2 \sqrt{eE} }{2\pi p } . \label{diver}
\end{equation}
This divergence comes from the fact that the energy cost to create massless particles is 0.
If there is an electric field for any strength, it creates massless particles endlessly.
It is adequate to treat a massless and 0-momentum mode as {\lq \lq}field" rather than {\lq \lq}particle".   
Thus, if one interprets this mode as particles and counts their number, it is infinite.
In this sense, the divergence \eqref{diver} is of the same quality with that of the Bose--Einstein distribution
of massless particle:
\begin{equation}
n_\mathbf{p} ^\mathrm{BE} = \frac{1}{e^{\omega _p /T} -1} \nyoro _{\begin{smallmatrix} m=0 \\ p\sim 0 \end{smallmatrix} }
 \frac{T}{p} . \label{BE}
\end{equation}  
Therefore, if one watches only the behavior around 0-momentum, 
the distribution of massless particles created in the constant
electric filed takes the same form as that of the Bose distribution. 
The electric field plays the role of a thermal bath, of which {\lq \lq}temperature" is 
$\left \{ \Gamma \left( \frac{3}{4} \right) \right \} ^2 \sqrt{eE} /2\pi $.

\begin{figure}[t]
 \begin{tabular}{ccc}
  \begin{minipage}{0.33\textwidth}
   \begin{center}
    \includegraphics[scale=0.34,bb=0 0 353 279]{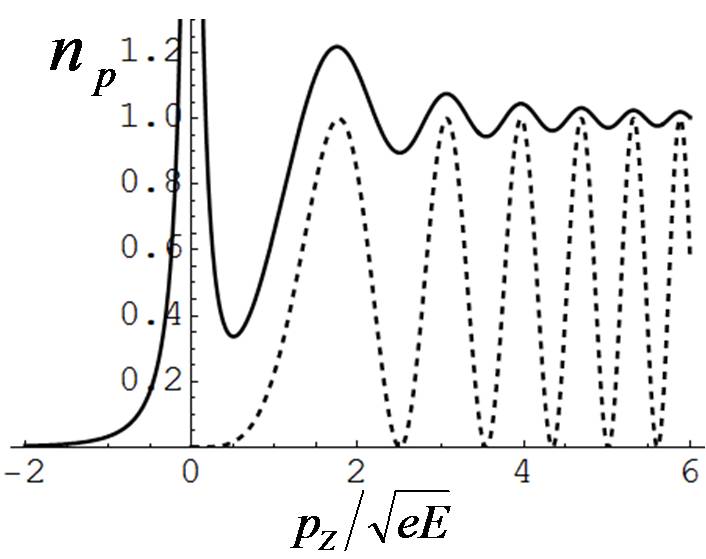} \\
    {\footnotesize (a) $a=0$ }
   \end{center}
  \end{minipage} &
  \begin{minipage}{0.33\textwidth}
   \begin{center}
    \includegraphics[scale=0.34,bb=0 0 353 291]{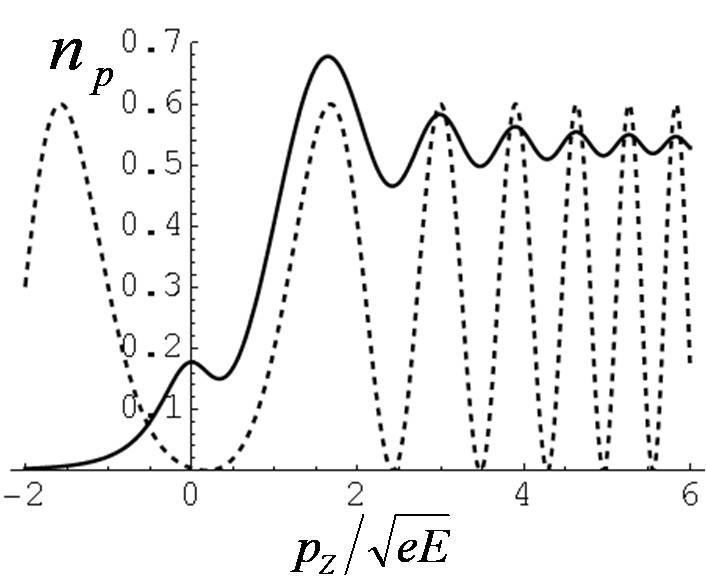} \\
    {\footnotesize (b) $a=0.1$ }
   \end{center}
  \end{minipage} &
  \begin{minipage}{0.33\textwidth}
   \begin{center}
    \includegraphics[scale=0.34,bb=0 0 353 291]{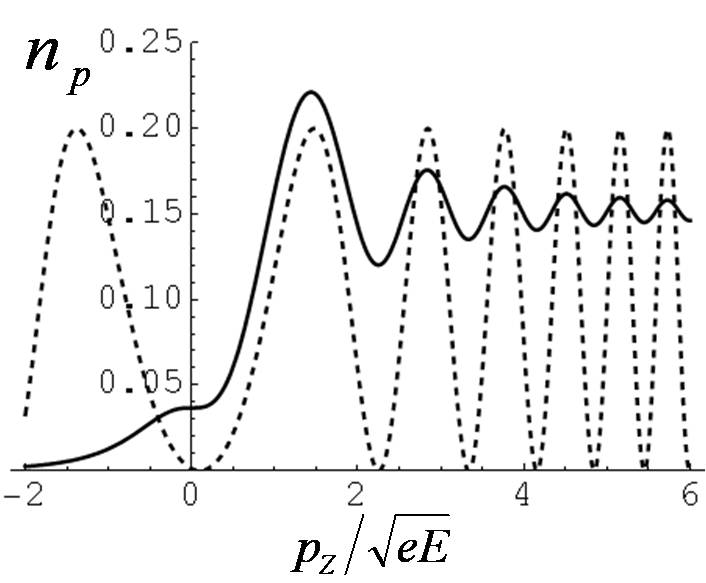} \\
    {\footnotesize (c) $a=0.3$ }
   \end{center}
  \end{minipage}
 \end{tabular}
 \caption{Comparison between the distribution \eqref{Bn_p} and the {\lq \lq}WKB interference term" \eqref{Binterference},
          which is plotted with an arbitrary vertical scale factor. }
 \label{fig:Binterference}
\end{figure}

Other notable point seen in the distribution is oscillation with respect to $p_z $.
Its periodicity is the order of $\sqrt{eE} $ and slightly depends on $a$.
This oscillation may be caused by interference between particles and antiparticles.
Now, particles and antiparticles are created everywhere in whole space and accelerated by the electric field;
particles are accelerated to the same direction with the electric field, while antiparticles to the opposite direction
with the electric field.
The WKB wave functions representing accelerated particles and antiparticles are $\exp \left[iS_{cl} (t) \right] $ and 
$\exp \left[-iS_{cl} (t) \right] $ respectively,
where $S_{cl} (t) $ is a classical action of a particle under the electric field \eqref{kotensayou}.
Because the concepts of particle and antiparticle are mixed by the Bogoliubov transformation in the electric field, 
particles and antiparticles can interfere with each other through the electric field.
These interference may be represented as
\begin{equation}
|e^{iS_{cl} (t)} -e^{-iS_{cl} (t)} |^2 = 2-2\cos \left[ 2S_{cl} (t)\right] . \label{Binterference}
\end{equation}
After changing the variable from $t$ to $p_z $ through the classical equation of motion $p_z =eEt$,
we plot Eq.\eqref{Binterference}
together with the distribution \eqref{Bn_p} in Fig.\ref{fig:Binterference}. 
The oscillatory behavior and its slight dependence on $a$ are very well agreed, 
although a physical origin of Eq.\eqref{Binterference}, 
especially the reason why the relative phase between $e^{iS_{cl} (t)} $ and $e^{-iS_{cl} (t)} $ is $-1$ is not clear.
This agreement is limited in the region of $p_z >0 $ because particles are created with approximately 0-momentum and 
accelerated to get positive momentum.

\begin{figure}[t]
 \begin{tabular}{cc}
  \begin{minipage}{0.5\textwidth}
   \begin{center}
    \includegraphics[scale=0.46,bb=0 0 356 220]{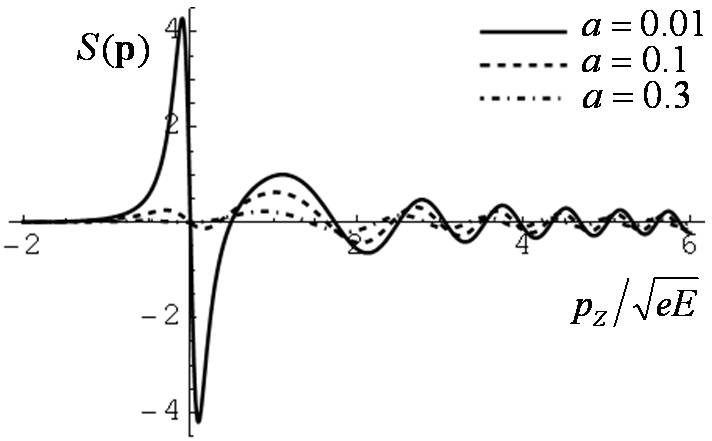}
   \end{center}
  \end{minipage} &
  \begin{minipage}{0.5\textwidth}
   \begin{center}
    \includegraphics[scale=0.46,bb=0 0 356 250]{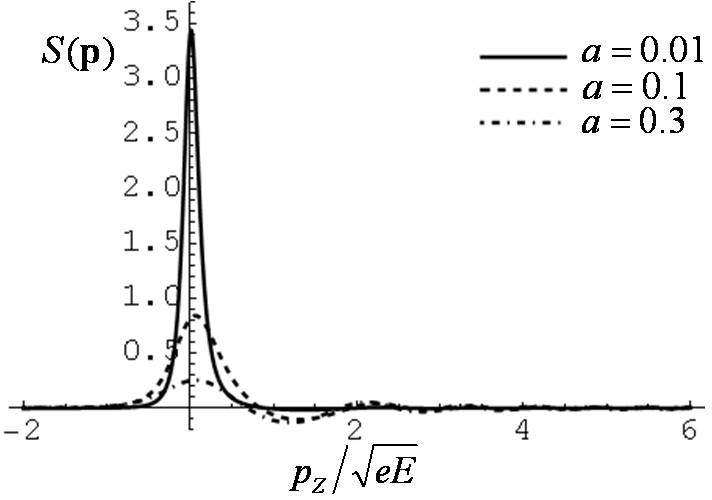}
   \end{center}
  \end{minipage} \\
  {\footnotesize (a) bosons } & {\footnotesize (b) fermions }
 \end{tabular}
 \caption{The source terms}
 \label{fig:source}
\end{figure}

\vspace{11pt}
Finally, let us derive a source term of the Boltzmann--Vlasov equation with the assumption that the distribution function
\eqref{Bn_p} satisfies the Boltzmann--Vlasov equation
\begin{equation}
\frac{\partial }{\partial t} n_\mathbf{p} (x) +\mathbf{v} \cdot \mathbf{\nabla } n_\mathbf{p} (x)
 +eE\frac{\partial }{\partial p_z } n_\mathbf{p} (x) = S(x,\mathbf{p} ) . \label{Boltzmann}
\end{equation}
Substituting the distribution \eqref{Bn_p} into Eq.\eqref{Boltzmann},
we can get the source term as follows: 
\begin{equation}
\begin{split}
S(\mathbf{p} ) &= -eE\frac{p_z }{\omega _p ^2 } n_\mathbf{p}
 +\sqrt{\frac{eE}{2} } \frac{e^{-\frac{\pi }{2} a} }{2\omega _p }
 \left\{ 2(\omega _p -p_z )\frac{p_z }{\omega _p }
 \left| D_{-ia-\frac{1}{2} } \right| ^2 
 -2\sqrt{2eE} \frac{p_z }{\omega _p } \mathrm{Re} \left[ e^{\frac{\pi }{4} i} 
 D_{-ia+\frac{1}{2} } ^* 
 D_{-ia-\frac{1}{2} }  \right] \right. \\
& \hspace{2cm} \left.
 +\sqrt{\frac{2}{eE} } \left[ (\omega _p -p_z )^2 +2(\omega _p -p_z )p_z -m_{\mathrm{T} } ^2 \right]
 \mathrm{Re} \left[ e^{\frac{\pi }{4} i} 
 D_{-ia-\frac{1}{2} } ^* D_{-ia+\frac{1}{2} }  \right] \right\} ,
\end{split}
\end{equation}
where the arguments of parabolic cylinder functions, $(-e^{\frac{\pi }{4} i} \sqrt{\frac{2}{eE} } p_z )$, are omitted.
The source terms for bosons are shown in Fig.\ref{fig:source}(a), in which oscillatory behavior appears.
Similar oscillatory source term has been obtained also in Ref.\cite{Kluger1998,Rau,Rau-Muller,Smolyansky}.

It will be convenient for a latter discussion to express the source term using the Bogoliubov coefficients.
Differentiating $\beta _\mathbf{p} ^* (t) = \frac{e^{-i\omega _\mathbf{p} t } }{\sqrt{2\omega _\mathbf{p} } }
\left[ \omega _\mathbf{p} {}_+ \! \phi _{\mathbf{p} -e\mathbf{E} t } ^\mathrm{in} (t)
-i{}_+ \! \dot{\phi } _{\mathbf{p} -e\mathbf{E} t } ^\mathrm{in} (t) \right] $ [Eq.\eqref{Bab}] 
by $\frac{d}{dt} =\frac{\partial }{\partial t} +eE\frac{\partial }{\partial p_z } $
and using the equation for ${}_\pm \! \phi _\mathbf{p} ^\mathrm{in} $ [Eq.\eqref{Schr}], one can verify
\begin{equation}
\begin{split}
\frac{d}{dt} \beta _\mathbf{p} ^* (t)
 &= \frac{\partial \beta _\mathbf{p} ^* }{\partial t} +eE\frac{\partial \beta _\mathbf{p} ^* }{\partial p_z } \\
 &= -ieE\frac{p_z }{\omega _\mathbf{p} } t \beta _\mathbf{p} ^* (t)
    +eE\frac{p_z }{2\omega _\mathbf{p} ^2 } e^{-2i\omega _\mathbf{p} t} \alpha _\mathbf{p} (t) .
\end{split}
\end{equation}
With this equation, we can derive the following expression for the source term:
\begin{equation}
\begin{split}
S(t,\mathbf{p} ) &= \frac{d}{dt} |\beta _\mathbf{p} (t) |^2 
          = 2\mathrm{Re} \left[ \beta _\mathbf{p} (t) \frac{d}{dt} \beta _\mathbf{p} ^* (t) \right] \\
         &= eE\frac{p_z }{\omega _\mathbf{p} ^2 } 
            \mathrm{Re} \left[ e^{-2i\omega _\mathbf{p} t} \alpha _\mathbf{p} (t) \beta _\mathbf{p} (t) \right] . \label{ab_sou}
\end{split}
\end{equation}
This is a general expression under a homogeneous and constant electric field. 

%%%%%%%%%%%%%%%%%%%%%%%%%
\subsubsection{Collinear electric and magnetic field} \label{subsubsec:PLboson-coll}
We derive a distribution function of created particles in a collinear electric and magnetic field 
in the same way as that in the pure electric field.

There is a difficulty to define instantaneous positive and negative frequency solutions:
The gauge potential at instant of time $t=t_0 $, namely $A(t_0 ,\mathbf{x} ) = (-By,0,-Et_0 ) $ is not pure gauge.
It gives the constant magnetic field. 
Mode solutions under the gauge $A(t_0 ,\mathbf{x} ) $ is not simple plane wave solutions: 
\begin{equation}
{}_\pm \! f_\mathbf{p} ^{(t_0 )} (x) = \frac{1}{\sqrt{2\omega _p } } e^{\mp i\omega _p t } 
           \frac{1}{2\pi } e^{ip_x x +i(p_z -eEt_0 )z}
           \sqrt{\frac{L}{2\pi } } \left( \frac{eB}{\pi } \right) ^\frac{1}{4} \frac{1}{\sqrt{n!} } D_n (\eta )
           . \label{t_0planeEB}
\end{equation} 
However, in spite of the presence of the magnetic field,
we can use these solutions as instantaneous positive and negative frequency solutions.
That is because the magnetic field does not affect separation of positive and negative frequency solutions
[$e^{-i\omega _p t } $ and $e^{+i\omega _p t } $], which means that a pure magnetic field cannot create particles.

The use of the instantaneous positive and negative frequency solution \eqref{t_0planeEB} instead of \eqref{t_0plane}
simply brings the replacement \eqref{EtoEB}.
Therefore, the resultant distribution function is \eqref{Bn_p} with the replacement 
$a\to a_n =\frac{m^2 +(2n+1)eB}{2eE} $.

%%%%%%%%%%%%%%%%%%%%%%%%%%%%%%%%%%%%%%%%%%
\subsection{Fermions in a constant electromagnetic field} \label{subsec:PLfermion}
\subsubsection{Pure electric field} \label{subsubsec:PLfermion-pure}
In this subsection, we calculate a pair distribution function of fermions in the pure electric field 
using the instantaneous particle picture.

The instantaneous positive and negative frequency solutions at time of $t=t_0 $ under the gauge potential 
$A^\mu =(0,0,0,-Et) $ is obtained from the free-field solutions \eqref{freeDirac} by the gauge transformation
$A^3 =0 \to -Et_0 $:
\begin{equation}
{}_\pm \! \psi _{\mathbf{p} s} ^{(t_0 )} (x) = {}_\pm \! \psi _{\mathbf{p} s} ^\mathrm{free} (x) e^{-ieEt_0 z} \ \ 
(s=\uparrow ,\downarrow ) .
\end{equation}
The inner products between these solutions and the in-solutions ${}_\pm \! \psi _{\mathbf{p} s} ^\mathrm{in} $ [Eq.\eqref{inpsi}] are
\begin{allowdisplaybreaks}
\begin{gather}
\begin{split}
& \!
({}_+ \! \psi _{\mathbf{p} s} ^{(t_0 )} ,{}_+ \! \psi _{\mathbf{q} s^\prime } ^\mathrm{in} ) 
  = \int \! d^3 x {}_+ \! \psi _{\mathbf{p} s} ^{(t_0 ) \dagger } (x) {}_+ \! \psi _{\mathbf{q} s^\prime } ^\mathrm{in} (x) \\
 &= \frac{e^{-\frac{\pi }{4} a} }{2\sqrt{eE} } \frac{e^{i\omega _p t_0 } }{\sqrt{\omega _p (\omega _p -p_z )} }
    \left[ m_{\mathrm{T} } ^2 D_{-ia-1} ^* (-e^{\frac{\pi }{4} i } \sqrt{\frac{2}{eE} } p_z )
    +e^{\frac{\pi }{4} i} \sqrt{2eE} (\omega _p -p_z ) D_{-ia} ^* (-e^{\frac{\pi }{4} i } \sqrt{\frac{2}{eE} } p_z ) \right] \\
 & \hspace{20pt} \times \delta _{s s^\prime } \delta ^3 (\mathbf{p} -\mathbf{q} -e\mathbf{E} t_0 ) \\
 &\equiv \alpha _\mathbf{p} (t_0 ) \delta _{s s^\prime } \delta ^3 (\mathbf{p} -\mathbf{q} -e\mathbf{E} t_0 )
\end{split} \notag \\
\begin{split}
& \!
({}_- \! \psi _{\mathbf{p} s} ^{(t_0 )} ,{}_+ \! \psi _{\mathbf{q} s^\prime } ^\mathrm{in} ) 
  = \int \! d^3 x {}_- \! \psi _{\mathbf{p} s} ^{(t_0 ) \dagger } (x) {}_+ \! \psi _{\mathbf{q} s^\prime } ^\mathrm{in} (x) \\
 &= \frac{e^{-\frac{\pi }{4} a} }{2\sqrt{eE} } \frac{e^{-i\omega _p t_0 } }{\sqrt{\omega _p (\omega _p +p_z )} }
    \left[ m_{\mathrm{T} } ^2 D_{-ia-1} ^* (-e^{\frac{\pi }{4} i } \sqrt{\frac{2}{eE} } p_z )
   -e^{\frac{\pi }{4} i} \sqrt{2eE} (\omega _p +p_z ) D_{-ia} ^* (-e^{\frac{\pi }{4} i } \sqrt{\frac{2}{eE} } p_z ) \right] \\
 & \hspace{20pt} \times \delta _{s s^\prime } \delta ^3 (\mathbf{p} -\mathbf{q} -e\mathbf{E} t_0 ) \\
 &\equiv -\beta _\mathbf{p} ^* (t_0 ) \delta _{s s^\prime } \delta ^3 (\mathbf{p} -\mathbf{q} -e\mathbf{E} t_0 )
\end{split} \label{Dab} \\
({}_+ \! \psi _{\mathbf{p} s} ^{(t_0 )} ,{}_- \! \psi _{\mathbf{q} s^\prime } ^\mathrm{in} )
 = \int \! d^3 x {}_+ \! \psi _{\mathbf{p} s} ^{(t_0 ) \dagger } (x) {}_- \! \psi _{\mathbf{q} s^\prime } ^\mathrm{in} (x)
 = \beta _\mathbf{p} (t_0 ) \delta _{s s^\prime } \delta ^3 (\mathbf{p} -\mathbf{q} -e\mathbf{E} t_0 ) \notag \\
({}_- \! \psi _{\mathbf{p} s} ^{(t_0 )} ,{}_- \! \psi _{\mathbf{q} s^\prime } ^\mathrm{in} )
 = \int \! d^3 x {}_- \! \psi _{\mathbf{p} s} ^{(t_0 ) \dagger } (x) {}_- \! \psi _{\mathbf{q} s^\prime } ^\mathrm{in} (x)
 = \alpha _\mathbf{p} ^* (t_0 ) \delta _{s s^\prime } \delta ^3 (\mathbf{p} -\mathbf{q} -e\mathbf{E} t_0 ) , \notag 
\end{gather}
\end{allowdisplaybreaks}
from which the relations
\begin{equation}
\begin{matrix}
{}_+ \! \psi _{\mathbf{p} s} ^\mathrm{in} (x) 
 = \alpha _{\mathbf{p} +e\mathbf{E} t_0 } (t_0 ) {}_+ \! \psi _{\mathbf{p} +e\mathbf{E} t_0 s} ^{(t_0 )} (x)
   -\beta _{\mathbf{p} +e\mathbf{E} t_0 } ^* (t_0 ) {}_- \! \psi _{\mathbf{p} +e\mathbf{E} t_0 s} ^{(t_0 )} (x) \\
{}_- \! \psi _{\mathbf{p} s} ^\mathrm{in} (x)
 = \alpha _{\mathbf{p} +e\mathbf{E} t_0 } ^* (t_0 ) {}_- \! \psi _{\mathbf{p} +e\mathbf{E} t_0 s} ^{(t_0 )} (x)
   +\beta _{\mathbf{p} +e\mathbf{E} t_0 } (t_0 ) {}_+ \! \psi _{\mathbf{p} +e\mathbf{E} t_0 s} ^{(t_0 )} (x)
\end{matrix} \label{DmBogo1}
\end{equation}
are followed\footnote{These expansions is definite by themselves and the additional condition \eqref{dttdBogo} is not 
necessary, because the Dirac equation is a first-order differential equation and the associated inner product 
\eqref{Dinner} does not contain time-differentiation.}.

The field expansions
\begin{equation}
\begin{split}
\psi (x) &= \sum _{s=\uparrow ,\downarrow } \int \! d^3 p 
  \left[ {}_+ \! \psi _{\mathbf{p} s } ^\mathrm{in} (x) a_{\mathbf{p} s } ^\mathrm{in}
  + {}_- \! \psi _{\mathbf{p} s } ^\mathrm{in} (x) b_{-\mathbf{p} s } ^{\mathrm{in} \dagger } \right] \\
 &= \sum _{s=\uparrow ,\downarrow } \int \! d^3 p 
  \left[ {}_+ \! \psi _{\mathbf{p} s} ^{(t_0 )} (x) a_{\mathbf{p} s} (t_0 )
    + {}_- \! \psi _{\mathbf{p} s} ^{(t_0 )} (x) b_{-\mathbf{p} s} ^{\dagger } (t_0 ) \right]
\end{split}
\end{equation}
and Eq.\eqref{DmBogo1} yield the relations between the creation/annihilation operator of the in-particle picture 
and that of the instantaneous particle picture of $t=t_0 $: 
\begin{equation}
\begin{matrix}
a_{\mathbf{p} s} (t_0 ) = \alpha _\mathbf{p} (t_0 ) a_{\mathbf{p}-e\mathbf{E} t_0 s } ^\mathrm{in}
                      +\beta _\mathbf{p} (t_0 ) b_{-\mathbf{p} +e\mathbf{E} t_0 s} ^{\mathrm{in} \dagger } \\[+7pt]
b_{-\mathbf{p} s} ^\dagger (t_0 ) = \alpha _\mathbf{p} ^* (t_0 ) b_{-\mathbf{p} +e\mathbf{E} t_0 s} ^{\mathrm{in} \dagger } 
                                -\beta _\mathbf{p} ^* (t_0 ) a_{\mathbf{p} -e\mathbf{E} t_0 s } ^\mathrm{in}
\end{matrix}
, \ (s,s^\prime =\uparrow ,\downarrow ) .
\end{equation}
Notice that this Bogoliubov transformation does not mix the two spin states and the Bogoliubov coefficients 
are common to both spin-$\uparrow $ and $\downarrow $ state because we use the same spin basis for both the in-solutions
and the $t_0 $-solutions.
Thus, we can get the same distribution function from any of
\begin{equation}
\begin{matrix}
\langle 0,\mathrm{in} |a_{\mathbf{p} \uparrow } ^\dagger (t) a_{\mathbf{p} \uparrow } (t)|0,\mathrm{in} \rangle & \ &
\langle 0,\mathrm{in} |a_{\mathbf{p} \downarrow } ^\dagger (t) a_{\mathbf{p} \downarrow } (t)|0,\mathrm{in} \rangle \\
\langle 0,\mathrm{in} |b_{-\mathbf{p} \uparrow } ^\dagger (t) b_{-\mathbf{p} \uparrow } (t)|0,\mathrm{in} \rangle & \ &
\langle 0,\mathrm{in} |b_{-\mathbf{p} \downarrow } ^\dagger (t) b_{-\mathbf{p} \downarrow } (t)|0,\mathrm{in} \rangle .
\end{matrix} \notag
\end{equation} 
Then, the pair distribution function at time $t$ is
\begin{equation}
\begin{split}
n_\mathbf{p} (t) &= |\beta _\mathbf{p} (t)|^2 \\
           &= \frac{e^{-\frac{\pi }{2} a} }{4eE\omega _p (\omega _p +p_z )}
              \left| m_{\mathrm{T} } ^2 D_{-ia-1} (-e^{\frac{\pi }{4} i } \sqrt{\frac{2}{eE} } p_z )
              -e^{-\frac{\pi }{4} i} \sqrt{2eE} (\omega _p +p_z ) 
              D_{-ia} (-e^{\frac{\pi }{4} i } \sqrt{\frac{2}{eE} } p_z ) \right| ^2 .
\end{split} \label{Dn_p}
\end{equation}
This is independent of time with the same reason as that of boson.

We plot the distribution \eqref{Dn_p} as a function of the longitudinal momentum $p_z $ for several values of 
$a=\frac{m_\mathrm{T} ^2 }{2eE} $ in Fig.\ref{fig:Fn_p}.

The fermion distribution shows little oscillation compared with boson. 
This is because of Pauli's exclusion principal, which makes collective behavior of fermions {\lq \lq}particle-like",
while bosons behave like as {\lq \lq}field-like". 
This {\lq \lq}particle-like" nature of fermions enhances the step function-like forms of the distributions.
Especially in the case of $a=0$, massless fermions are created as many as possible 
under the restriction of Pauli's exclusion principal, therefore, 
the distribution becomes the step function exactly: 
\begin{equation}
n_\mathbf{p} \wa _{a=0} \frac{|p_z |+p_z }{2|p_z |} = \theta (p_z ) .
\end{equation}
On the other hand, bosons which have {\lq \lq}field-like" nature interfere during acceleration 
and their distribution shows oscillation. 

Asymptotic behavior of the distribution \eqref{Dn_p} at $p_z \to \pm \infty $ are just the same with those of bosons:
\begin{equation}
\begin{matrix}
n_\mathbf{p} \rightarrow e^{-2\pi a} & \ (p\to +\infty ) \ \\
n_\mathbf{p} \rightarrow 0 & \ (p\to -\infty ). 
\end{matrix}
\end{equation}
This is because, as already noted on Sec.\ref{subsec:fermion}, high energy limit ($p_z \to \pm \infty $) blinds
spin statistics. 
Meantime, low momentum region greatly reflects spin statistics.

\vspace{11pt}
The source term obtained by substituting the distribution \eqref{Dn_p} into the Boltzmann--Vlasov equation 
\eqref{Boltzmann} is
\begin{equation}
\begin{split}
S(\mathbf{p} ) &= eE\frac{p_z }{\omega _p ^2 } \left( \frac{1}{2} -n_\mathbf{p} \right)
 +eE\frac{e^{-\frac{\pi }{2} a} }{2\omega _p }
 \left\{ \left| D_{-ia} (-e^{\frac{\pi }{4} i} \sqrt{\frac{2}{eE} } p_z )\right| ^2
 -a\left| D_{-ia-1} (-e^{\frac{\pi }{4} i} \sqrt{\frac{2}{eE} } p_z )\right| ^2 \right\} ,
\end{split}
\end{equation}
which is plotted in Fig.\ref{fig:source}(b).
The source terms for fermions have a peak at $p_z =0 $. 
Especially if $a=0 $, the source term becomes
\begin{equation}
S(\mathbf{p} ) \wa _{a=0 } eE\delta (p_z ) .
\end{equation}
This delta function source term has been introduced phenomenologically in 
Ref.\cite{Kajantie,Gatoff}\footnote{However, the argument of delta function is not momentum but rapidity 
in these reference.}.
On the other hand, when $a\neq 0 $, the peak at $p_z =0 $ is broadened by quantum fluctuation.

Finally let us derive a general expression of the source term under a homogeneous electric field. 
Using expressions
\begin{equation}
{}_\pm \! \psi _\mathbf{p} ^{(t_0 )} (x) = \frac{1}{\sqrt{2(\omega _p \mp p_z )} } \left[m\Gamma _1
 -(p_x +ip_y )\Gamma _2 +(\pm \omega _p -p_z )\Gamma _3 \right] {}_\pm \! \phi _\mathbf{p} ^{(t_0 )} (x) ,
\end{equation}
and
\begin{equation}
{}_+ \! \psi _\mathbf{p} ^\mathrm{in} (x) = \mathcal{N} _\mathbf{p} \left[ m\Gamma _1 
 -(p_x +ip_y )\Gamma _2 +(i\partial _0 -p_z -eEt) \Gamma _3 \right] 
 {}_+ \! \phi _\mathbf{p} ^\mathrm{in} (t) \frac{e^{i\mathbf{p} \cdot \mathbf{x} } }{\sqrt{(2\pi )^3 } } ,
\end{equation}
in which ${}_+ \! \phi _\mathbf{p} ^\mathrm{in} (t) $ is the positive frequency solution of {\lq \lq}Klein--Gordon equation" \eqref{2ndDirac}
and $\mathcal{N} _\mathbf{p} $ is a normalization constant,
one can get the equations
\begin{gather}
\alpha _\mathbf{p} (t) = \frac{\mathcal{N} _{\mathbf{p} -e\mathbf{E} t} e^{i\omega _p t} }{\sqrt{\omega _p (\omega _p -p_z )} }
 \left[ (\omega _p -p_z )(i{}_+ \! \dot{\phi } _{\mathbf{p} -e\mathrm{E} t} ^\mathrm{in} -p_z {}_+ \! \phi _{\mathbf{p} -e\mathrm{E} t} ^\mathrm{in} )
 +m_\mathrm{T} ^2 {}_+ \! \phi _{\mathbf{p} -e\mathrm{E} t} ^\mathrm{in} ) \right] \label{gDa} \\
\beta _\mathbf{p} ^* (t) = \frac{\mathcal{N} _{\mathbf{p} -e\mathbf{E} t} e^{-i\omega _p t} }{\sqrt{\omega _p (\omega _p +p_z )} }
 \left[ (\omega _p +p_z )(i{}_+ \! \dot{\phi } _{\mathbf{p} -e\mathrm{E} t} ^\mathrm{in} -p_z {}_+ \! \phi _{\mathbf{p} -e\mathrm{E} t} ^\mathrm{in} )
 -m_\mathrm{T} ^2 {}_+ \! \phi _{\mathbf{p} -e\mathrm{E} t} ^\mathrm{in} ) \right] \label{gDb} . 
\end{gather}
Differentiating \eqref{gDb} with respect to $t$ and using Eq.\eqref{2ndDirac} yield
\begin{equation}
\begin{split}
\frac{d}{dt} \beta _\mathbf{p} ^* (t) 
 &= \frac{\partial }{\partial t} \beta _\mathbf{p} ^* (t) +eE\frac{\partial }{\partial p_z } \beta _\mathbf{p} ^* (t) \\
 &= -ieE\frac{p_z }{\omega _p } t\beta _\mathbf{p} ^* (t) +eE\frac{m_\mathrm{T} }{2\omega _p ^2 }
    e^{-2i\omega _p t} \alpha _\mathbf{p} (t) .
\end{split}
\end{equation}
Then, the source term 
\begin{equation}
S(t,\mathbf{p} ) = \frac{d}{dt} |\beta _\mathbf{p} (t) |^2
           = eE\frac{m_\mathrm{T} }{2\omega _p ^2 } \mathrm{Re} 
             \left[ e^{-2i\omega _p t} \alpha _\mathbf{p} (t) \beta _\mathbf{p} (t) \right]
\end{equation}
is derived.
This will be useful for a latter discussion on a polarization current.

%%%%%%%%%%%%%%%%%%%%%%%%%
\subsubsection{Collinear electric and magnetic field} \label{subsubsec:PLfermion-coll}
Similarly to the boson case, we use solutions under the gauge potential $\mathbf{A} (t_0 ,\mathbf{x} ) = (-By,0,-Et_0 ) $
as instantaneous positive and negative frequency solutions at time of $t=t_0 $.
They are given as follows: 
\begin{equation}
{}_\pm \! \psi _{\mathbf{p} s} ^{(t_0 )} (x) = {}_\pm \! \psi _{\mathbf{p} s} ^B (x) e^{-iEt_0 z} \ \ 
(s=\uparrow ,\downarrow ), \label{psi^t0_EM}
\end{equation}
in which ${}_\pm \! \psi _{\mathbf{p} s} ^B $ is solutions under the gauge $\mathbf{A} = (-By,0,0) $ [Eq.\eqref{psi_pM}].
The pair distribution function can be obtained by taking the inner product between the in-solutions \eqref{psi^in_EM}
and the instantaneous solutions \eqref{psi^t0_EM}. 
In the same way as Sec.\ref{subsubsec:fermion-coll}, the longitudinal magnetic field results in the replacement \eqref{F:EtoEB}.

%%%%%%%%%%%%%%%%%%%%%%%%%%%%%%%%%%%%%%%%%%%%%%%%%%%%%%%%%%%%%%%%%%%%%%%%%%%%%%%%%%%%%%%%%
\section{Non-steady electric fields} \label{sec:Nonste}
In the previous section, we revealed the momentum distribution of created particle using the instantaneous particle picture.
However, if one integrates the distribution (for example Eq.\eqref{Bn_p}) with respect to the momentum,
it diverges although it represents the total particle number per unit volume.
This is not a defect of the instantaneous particle picture but a natural result of the constant electric field;
particles have been continued to be created from infinite past and their number has already reached infinity.

To get a finite particle number per unit volume, we must treat a non-steady case in which 
an electric field strength is 0 up to some time. 
It will give us the information how the distribution evolve in time.

%%%%%%%%%%%%%%%%%%%%%%%%%%%%%%%%%%%%%%%%
\subsection{Sudden switch-on} \label{subsec:sudden}
As the simplest example of non-steady electric fields, we first deal with the {\lq \lq}sudden switch-on" electric field: 
\begin{equation}
\mathbf{E} = \bigr( 0,0,E\theta (t-t_0 ) \bigl) . \label{suddenE}
\end{equation}
Although realizing this space--time configuration is impossible due to causality, 
we use it because it is the simplest model revealing the dynamical process of pair creation
and a distribution function in this electric field can be calculated analytically.

The gauge potential giving the electric field \eqref{suddenE} is
\begin{equation}
A^3 = \begin{cases}
       -Et & (t\geq t_0 ) \ \\
       -Et_0 & (t<t_0 ). 
      \end{cases} 
\end{equation}

We show only the derivation of the boson distribution function since the fermion distribution can be obtained
in a similar way.

In this case, we do not need to use the asymptotic WKB criterion because the electric field does not exist before the 
time $t_0 $. 
The in-solutions at $t<t_0 $ is obvious: 
\begin{equation}
{}_{\pm } \! f_\mathbf{p} ^{\mathrm{in} } (x) = {}_{\pm} \! f_\mathbf{p} ^{(t_0 )} (x) \ \ (\text{for} \ t<t_0 ) ,
\label{t<t_0}
\end{equation}
where ${}_{\pm} \! f_\mathbf{p} ^{(t_0 )} $ is given by Eq.\eqref{t_0plane}.
One can get the in-solution at $t>t_0 $ by solving the field equation with Eq.\eqref{t<t_0} as an initial condition
at $t=t_0 $.
However, we do not need to re-solve the field equation. 
We can construct the in-solutions at $t>t_0 $ using the result in Sec.\ref{subsubsec:PLboson-pure}.
Rewriting ${}_\pm \! f_\mathbf{p} ^\mathrm{in} $ in Sec.\ref{subsubsec:PLboson-pure} as ${}_\pm \! \tilde{f} _\mathbf{p} ^\mathrm{in} $,
we can obtain the in-solutions at $t>t_0 $ as follows: 
\begin{equation}
\begin{matrix}
{}_+ \! f^{\mathrm{in} } _\mathbf{p} (x) 
 = \alpha _\mathbf{p} ^* (t_0 ) {}_+ \! \tilde{f}^{\mathrm{in} } _{\mathbf{p} -e\mathbf{E} t_0 } (x)
   -\beta _\mathbf{p} ^* (t_0 ) {}_- \! \tilde{f}^{\mathrm{in} } _{\mathbf{p} -e\mathbf{E} t_0 } (x) \\
{}_- \! f^{\mathrm{in} } _\mathbf{p} (x) 
 = \alpha _\mathbf{p} (t_0 ) {}_- \! \tilde{f}^{\mathrm{in} } _{\mathbf{p} -e\mathbf{E} t_0 } (x)
   -\beta _\mathbf{p} (t_0 ) {}_+ \! \tilde{f}^{\mathrm{in} } _{\mathbf{p} -e\mathbf{E} t_0 } (x)
\end{matrix}
\hspace{1cm} (\text{for} \ t>t_0 ), \label{t>t_0}
\end{equation}
where $\alpha _\mathbf{p} (t_0 ) $ and $\beta _\mathbf{p} (t_0 ) $ are defined by Eq.\eqref{Bab}.
One can confirm that this solutions satisfy the initial condition at $t=t_0 $ by substituting Eq.\eqref{tenkai1}
(which holds only at instant of time $t_0 $) into ${}_\pm \! \tilde{f} _\mathbf{p} ^\mathrm{in} $.
We can get the relation between the in-solutions \eqref{t>t_0} and the instantaneous positive and negative frequency 
solutions at time $t_1 (>t_0 )$ by
substituting Eq.\eqref{tenkai1} with the replacement $t_0 \to t_1 $ into ${}_\pm \! \tilde{f} _\mathbf{p} ^\mathrm{in} $
of Eq.\eqref{t>t_0}:
\begin{equation}
\begin{matrix}
\begin{split}
{}_+ \! f_\mathbf{p} ^\mathrm{in} (x) 
 &= A_\mathbf{p} (t_0 ,t_1 ) {}_+ \! f_{\mathbf{p} +e\mathbf{E} \Delta t} ^{(t_1 )} (x)
    +B_\mathbf{p} ^* (t_0 ,t_1 ) {}_- \! f_{\mathbf{p} +e\mathbf{E} \Delta t} ^{(t_1 )} (x)
\end{split} \\[+7pt]
\begin{split}
{}_- \! f_\mathbf{p} ^\mathrm{in} (x) 
 &= A_\mathbf{p} ^* (t_0 ,t_1 ) {}_- \! f_{\mathbf{p} +e\mathbf{E} \Delta t} ^{(t_1 )} (x)
    +B_\mathbf{p} (t_0 ,t_1 ) {}_+ \! f_{\mathbf{p} +e\mathbf{E} \Delta t} ^{(t_1 )} (x) ,
\end{split}
\end{matrix} \label{tenkai2}
\end{equation}
where
\begin{equation}
\begin{matrix}
A_\mathbf{p} (t_0 ,t_1 ) = \alpha _\mathbf{p} ^* (t_0 ) \alpha _{\mathbf{p} +e\mathbf{E} \Delta t} (t_1 )
                     -\beta _\mathbf{p} ^* (t_0 ) \beta _{\mathbf{p} +e\mathbf{E} \Delta t} (t_1 ) \\[+7pt]
B_\mathbf{p} (t_0 ,t_1 ) = \alpha _\mathbf{p} (t_0 ) \beta _{\mathbf{p} +e\mathbf{E} \Delta t} (t_1 )
                     -\beta _\mathbf{p} (t_0 ) \alpha _{\mathbf{p} +e\mathbf{E} \Delta t} (t_1 )
\end{matrix}
\end{equation}
and $\Delta t=t_1 -t_0 $.

To define an in-particle picture and the associated vacuum, we expand the field operator with ${}_\pm \! f_\mathbf{p} ^\mathrm{in} $
which is given by Eq.\eqref{t<t_0} and \eqref{t>t_0}: 
\begin{equation}
\phi (x) = \int \! dp [{}_+ \! f^{\mathrm{in} } _{\mathbf{p} } (x) a^{\mathrm{in} } _{\mathbf{p} } 
              +{}_- \! f^{\mathrm{in} } _{\mathbf{p} } (x) b ^{\mathrm{in} \dagger } _{-\mathbf{p} } ] . \label{ba}
\end{equation}
The Bogoliubov relation between the creation/annihilation operator of an in-particle and that of a $t_1$-particle
can be found by substituting Eq.\eqref{tenkai2} into Eq.\eqref{ba}: 
\begin{equation}
\phi (x) = \int \! d^3 p \left[ a_\mathbf{p} (t_1 ) {}_+ \! f_\mathbf{p} ^{(t_1 )} (x)
           + b_\mathbf{p} ^\dagger (t_1 ) {}_- \! f_\mathbf{p} ^{(t_1 )} (x) \right] \label{t1-exp}
\end{equation}
\begin{equation}
\begin{matrix}
a_\mathbf{p} (t_1 ) = A _{\mathbf{p} -e\mathbf{E} \Delta t} (t_0 ,t_1 ) a_{\mathbf{p} -e\mathbf{E} \Delta t} ^\mathrm{in} +
                  B _{\mathbf{p} -e\mathbf{E} \Delta t} (t_0 ,t_1 ) b_{-\mathbf{p} +e\mathbf{E} \Delta t} ^{\mathrm{in} \dagger } \\[+7pt]
b_\mathbf{p} ^\dagger (t_1 ) = A _{\mathbf{p} -e\mathbf{E} \Delta t} ^* (t_0 ,t_1 ) b_{-\mathbf{p} +e\mathbf{E} \Delta t} ^{\mathrm{in} \dagger } +
                  B _{\mathbf{p} -e\mathbf{E} \Delta t} ^* (t_0 ,t_1 ) a_{\mathbf{p} -e\mathbf{E} \Delta t} ^\mathrm{in}
\end{matrix}
\ \ (\text{for} \ t_1 >t_0 ).
\end{equation}

\begin{figure}[t]
 \begin{tabular}{cc}
  \begin{minipage}{0.5\textwidth}
   \begin{center}
    \includegraphics[scale=0.386,bb=0 0 563 458]{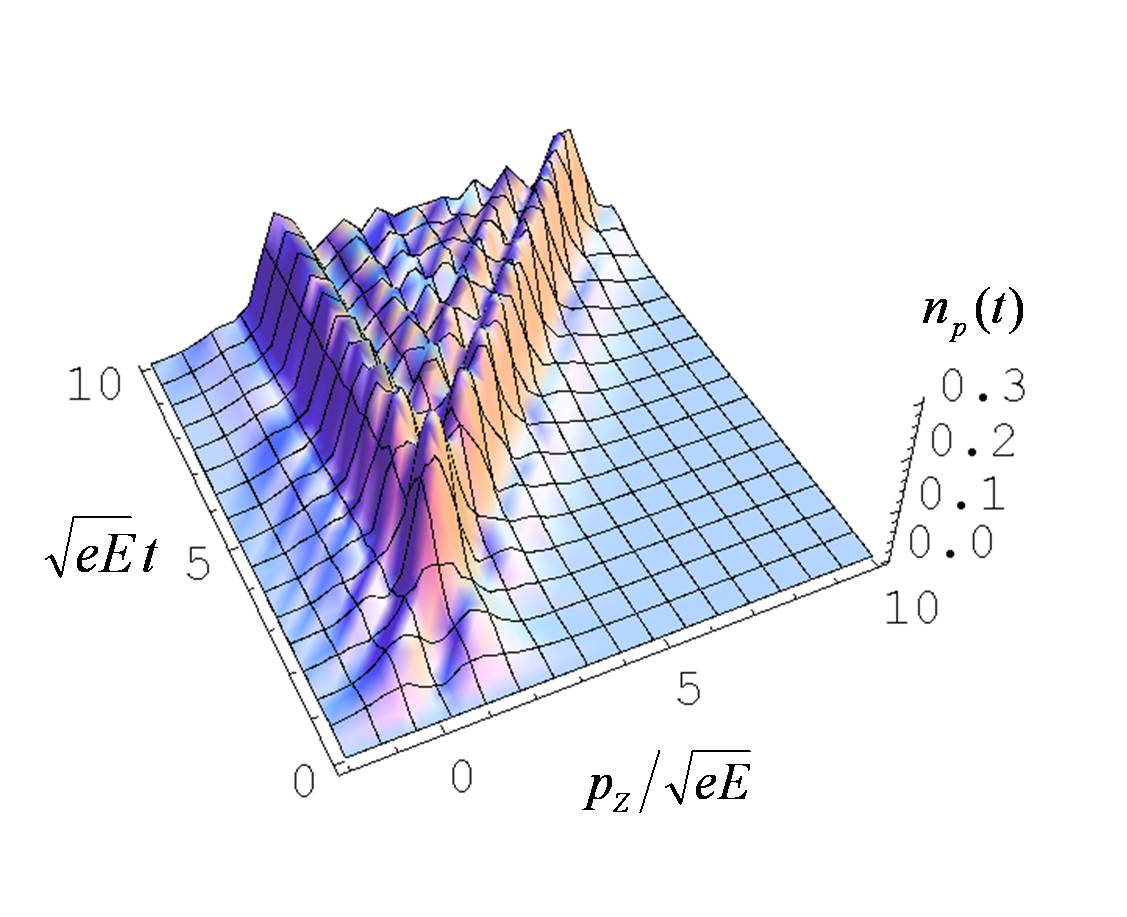}
    {\footnotesize (a) boson $a=0.3$ }
   \end{center}
  \end{minipage} &
  \begin{minipage}{0.5\textwidth}
   \begin{center}
    \includegraphics[scale=0.386,bb=0 0 563 445]{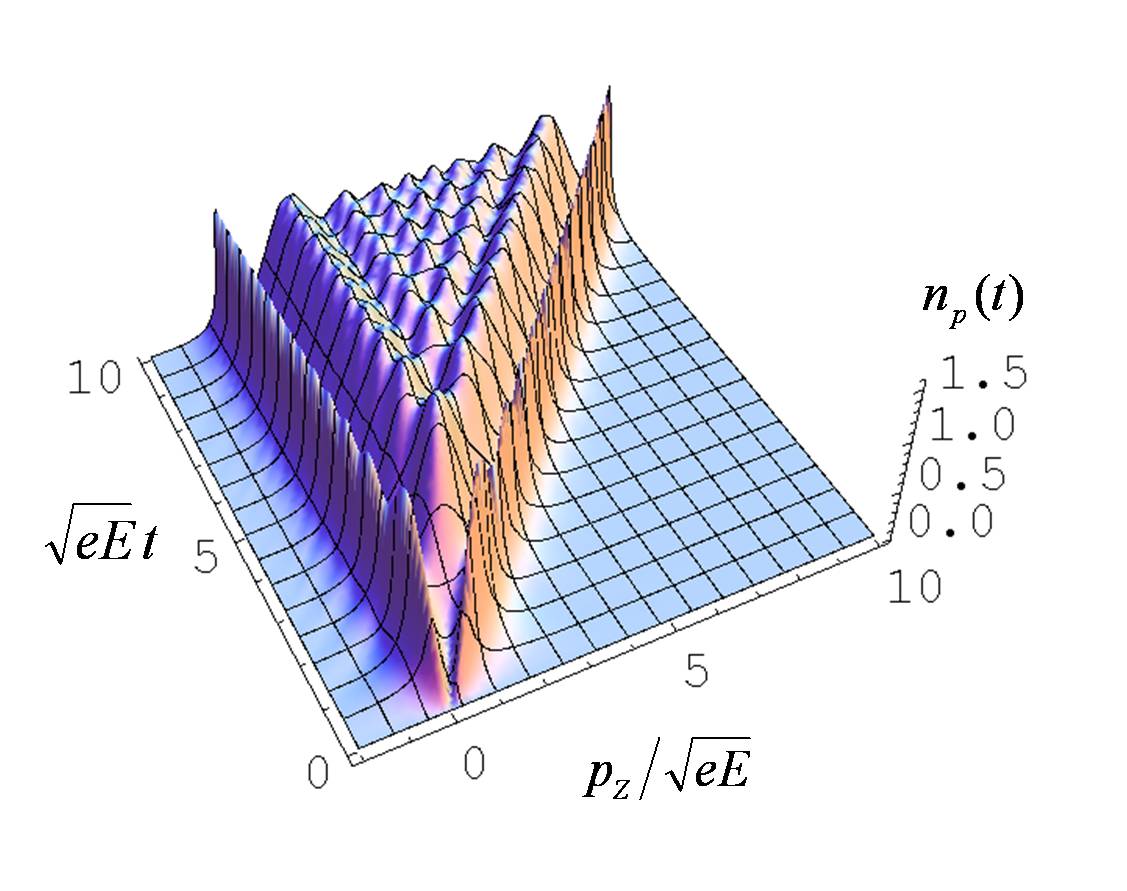}
    {\footnotesize (b) boson $a=0.01$ }
   \end{center}
  \end{minipage} \\

  \begin{minipage}{0.5\textwidth}
   \begin{center}
    \includegraphics[scale=0.386,bb=0 0 563 430]{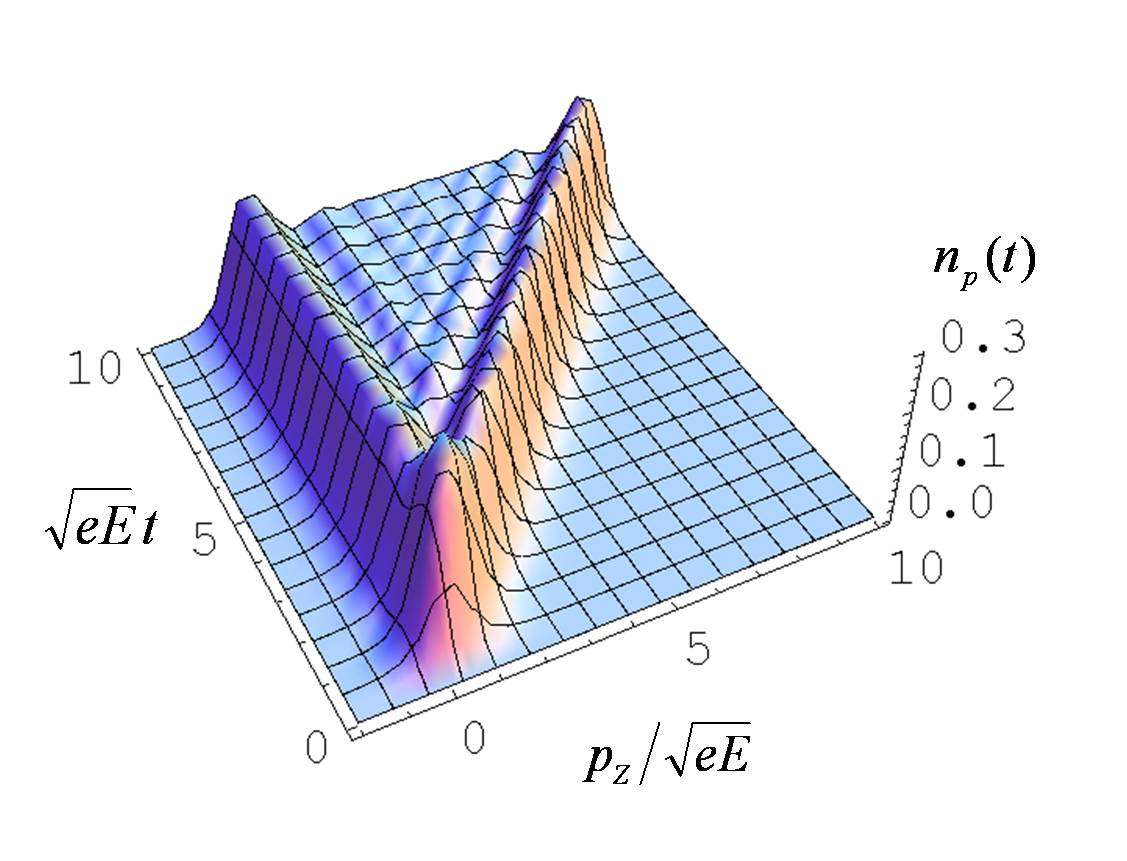}
    {\footnotesize (c) fermion $a=0.3$ }
   \end{center}
  \end{minipage} &
  \begin{minipage}{0.5\textwidth}
   \begin{center}
    \includegraphics[scale=0.386,bb=0 0 563 433]{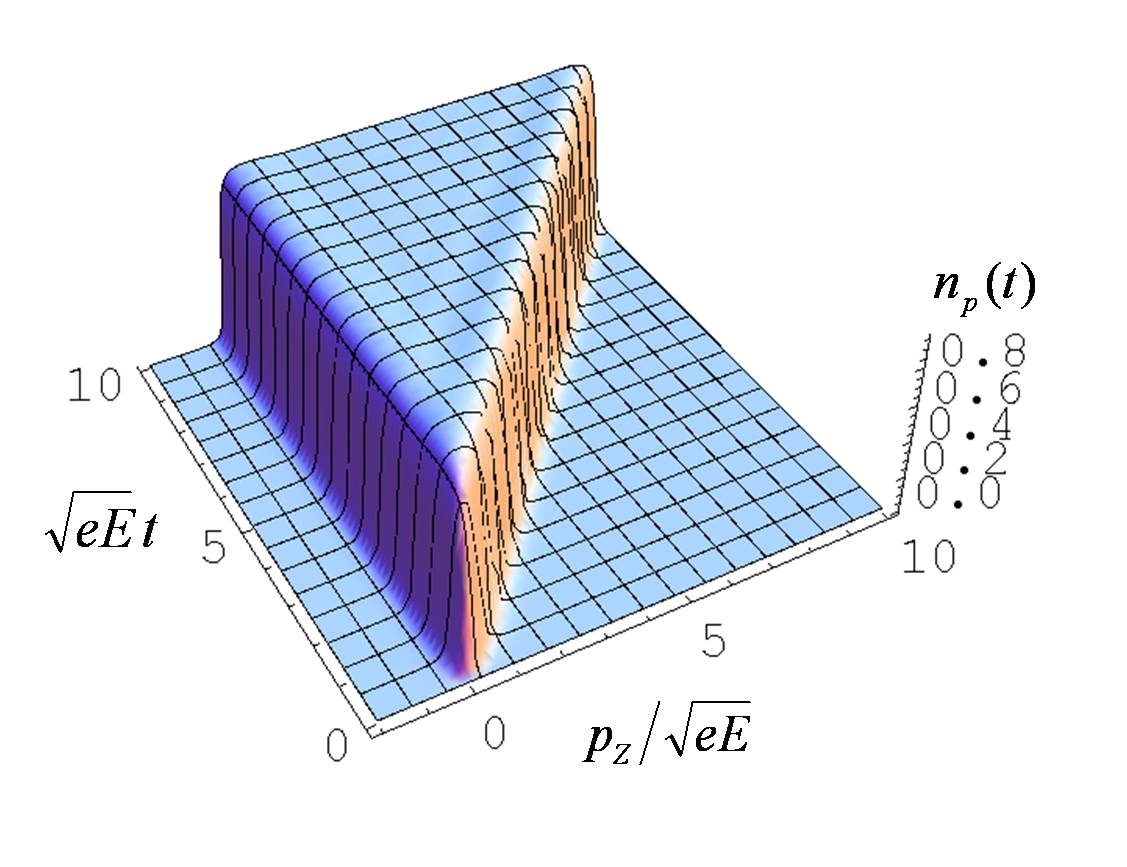}
    {\footnotesize (d) fermion $a=0.01$ }
   \end{center}
  \end{minipage}
 \end{tabular}
 \caption{The time evolution of the longitudinal momentum distributions }
 \label{fig:BFn_p(t)}
\end{figure}

Now we can calculate the pair distribution function at time $t_1 (>t_0 )$ as follows: 
\begin{equation}
\begin{split}
n_\mathbf{p} (\Delta t) &= \langle 0,\mathrm{in} | a_\mathbf{p} ^\dagger (t_1 ) a_\mathbf{p} (t_1 ) |0,\mathrm{in} \rangle \frac{(2\pi )^3 }{V}
            = \langle 0,\mathrm{in} | b_{-\mathbf{p} } ^\dagger (t_1 ) b_{-\mathbf{p} } (t_1 ) |0,\mathrm{in} \rangle \frac{(2\pi )^3 }{V} \\
           &= |B _{\mathbf{p} -e\mathbf{E} \Delta t } (t_0 ,t_1 ) |^2 \\
           &= |\alpha _{\mathbf{p} -e\mathbf{E} \Delta t } (t_0 ) \beta _\mathbf{p} (t_1 )
               -\beta _{\mathbf{p} -e\mathbf{E} \Delta t } (t_0 ) \alpha _\mathbf{p} (t_1 ) |^2 .
\end{split} \label{Bn_p(t)}
\end{equation}
This depends on time through only $\Delta t$, 
because $\alpha _\mathbf{p} (t) $ and $\beta _\mathbf{p} (t) $ depend on $t$ through only the phase factor $e^{\pm i\omega _p t} $.  
Of course, $n_\mathbf{p} (\Delta t) =0$ when $t_1<t_0 $.

The distribution of fermions can be derived in a similar way.
The result is the same form as that of bosons [Eq.\eqref{Bn_p(t)}],
however $\alpha _\mathbf{p} (t_0 ) $ and $\beta _\mathbf{p} (t_0 ) $ have been given by Eq.\eqref{Dab} instead of Eq.\eqref{Bab}.

\begin{figure}[t]
 \begin{tabular}{cc}
  \begin{minipage}{0.5\textwidth}
   \begin{center}
    \includegraphics[scale=0.37,bb=0 0 563 433]{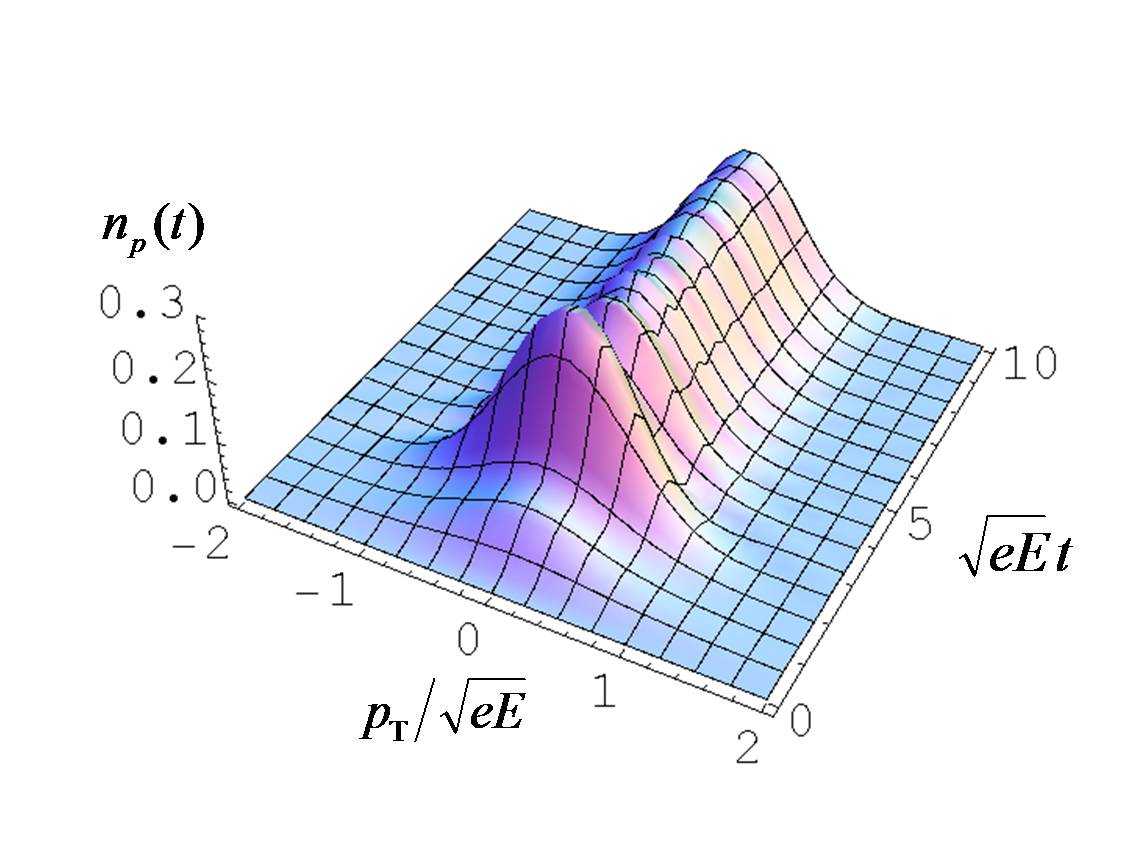}
   \end{center}
  \end{minipage} &
  \begin{minipage}{0.5\textwidth}
   \begin{center}
    \includegraphics[scale=0.37,bb=0 0 563 461]{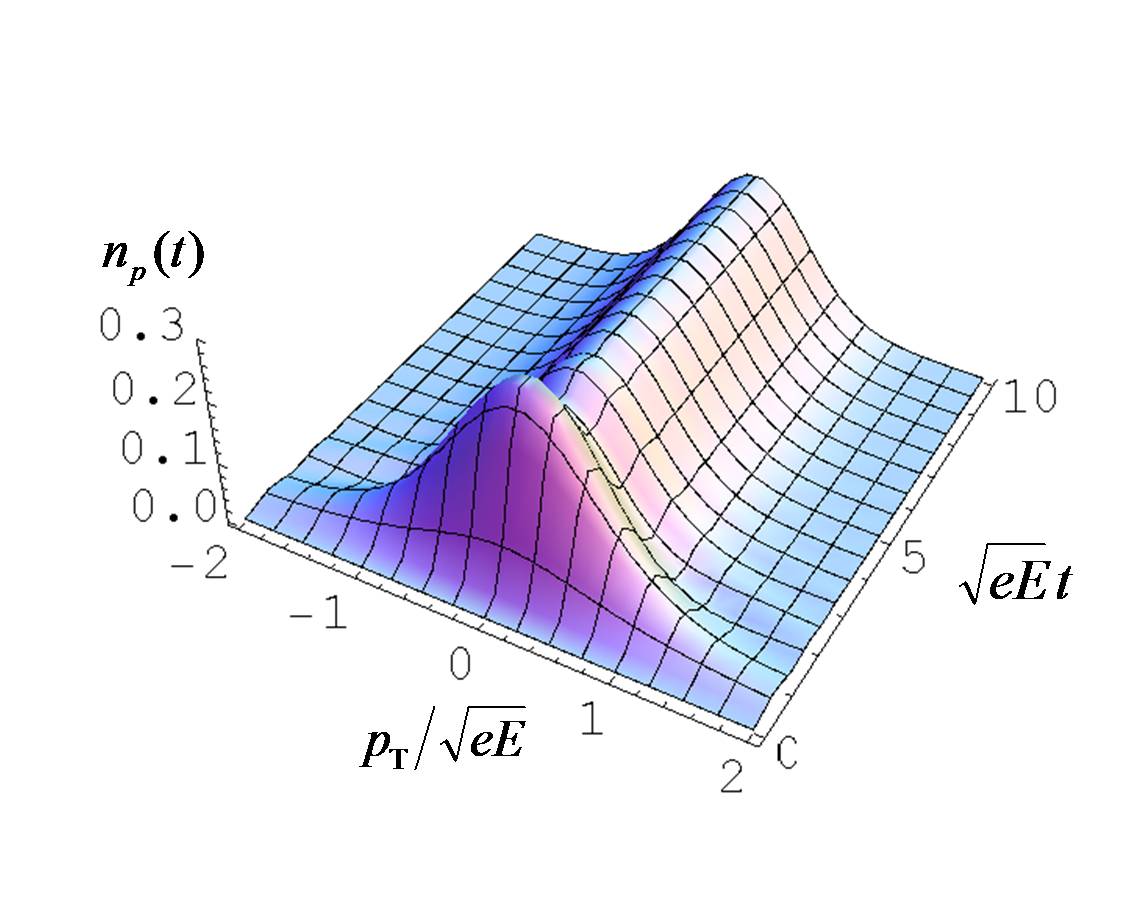}
   \end{center}
  \end{minipage} \\
   {\footnotesize (a) bosons \ $\bar{a} =\frac{m^2 }{2eE} =0.3, p_z /\sqrt{eE} = 1.45 $ } &
   {\footnotesize (b) fermions \ $\bar{a} =\frac{m^2 }{2eE} =0.3, p_z /\sqrt{eE} =0.77 $ }
 \end{tabular}
 \caption{The time evolution of the transverse momentum distributions }
 \label{fig:PT(t)}
\end{figure}

The time evolution of the longitudinal momentum distribution, $n_\mathbf{p} (t) $ with fixed $a$, and that of
the transverse momentum distribution, $n_\mathbf{p} (t) $ with fixed $p_z $, are shown for both bosons and fermions
in Fig.\ref{fig:BFn_p(t)} and Fig.\ref{fig:PT(t)}.

The longitudinal momentum distributions rise up first around 0-momentum after the field is switched on. 
As already expected in Sec.\ref{sec:PL}, 
only particles with approximately 0-momentum can be created.
However, not strictly 0 but broadened around 0 due to quantum fluctuation. 

Comparing the two longitudinal momentum distributions for different values of $a$, we can notice that
the time when the distribution peaks first is earlier for smaller $a$.
This is to say, the lighter the mass or the stronger the electric field, then the shorter time the electric field takes 
to create particles, or for fixed $eE$ and $m$, particles with $p_\text{T} =0 $ are created first
and those with higher $p_\text{T} $ are later.

After particles are created with about 0-momentum, they are accelerated by the electric field 
according to the classical equation of motion: $p_z =eEt +\mathrm{const.} $
This process makes triangle shaped distributions in $t$-$p_z $ space. 
On the other hand, the transverse momentum distributions take a saddle-like shape in $t$-$p_\mathrm{T} $ space.
This is because the transverse momentum distribution can be approximated as a Gaussian at any time and
settles in a fixed distribution after sufficiently long time; 
especially $e^{-2\pi a} $ in the limit of $p_z \to \infty $. 

Both of the longitudinal and the transverse momentum distribution of bosons show oscillations.
These oscillations are the same as that has seen in the distribution in the constant electric field 
[Fig.\ref{fig:Bn_p}].

One can recover the distribution in the constant electric field [Eq.\eqref{Bn_p} or \eqref{Dn_p}]
taking the limit of $\Delta t \to +\infty $ in Eq.\eqref{Bn_p(t)}.
Therefore, the use of the asymptotic WKB criterion in a constant electric field 
is consistent with the treatment in this section, where the in-vacuum is defined in absence of an electric field
and the asymptotic WKB criterion is not used.

Let us comment on a symmetry of the distribution \eqref{Bn_p(t)}. 
The distribution $n_\mathbf{p} (t) $ is symmetric by reflection with respect to the axis 
$p_z =\frac{1}{2} eEt $ in $p_z $ space; 
$n_{\frac{1}{2} e\mathbf{E} t+ \mathbf{p} } (t) = n_{\frac{1}{2} e\mathbf{E} t- \mathbf{p} } (t) $ for any $t$. 
This symmetry originates in the time reversal symmetry: 
(i) Because the argument $(t)$ of $\alpha _\mathbf{p} (t) $ and $\beta _\mathbf{p} (t) $ is irrelevant to $n_\mathbf{p} (t) $,
\begin{equation}
n_{e\mathbf{E} t-\mathbf{p} } (t) 
 = |\alpha _{-\mathbf{p} } \beta _{e\mathbf{E} t-\mathbf{p} } -\beta _{-\mathbf{p} } \alpha _{e\mathbf{E} t-\mathbf{p} } |^2 
 = \check{n} _{-\mathbf{p} } (-t) \label{sym1}
\end{equation} 
holds formally for $t>0$. 
We put $\check{} $ on $n_{-\mathbf{p} } (-t) $ because originally Eq.\eqref{Bn_p(t)} gives $n_\mathbf{p} (t) $ only for $t>0 $.
(ii) Seeing the derivation of Eq.\eqref{Bn_p(t)}, one can notice that $n_\mathbf{p} (t) $ of Eq.\eqref{Bn_p(t)} for negative $t$ 
represents the distribution in the time-reversed system, namely the system with the boundary condition $n_\mathbf{p} (0)=0$
in the electric field which is switched on at infinite past and switched off at $t=0$.
Thus,
\begin{equation}
\check{n} _{-\mathbf{p} } (-t) = \text{T} {n} _{-\mathbf{p} } (-t) \label{sym2}
\end{equation}
is satisfied, where T is a time-reversal operator.
(iii) Because this system is invariant under time reversal,
\begin{equation}
\text{T} {n} _{-\mathbf{p} } (-t) = n_\mathbf{p} (t) \label{sym3}
\end{equation}
holds. 
Eqs.\eqref{sym1},\eqref{sym2} and \eqref{sym3} yield the symmetry in question: 
$n_\mathbf{p} (t) = n_{e\mathbf{E} t-\mathbf{p} } (t) $.

\begin{figure}[t]
 \begin{tabular}{cc}
  \begin{minipage}{0.5\textwidth}
   \begin{center}
    \includegraphics[scale=0.5,bb=0 0 429 248]{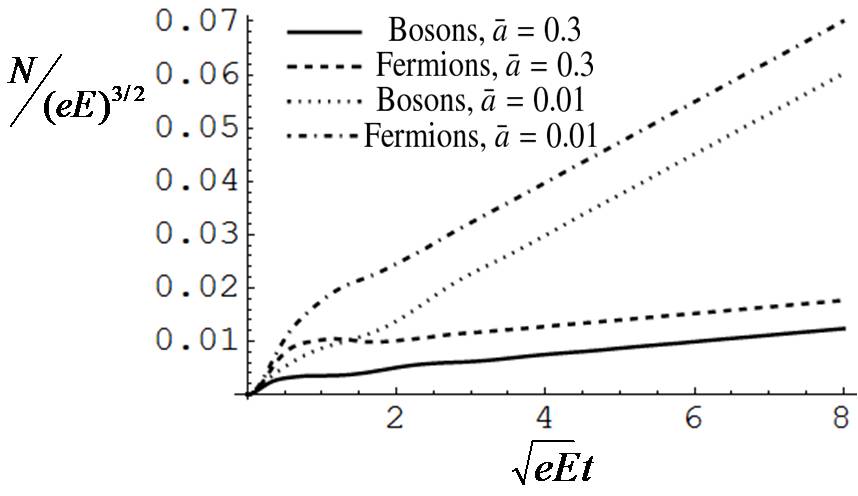} \\
    {\footnotesize (a) the total particle number density }
   \end{center}
  \end{minipage} &
  \begin{minipage}{0.5\textwidth}
   \begin{center}
    \includegraphics[scale=0.5,bb=0 0 446 248]{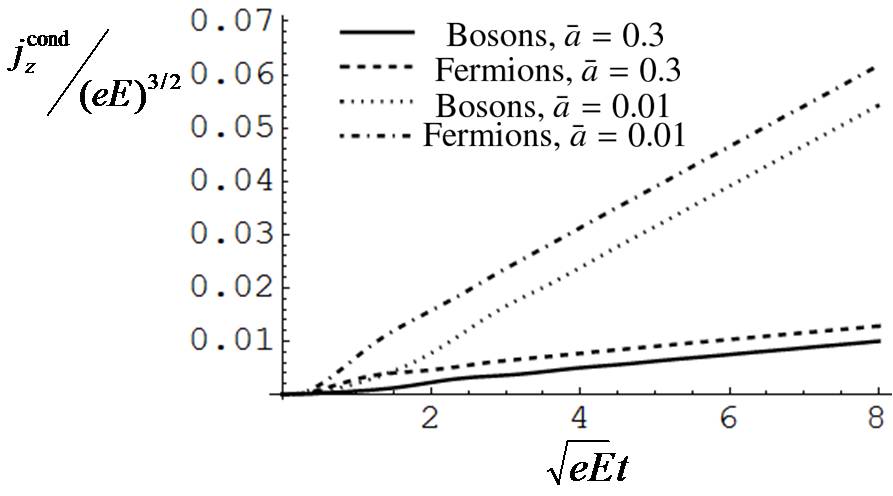} \\
    {\footnotesize (b) the conduction current }
   \end{center}
  \end{minipage} \\
  {} & {} \\
  \begin{minipage}{0.5\textwidth}
   \begin{center}
    \includegraphics[scale=0.5,bb=0 0 440 233]{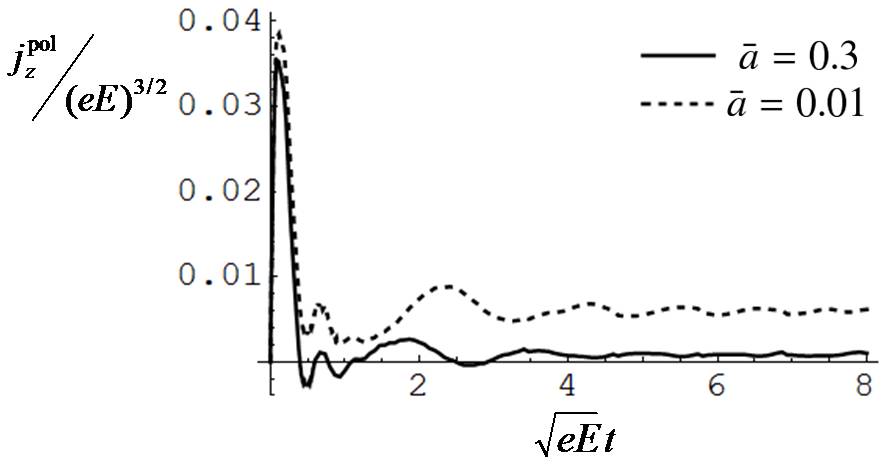} \\
    {\footnotesize (c) the polarization current of bosons }
   \end{center}
  \end{minipage} &
  \begin{minipage}{0.5\textwidth}
   \begin{center}
    \includegraphics[scale=0.5,bb=0 0 440 232]{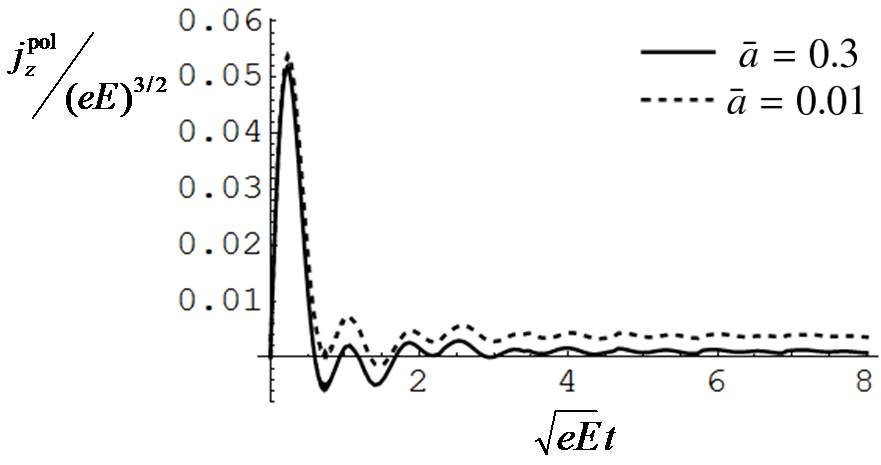} \\
    {\footnotesize (d) the polarization current of fermions }
   \end{center}
  \end{minipage}
 \end{tabular}
 \caption{The time evolution of the total particle number density and the current density 
          for several values of $\bar{a} = \frac{m^2 }{2eE} $.
          For fermions, these values have been averaged over the spin states. $e=1$.}
 \label{fig:NandJ} 
\end{figure}

\vspace{11pt}
Finite time imposition of the electric field offers us finite expectation values of the total particle number density,
the charge current density and so on. 
In Fig.\ref{fig:NandJ}(a) we show the time evolution of the total particle number density 
$N=\int \! \frac{d^3 p}{(2\pi )^3 } n_\mathbf{p} $ for several values of $\bar{a} \equiv \frac{m^2 }{2eE} $.
In Fig.\ref{fig:NandJ}(b),(c),(d) the current density, 
which is separated into a conduction current and a polarization current is plotted. 
For fermions, the values averaged over the two spin states are shown.

The Noether current associated with $U(1) $ symmetry in Lagrangian \eqref{KGlag} is
\begin{equation}
\hat{j} ^\mu = ie\left[ \phi ^\dagger (\partial ^\mu +ieA^\mu )\phi 
               -(\partial ^\mu -ieA^\mu )\phi ^\dagger \cdot \phi \right] . \label{current-ope}
\end{equation}
Taking the in-vacuum expectation of the $z$-component of this current operator with the $t_1 $-field expansion
\eqref{t1-exp} and the gauge potential at instant of time $t=t_1 $: $A^3 =-Et_1 $,
we get the mean current density
\begin{equation}
\begin{split}
j_z (t_1 ) &= \langle 0,\mathrm{in} |\hat{j} _z |0,\mathrm{in} \rangle \\
 &= e\int \! d^3 p \int \! d^3 q \frac{p_z +q_z }{(2\pi )^3 \sqrt{2\omega _p 2\omega _q } }
    \langle 0,\mathrm{in} | \left[ e^{i(\omega _p -\omega _q )t_1 } a_\mathbf{p} ^\dagger (t_1 ) a_\mathbf{q} (t_1 ) \right. \\
 & \left.
    +e^{-i(\omega _p -\omega _q )t_1 } b_{-\mathbf{p} } (t_1 ) b_{-\mathbf{q} } ^\dagger (t_1 )
    +e^{i(\omega _p +\omega _q )t_1 } a_\mathbf{p} ^\dagger (t_1 ) b_{-\mathbf{q} } ^\dagger (t_1 ) 
    +e^{-i(\omega _p +\omega _q )t_1 } b_{-\mathbf{p} } (t_1 ) a_\mathbf{q} (t_1 ) \right] |0,\mathrm{in} \rangle \\
 &= 2e\int \! \frac{d^3 p }{(2\pi )^3 } \frac{p_z }{\omega _p } |\beta _\mathbf{p} (t_1 ) |^2 
    +2e\int \! \frac{d^3 p }{(2\pi )^3 } \frac{p_z }{\omega _p }
    \mathrm{Re} \left[ e^{-2i\omega _p t_1 } \alpha _\mathbf{p} (t_1 )\beta _\mathbf{p} (t_1 ) \right] \\
 &= 2e\int \! \frac{d^3 p }{(2\pi )^3 } \frac{p_z }{\omega _p } n_\mathbf{p} (t_1 )
    +2e\int \! \frac{d^3 p }{(2\pi )^3 } \frac{\omega _\mathbf{p} }{eE} S(t_1 ,\mathbf{p} ) \hspace{30pt} \text{\textit{bosons}}.
 \label{Bcurrent}
\end{split}
\end{equation} 
The first term is a conduction current, which is caused by movement of real particles 
and the second is a polarization current, which originates in pair creation process.

\begin{figure}[t]
\begin{center}
\includegraphics[scale=0.34,bb=0 0 497 402]{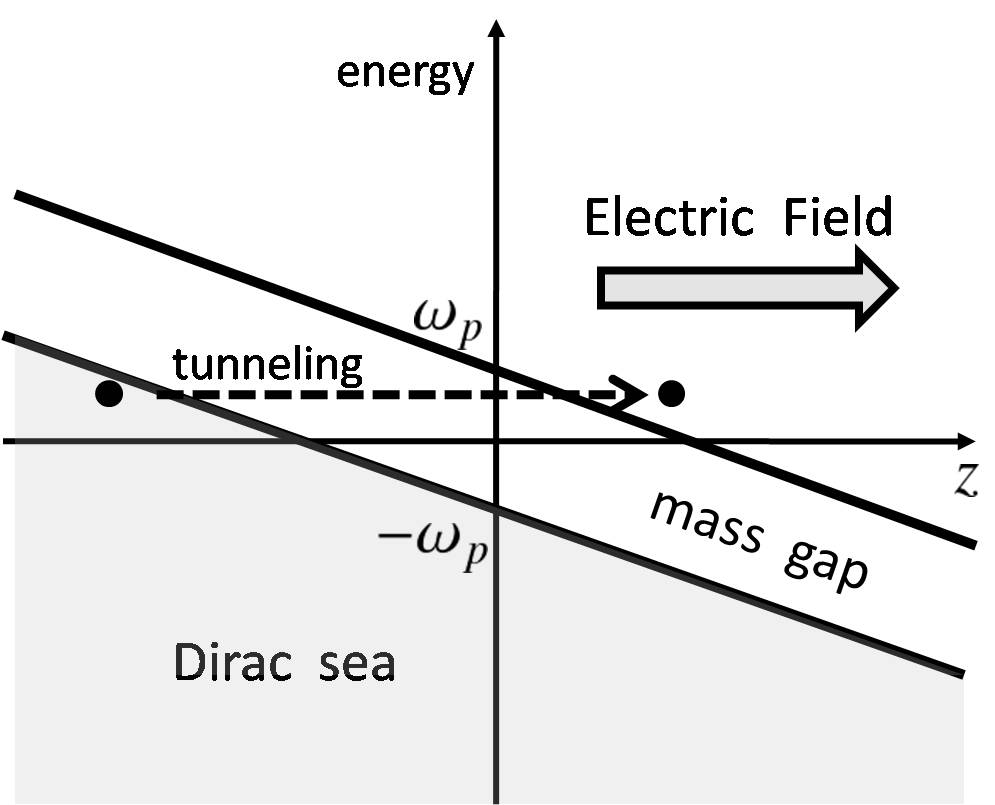} \\
\end{center}
\vskip -\lastskip \vskip -3pt
\caption{Schematic picture of a tunneling process}
\label{fig:tunnering}
\end{figure}

A schematic picture of a tunneling process, Fig.\ref{fig:tunnering}, may help us to understand 
a meaning of the polarization current.
Figure \ref{fig:tunnering} itself is applicable to only fermions,
however, the following explanation is also relevant to bosons.  
Suppose that a pair of a particle with momentum $\mathbf{p} $ and an antiparticle with momentum $-\mathbf{p} $ is created.
The distance between them is $2\frac{\omega _\mathbf{p} }{eE} $ when they become on-shell.
Thus, this process can be understood as a creation of an electric dipole $2e\frac{\omega _\mathbf{p} }{eE} $.
The polarization current is induced by the variation of an electric dipole
and now the creation rate of a dipole is $\frac{d}{dt} n_\mathbf{p} (t) = S(t,\mathbf{p} ) $.
Therefore, the variation rate of the total electric dipole, or the polarization current is 
\begin{equation}
j_z ^\mathrm{pol} (t) 
 = \int \! \frac{d^3 p }{(2\pi )^3 } 2e\frac{\omega _\mathbf{p} }{eE} S(t,\mathbf{p} ) ,
\end{equation}
which is identical to the second term of Eq.\eqref{Bcurrent}.

For fermions, the $z$-component of the mean current density is
\begin{equation}
\begin{split}
j_z (t) 
 &= -\langle 0,\mathrm{in} |e\bar{\psi } \gamma ^3 \psi |0,\mathrm{in} \rangle \\
 &= 2e\sum _{s=\uparrow ,\downarrow } \int \! \frac{d^3 p }{(2\pi )^3 } \frac{p_z }{\omega _p } |\beta _{\mathbf{p} s} (t)|^2
    +2e\sum _{s=\uparrow ,\downarrow } \int \! \frac{d^3 p }{(2\pi )^3 } \frac{m_\mathrm{T} }{\omega _p }
    \mathrm{Re} \left[ e^{-2i\omega _p t} \alpha _{\mathbf{p} s} (t)\beta _{\mathbf{p} s} (t) \right] \\
 &= 2e\sum _{s=\uparrow ,\downarrow } \int \! \frac{d^3 p }{(2\pi )^3 } \frac{p_z }{\omega _p } n_{\mathbf{p} s} (t)
    +2e\sum _{s=\uparrow ,\downarrow } \int \! \frac{d^3 p }{(2\pi )^3 } \frac{\omega _p }{eE} S(t,\mathbf{p} ) 
 \hspace{30pt} \text{\textit{fermions}}, \label{Fcurrent}
\end{split}
\end{equation}
of which last expression is identical to that of bosons except spin summations.

The polarization current has a pulse just after the switch-on. 
This is because the particle number undergoes a rapid increase from the initial state with no particle.
Some time after the switch-on, the increase of the particle number gets equilibrated to a fixed rate.
Therefore, the polarization current settles in a constant value and the conduction current becomes proportional to time.

Although we have the analytic form of the distribution functions, carrying out the integral analytically 
in the total particle number density $N=\int \! \frac{d^3 p}{(2\pi )^3 } n_\mathbf{p} (t) $
and the current density \eqref{Bcurrent} is impossible and the parameter dependence of these values is not clear. 
Thus, we model the distribution function in a simple form 
\begin{equation}
n_\mathbf{p} (t) = e^{-2\pi a } \theta (p_z )\theta(eEt-p_z ) , \label{model}
\end{equation}
and investigate the parameter dependence.
This approximated distribution is exact for fermions with $a=0$ and valid for other cases except
the time just after the switch-on and bosons with a small $a$. 
Eq.\eqref{model} corresponds to the delta function source term: 
\begin{equation}
S(t,\mathbf{p} ) = eEe^{-2\pi a } \delta (p_z =0) \theta (t) . \label{model-S}
\end{equation}

With the help of Eq.\eqref{model} and approximations which are valid under $m^2 \ll E $, 
the integrals in the total particle number density and the current density can be easily done: 
\begin{gather}
N(t) \simeq \frac{2(eE)^2 }{(2\pi )^3 } e^{-\frac{\pi m^2 }{eE} } t \ \theta (t) \label{approxi-Num} \\
j_z ^\mathrm{cond} (t) \simeq \frac{2e(eE)^2 }{(2\pi )^3 } e^{-\frac{\pi m^2 }{eE} } t \ \theta (t) \label{approxi-jcond} \\
j_z ^\mathrm{pol} (t) \simeq \frac{e(eE)^{3/2} }{(2\pi )^3 } e^{-\frac{\pi m^2 }{eE} }\theta (t) \label{approxi-jpol} .
\end{gather}
These are common to bosons, fermions with spin-$\uparrow $ and with spin-$\downarrow $ mode.

The approximated values \eqref{approxi-Num}, \eqref{approxi-jcond} and \eqref{approxi-jpol} succeed in reproducing 
the slope of the linear increase of $N(t) $ and $j_z ^\mathrm{cond} (t) $ obtained by the numerical calculation
and the constant value that $j_z ^\mathrm{pol} (t) $ approaches, 
except (i) the time just after the switch-on, in which the system
remembers the information of the initial state, on the other hand, after sufficiently long time from the switch-on,
the pair creation process settles in a steady state, which is well expressed by the approximation \eqref{model-S},
and (ii) bosons with a small $a$ because its distribution has large peaks on the $p_z =0 $ and $p_z =eEt $ lines
which is not expressed in the approximation \eqref{model}.
These approximated expressions will be useful to analyze a time scale of back reaction (Sec.\ref{sec:BR}). 

\begin{figure}[t]
 \begin{tabular}{cc}
  \begin{minipage}{0.5\textwidth}
   \begin{center}
    \includegraphics[scale=0.5,bb=0 0 429 243]{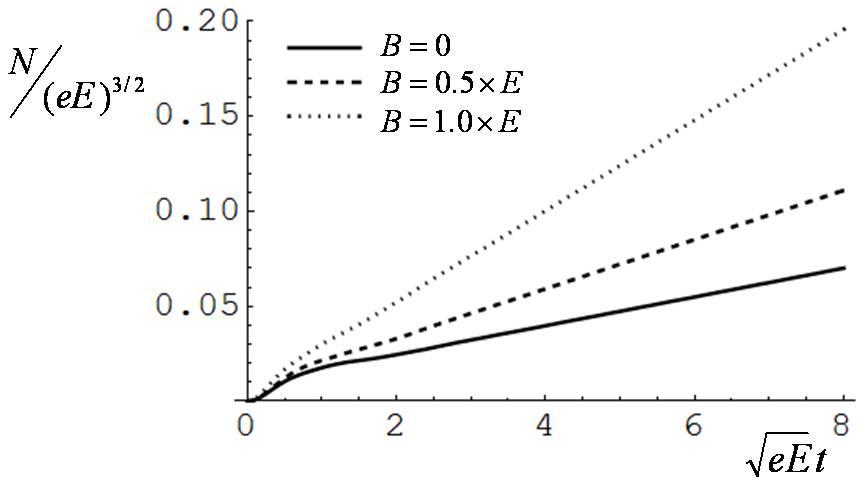} \\
    {\footnotesize (a) the total particle number density }
   \end{center}
  \end{minipage} 
  \begin{minipage}{0.5\textwidth}
   \begin{center}
    \includegraphics[scale=0.5,bb=0 0 429 243]{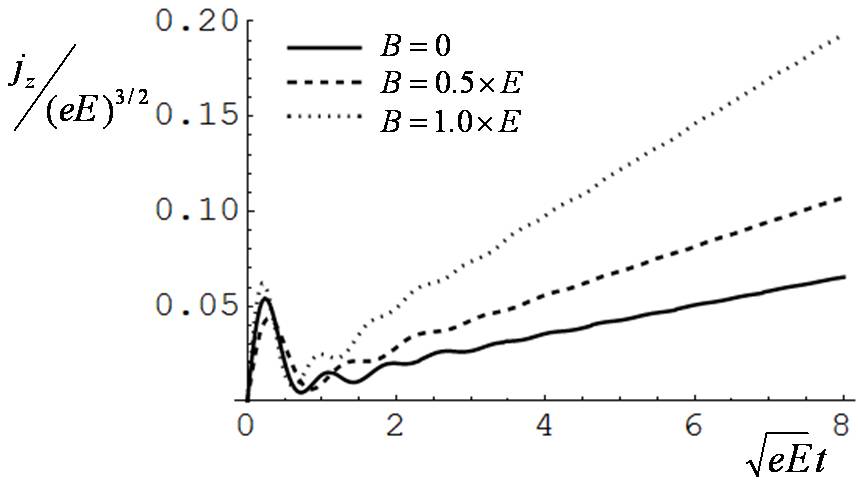} \\
    {\footnotesize (b) the current density }
   \end{center}
  \end{minipage}
 \end{tabular}
 \caption{The total particle number density and the current density of fermions
          in the collinear electromagnetic field ($\bar{a} =\frac{m^2 }{2eE} =0.01, e=1 $).
          The values averaged over the spin states are shown. }
 \label{fig:NandJinB}
\end{figure}

\vspace{11pt}
In Fig.\ref{fig:NandJinB} the time evolution of the total particle number density
and the current density of fermions in the collinear electric and magnetic field are shown.
(We do not plot those of bosons because they are strongly suppressed by an intense magnetic field.)
As already noted in Sec.\ref{subsubsec:fermion-coll}, the longitudinal magnetic field increases balk quantities of fermions.

In the collinear electromagnetic field, the approximated particle number and currents are
\begin{gather}
N (t) \simeq \frac{2e^2 EB}{(2\pi )^2 } e^{-\frac{\pi m^2 }{eE} } 
 \frac{e^{-\pi \kappa \frac{B}{E} } }{1-e^{-2\pi \frac{B}{E} } } t \ \theta(t) \label{NinBappro} \\
j_z ^\mathrm{cond} (t) \simeq \frac{2e^3 EB}{(2\pi )^2 } e^{-\frac{\pi m^2 }{eE} } 
 \frac{e^{-\pi \kappa \frac{B}{E} } }{1-e^{-2\pi \frac{B}{E} } } t \ \theta(t) \\
j_z ^\mathrm{pol} (t) \simeq \frac{2e(eB)^{3/2} }{(2\pi )^2 } e^{-\frac{\pi m^2 }{eE} } 
 \sum _{n=0} ^\infty \sqrt{2n+\kappa } e^{-\pi (2n+\kappa ) \frac{B}{E} } \theta(t) ,
\end{gather}
where
\begin{equation}
\kappa = \left \{ 
 \begin{array}{cc}
  1 & \text{for bosons} \ \\
  0 & \text{for fermions with spin-}\uparrow \ \\
  2 & \text{for fermions with spin-}\downarrow .
 \end{array} \right.
\end{equation}
Also in this case, the approximated values quite well agree with the exact ones
with the same exception as before.
Eq.\eqref{NinBappro} tells us that the longitudinal magnetic field increases the particle number of
fermions with spin-$\uparrow$ approximately in proportion to $B$.

\vspace{11pt}
So far, we have employed the in-vacuum as an initial state,
in which case there is no particle before the electric field is turned on.
We can also use the state where particle pairs are condensed,
in which case there are initial particles and they evolve under the influence of the electric field after the switch-on.

The state with an arbitrary initial distribution $f(\mathbf{p} )$ is constructed as follows:
\begin{equation}
|f \rangle = \prod _\mathbf{p} \mathcal{N} _\mathbf{p} ^{-1/2}
             \exp \left[ \frac{(2\pi )^3 }{V} F(\mathbf{p} ) a_\mathbf{p} ^{\mathrm{in} \dagger } b_{-\mathbf{p} } ^{\mathrm{in} \dagger } \right]
             |0,\mathrm{in} \rangle ,
\end{equation}
where $F(\mathbf{p} ) $ and the normalization constant $\mathcal{N} _\mathbf{p} $ are related with the initial distribution
$f(\mathbf{p} )$ as
\begin{equation}
|F(\mathbf{p} )|^2 = \frac{f(\mathbf{p} )}{1\pm f(\mathbf{p} )} , \
|\mathcal{N} _\mathbf{p} |^2 = \left \{ 1\pm f(\mathbf{p} ) \right \} ^\pm \ \
\left( \begin{array}{cc} + : & \text{bosons} \\ - : & \text{fermions} \end{array} \right) .
\end{equation}
One can confirm that the expectation of the in-particle number operator with the state $|f\rangle $ is surely
equal to the initial distribution $f(\mathbf{p} ) $: 
\begin{equation}
\langle f|a_\mathbf{p} ^{\mathrm{in} \dagger } a_\mathbf{p} ^\mathrm{in} |f\rangle \frac{(2\pi )^3 }{V}
 = \langle f|b_{-\mathbf{p} } ^{\mathrm{in} \dagger } b_{-\mathbf{p} } ^\mathrm{in} |f\rangle \frac{(2\pi )^3 }{V} 
 = f(\mathbf{p} ) .
\end{equation}
The distribution at time $t$ with the state $|f\rangle $ is
\begin{equation}
\begin{split}
\tilde{n} _\mathbf{p} (t) &= \langle f|a_\mathbf{p} ^\dagger (t) a_\mathbf{p} (t) |f\rangle \frac{(2\pi )^3 }{V}
  = \langle f|b_{-\mathbf{p} } ^\dagger (t) b_{-\mathbf{p} } (t) |f\rangle \frac{(2\pi )^3 }{V} \\
 &= \left[ 1\pm 2n_\mathbf{p} (\Delta t) \right] f(\mathbf{p} -e\mathbf{E} \Delta t) +n_\mathbf{p} (\Delta t) \\
 &= \left[ 1\pm 2f(\mathbf{p} -e\mathbf{E} \Delta t) \right] n_\mathbf{p} (\Delta t) +f(\mathbf{p} -e\mathbf{E} \Delta t) , \label{ini-n}
\end{split}
\end{equation}
in which $n_\mathbf{p} (\Delta t) $ is the distribution with the state $|0,\mathrm{in} \rangle $ [Eq.\eqref{Bn_p(t)}].
This distribution can be interpreted as the summation of the initial distribution accelerated according to the classical
equation of motion $f(\mathbf{p} -e\mathbf{E} \Delta t) $ and the usual distribution of created particles $n_\mathbf{p} (\Delta t) $,
however, there is a quantum correction; Bose enhancement or Pauli blocking factor $1\pm 2f(\mathbf{p} -e\mathbf{E} \Delta t) $
[or one can consider $1\pm 2n_\mathbf{p} (\Delta t) $ are that factors.]

\begin{figure}[t]
 \begin{tabular}{cc}
  \begin{minipage}{0.5\textwidth}
   \begin{center}
    \includegraphics[scale=0.36,bb=0 0 577 461]{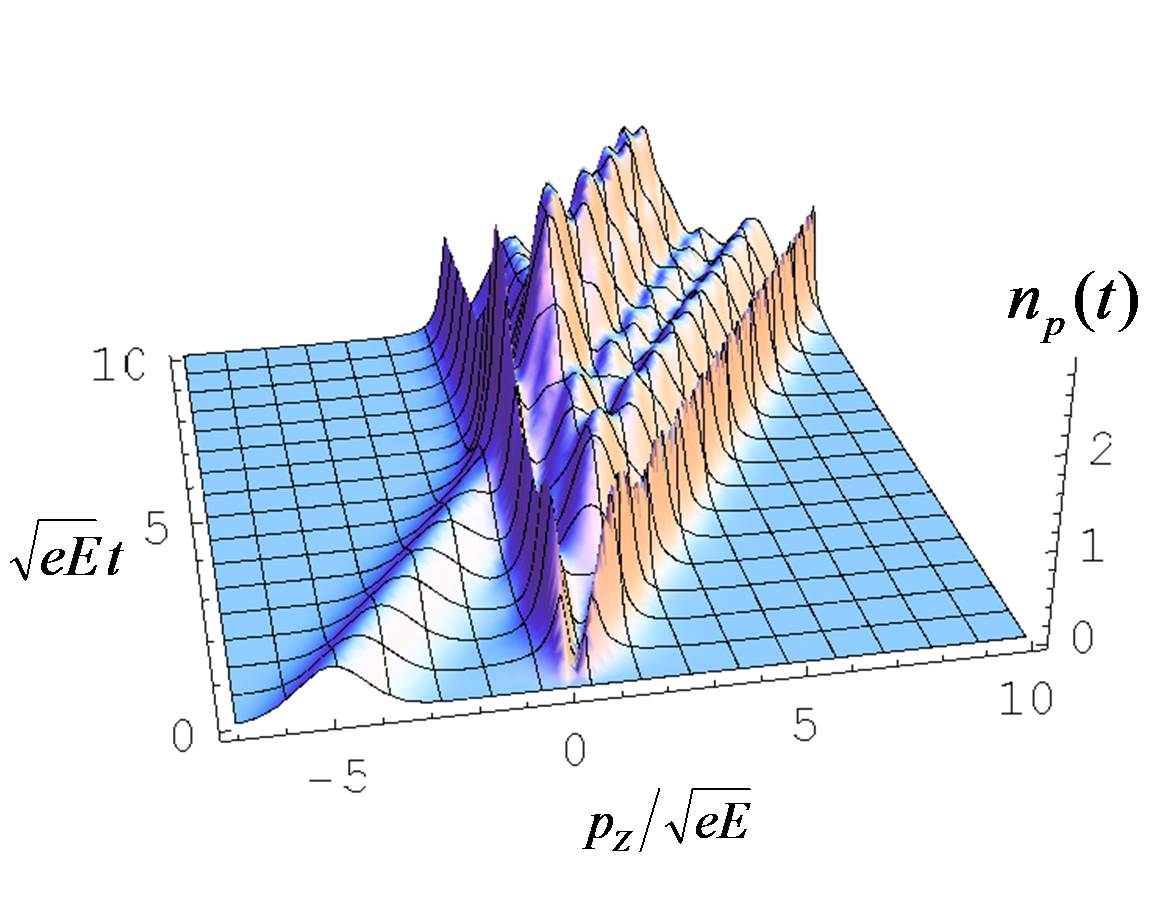}
   \end{center}
  \end{minipage} &
  \begin{minipage}{0.5\textwidth}
   \begin{center}
    \includegraphics[scale=0.36,bb=0 0 563 433]{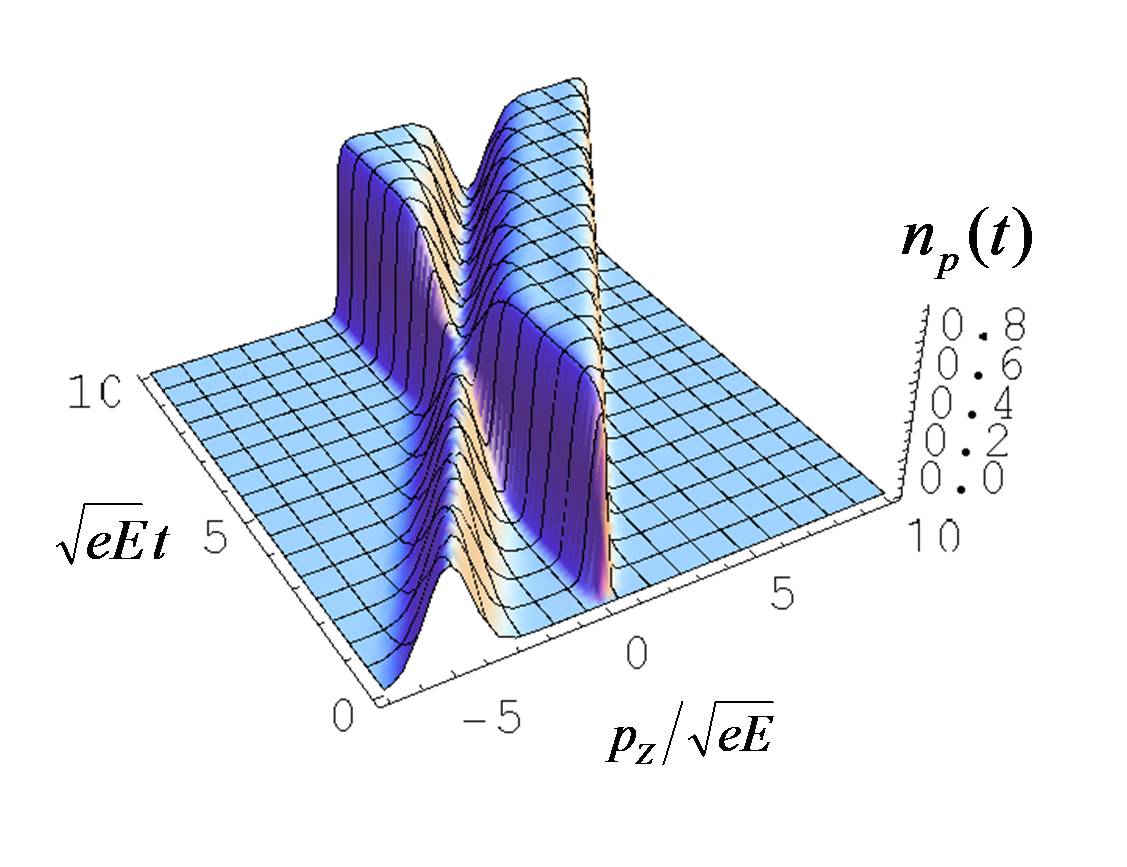}
   \end{center}
  \end{minipage} \\
    {\footnotesize (a) bosons, $a=0.01$ } &
    {\footnotesize (b) fermions, $a=0.01$ }
 \end{tabular}
 \caption{The longitudinal momentum distributions with the initial condition  
          $f(\mathbf{p} )=0.5 e^{-(p/\sqrt{eE} +5)^2 } $. }
 \label{fig:initial}
\end{figure}

The Bose enhancement factor tells us that if there are initial particles, the pair creation at the point
where the initial particles are located in momentum space is enhanced.
On the other hand, for fermion system, initial particles suppress the subsequent pair creation. 
Especially, if $n_\mathbf{p} (\Delta t) >0.5 $, an initial distribution reduces particle number; 
$\tilde{n} _\mathbf{p} (\Delta t) <n_\mathbf{p} (\Delta t) $. 
As an illustration of these phenomena, we show the time evolution of the distribution ($a=0.01$) 
with an initial condition of $f(\mathbf{p} )=0.5 e^{-(p/\sqrt{eE} +5)^2 } $ in Fig.\ref{fig:initial}. 
Actually, the boson distribution increases more than double when the initial particles merge into the region of
$n_\mathbf{p} (\Delta t) $. 
Meanwhile the fermion distribution gets dented because $n_\mathbf{p} (\Delta t) >0.5 $ holds for $a=0.01$.
These effects will be conspicuous in the evolution of a distribution with back reaction (Sec.\ref{sec:BR}). 

If one sets $|F(\mathbf{p} )|^2 =e^{-\omega _p /T} $, Eq.\eqref{ini-n} describes 
the Schwinger mechanism at finite temperature. 
Our result is slightly different from those in Ref.\cite{Hallin,Kim2007b}, 
where the Schwinger mechanism at finite temperature is studied by means of the density operator. 
In Ref.\cite{Hallin,Kim2007b}, thermal distribution is set to be free from acceleration by the electric field. 

%%%%%%%%%%%%%%%%%%%%%%%%%%%%%%%%%%%%%%%%
\subsection{Damping electric field} \label{subsec:damp}
In the previous subsection, we have dealt with the electric field which is kept constant after the switch-on.
However, in most of realistic situation, an electric field is damped away after sufficiently long time,
because (i) its energy are taken away by particles due to pair creation and subsequent acceleration,
and (ii) energy of particles would dissipate owing to collision among particles. 
The reason (i) will be treated in the next section, however, (ii) is not considered in this article. 
Furthermore, in heavy-ion collisions, there is another mechanism which makes an electric field decrease;
classical time evolution of gauge fields following expansion of the system \cite{Lappi,Fujii}.

Instead of studying these effects directly, we treat a damping electric field, which is controlled by hands.
As an example, we take an exponentially decreasing electric filed: 
\begin{equation}
E_z (t) = \begin{cases} 0 & (t<0) \\ E_0 e^{-t/\tau } & (t\geq 0) \end{cases} , \
A^3 (t) = \begin{cases} E_0 \tau & (t<0) \\ E_0 \tau e^{-t/\tau } & (t\geq 0). \end{cases}
 \label{dampingE}
\end{equation}
We have solved the Klein--Gordon equation and the Dirac equation numerically.

\begin{figure}[t]
 \begin{tabular}{cc}
  \begin{minipage}{0.5\textwidth}
   \begin{center}
    \includegraphics[scale=0.35,bb=0 0 603 436]{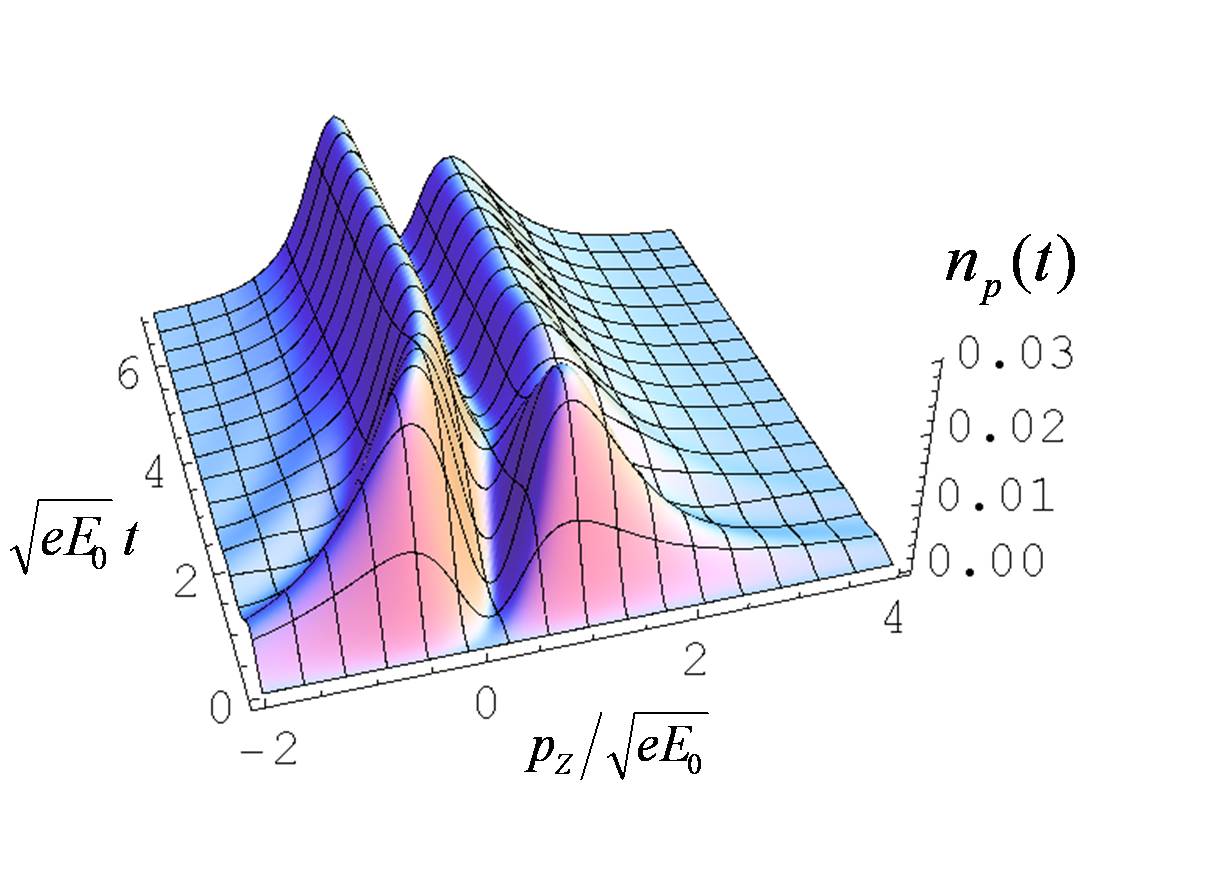} \\
    {\footnotesize (a) bosons $a=0.3$ }
   \end{center}
  \end{minipage} 
  \begin{minipage}{0.5\textwidth}
   \begin{center}
    \includegraphics[scale=0.35,bb=0 0 602 448]{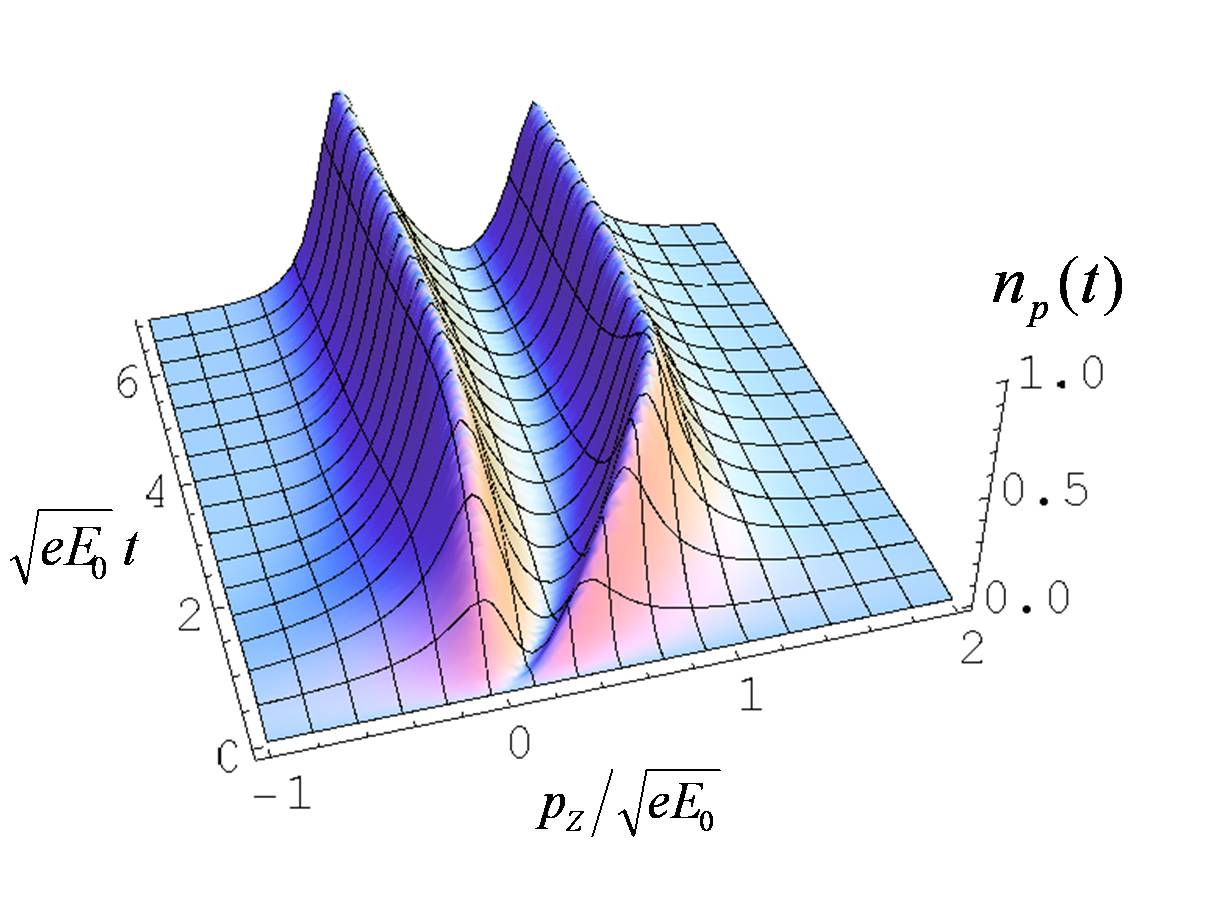} \\
    {\footnotesize (b) bosons $a=0.01$ }
   \end{center}
  \end{minipage} \\

  \begin{minipage}{0.5\textwidth}
   \begin{center}
    \includegraphics[scale=0.35,bb=0 0 603 440]{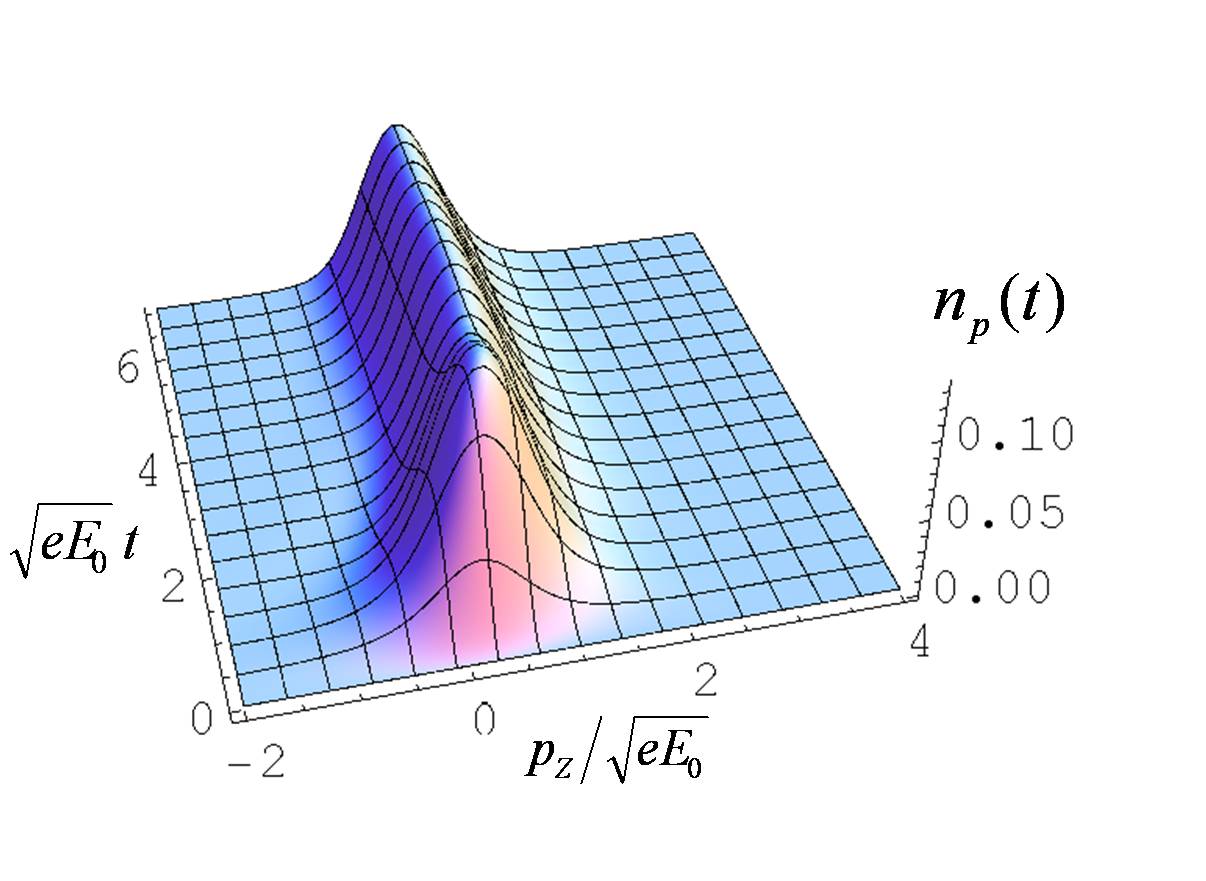} \\
    {\footnotesize (c) fermions $a=0.3$ }
   \end{center}
  \end{minipage} 
  \begin{minipage}{0.5\textwidth}
   \begin{center}
    \includegraphics[scale=0.35,bb=0 0 614 450]{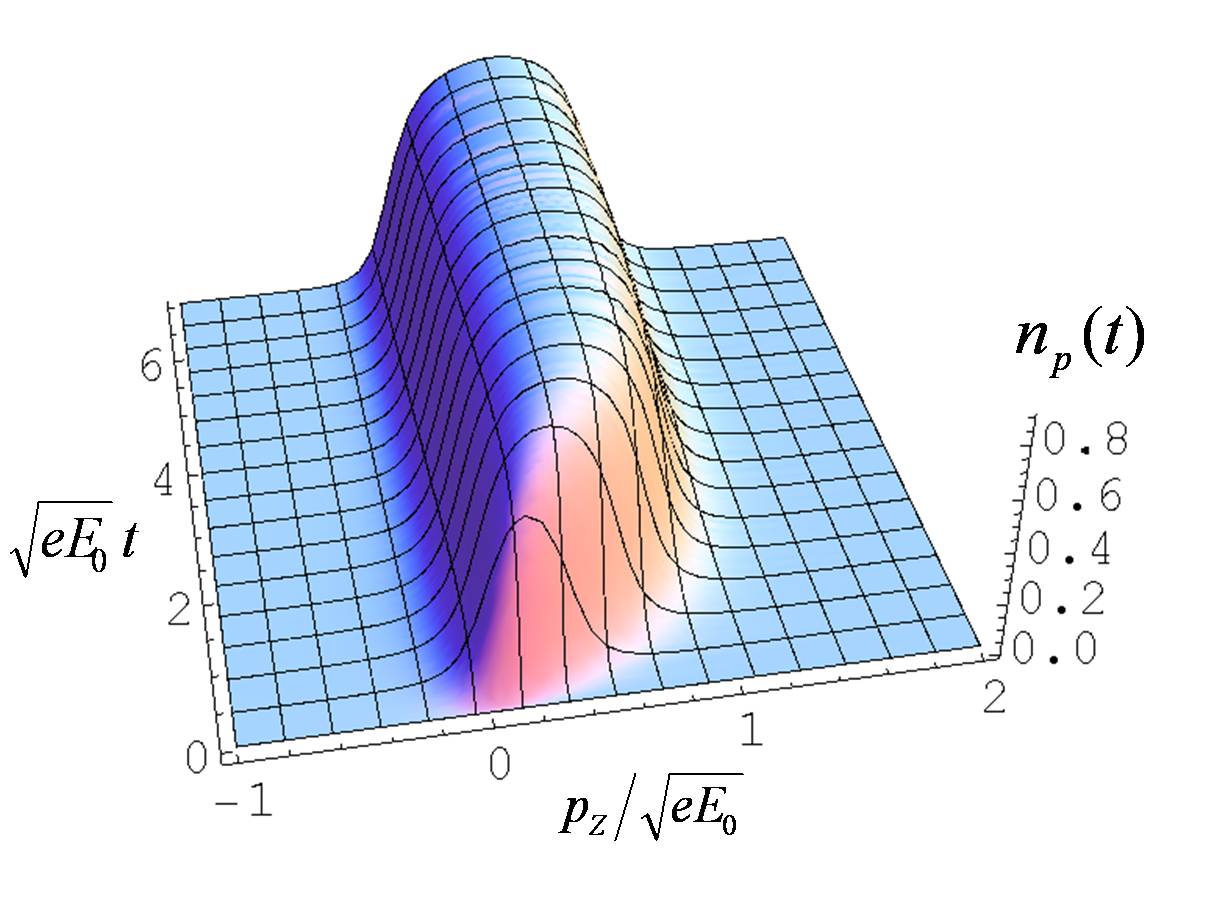} \\
    {\footnotesize (d) fermions $a=0.01$ }
   \end{center}
  \end{minipage}
 \end{tabular}
 \caption{The time evolution of the longitudinal distributions in the damping electric field
          \eqref{dampingE} with $\tau = 1.0/\sqrt{eE_0 } $. }
 \label{fig:dam-dis}
\end{figure}

\begin{figure}[t] 
  \begin{tabular}{cc}
  \begin{minipage}{0.5\textwidth}
   \begin{center}
    \includegraphics[scale=0.46,bb=0 0 440 238]{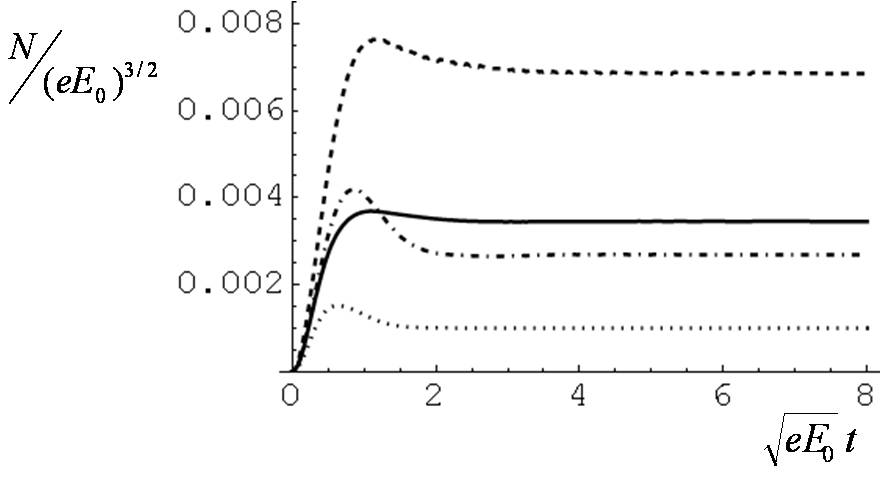} \\
  {\footnotesize (a) the total particle number }
   \end{center}
  \end{minipage} 
  \begin{minipage}{0.2\textwidth}
   \begin{center}
    \includegraphics[scale=0.52,bb=0 0 202 74]{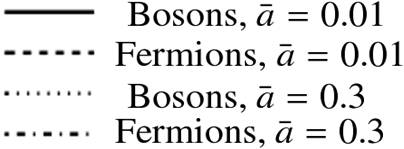}
   \end{center}
  \end{minipage} \\
  {} & {} \\
  \begin{minipage}{0.5\textwidth}
   \begin{center}
    \includegraphics[scale=0.46,bb=0 0 449 236]{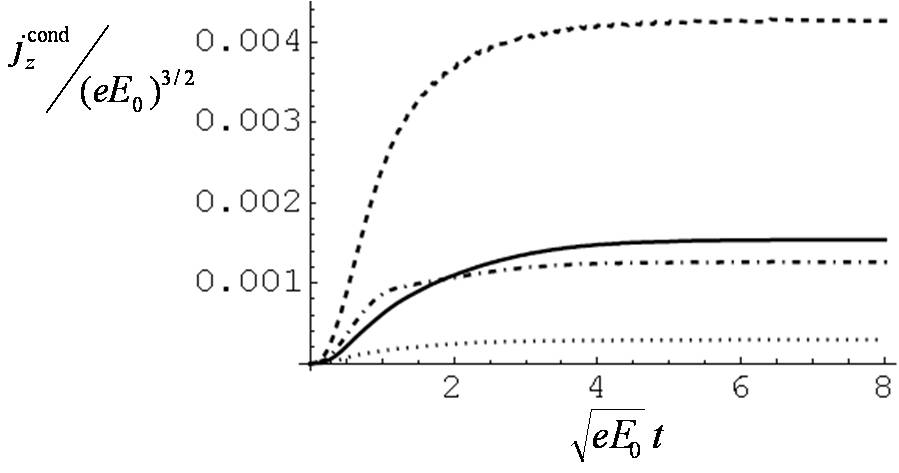} \\
    {\footnotesize (b) the conduction current }
   \end{center}
  \end{minipage}
  \begin{minipage}{0.5\textwidth}
   \begin{center}
    \includegraphics[scale=0.46,bb=0 0 432 189]{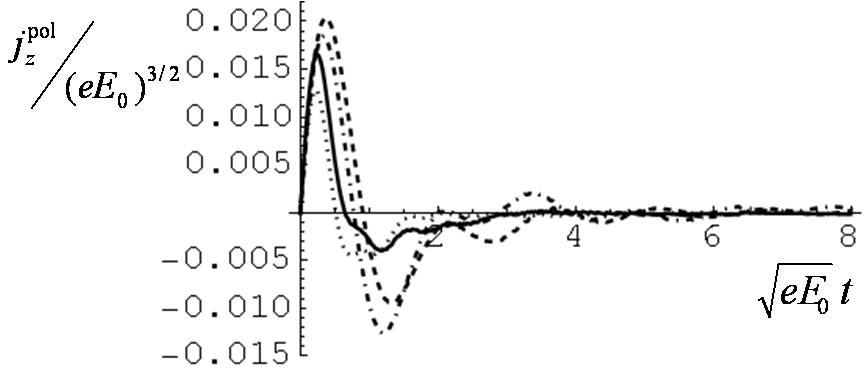} \\
    {\footnotesize (c) the polarization current }
   \end{center}
  \end{minipage} 
 \end{tabular}
 \caption{The time evolution of the total particle number density and the current density
          in the damping electric field \eqref{dampingE} with $\tau = 1.0/\sqrt{eE_0 }, e=1 $.
          For fermions, these values are plotted after averaged over the spin states. }
 \label{fig:damNandJ}
\end{figure} 

A generalized expression of the Bogoliubov coefficients under the gauge $\mathbf{A} (t)=(0,0,A^3 (t))$ is
\begin{equation}
\begin{matrix}
\alpha _\mathbf{p} (t) = \frac{e^{i\omega _\mathbf{p} t } }{\sqrt{2\omega _\mathbf{p} } }
\left[ \omega _\mathbf{p} {}_+ \! \phi _{\mathbf{p} +e\mathbf{A} } ^\mathrm{in} (t)
+i{}_+ \! \dot{\phi } _{\mathbf{p} +e\mathbf{A} } ^\mathrm{in} (t) \right] \\
\beta _\mathbf{p} ^* (t) = \frac{e^{-i\omega _\mathbf{p} t } }{\sqrt{2\omega _\mathbf{p} } }
\left[ \omega _\mathbf{p} {}_+ \! \phi _{\mathbf{p} +e\mathbf{A} } ^\mathrm{in} (t)
-i{}_+ \! \dot{\phi } _{\mathbf{p} +e\mathbf{A} } ^\mathrm{in} (t) \right] 
\end{matrix}
\label{gBab}
\end{equation}
for bosons and
\begin{equation}
\begin{matrix}
\alpha _\mathbf{p} (t) = \frac{e^{i\omega _p t} }{\sqrt{2\omega _p (\omega _p -p_z )
 (\omega _{\mathbf{p} +e\mathbf{A} } -p_z -eA^3 ) } }
 \left[ (\omega _p -p_z )(i{}_+ \! \dot{\phi } _{\mathbf{p} +e\mathbf{A} } ^\mathrm{in} -p_z {}_+ \! \phi _{\mathbf{p} +e\mathbf{A} } ^\mathrm{in} )
 +m_\mathrm{T} ^2 {}_+ \! \phi _{\mathbf{p} +e\mathbf{A} } ^\mathrm{in} ) \right] \\
\beta _\mathbf{p} ^* (t) = \frac{e^{-i\omega _p t} }{\sqrt{2\omega _p (\omega _p +p_z )
 (\omega _{\mathbf{p} +e\mathbf{A} } -p_z -eA^3 )} }
 \left[ (\omega _p +p_z )(i{}_+ \! \dot{\phi } _{\mathbf{p} +e\mathbf{A} } ^\mathrm{in} -p_z {}_+ \! \phi _{\mathbf{p} +e\mathbf{A} } ^\mathrm{in} )
 -m_\mathrm{T} ^2 {}_+ \! \phi _{\mathbf{p} +e\mathbf{A} } ^\mathrm{in} ) \right] ,
\end{matrix}
\label{gDab}
\end{equation}
for fermions, 
where ${}_\pm \! \phi _\mathbf{p} ^\mathrm{in} (t) $ are the solutions of Eq.\eqref{originalKG} for bosons 
or Eq.\eqref{2ndDirac} for fermions
with the initial condition ${}_\pm \! \phi _\mathbf{p} ^\mathrm{in} (t) =\frac{1}{\sqrt{2\omega _p } } e^{\mp i\omega _p t}$ 
for $t\leq 0$, respectively.
Using these equations, we can calculate the distribution function, the total particle density and the charge current.

The results of the numerical calculations are shown in Figs.\ref{fig:dam-dis} and \ref{fig:damNandJ}. 
The distribution functions evolve in time naturally as one can expect from 
the results in the previous subsection (Fig.\ref{fig:BFn_p(t)}): 
(i) rise up in a similar way to Fig.\ref{fig:BFn_p(t)}
(two peaks structure in Fig.\ref{fig:dam-dis}(a) and (b) are also seen in Fig.\ref{fig:BFn_p(t)}(a) and (b) 
just after the switch-on),
and (ii) accelerated to get higher momentum, however, 
(iii) pair creation and acceleration is stopped after the electric field
is damped away and there left a steady distribution. 

Before the switch-on and after the electric field is damped away, our particle picture is absolutely reliable
owing to absence of the electric field. 
Furthermore, the instantaneous particle picture gives us the proper description of 
the time evolution of the system during intermediate time even in the presence of the electric field. 
During the evolution, there is no pathologic behavior in the distribution or the total particle density 
unlike the case in cosmological particle creation \cite{Fulling-textbook}. 
This observation as well as the results in the previous subsection would support the validity to use the particle picture
associated with the instantaneous positive and negative frequency, and to study time evolution of a distribution
in the presence of an electric field.

%%%%%%%%%%%%%%%%%%%%%%%%%%%%%%%%%%%%%%%%%%%%%%%%%%%%%%%%%%%%%%%%%%%%%%%%%%%%%%%%%%%%%%%%%
\section{Back reaction} \label{sec:BR}
So far, we have controlled the electromagnetic fields by hand and neglected back reaction; 
an electromagnetic field which is generated
by particles and antiparticles created by an original electric field.
To take account of back reaction, we treat an electromagnetic field as a dynamical variable.
However, we does not quantize it and treat it as a classical background.
This approximation may be valid for a strong electromagnetic field, quantum fluctuation of which is negligible.  

\begin{figure}[t]
 \begin{tabular}{cc}
  \begin{minipage}{0.5\textwidth}
   \begin{center}
    \includegraphics[scale=0.33,bb=0 0 636 403]{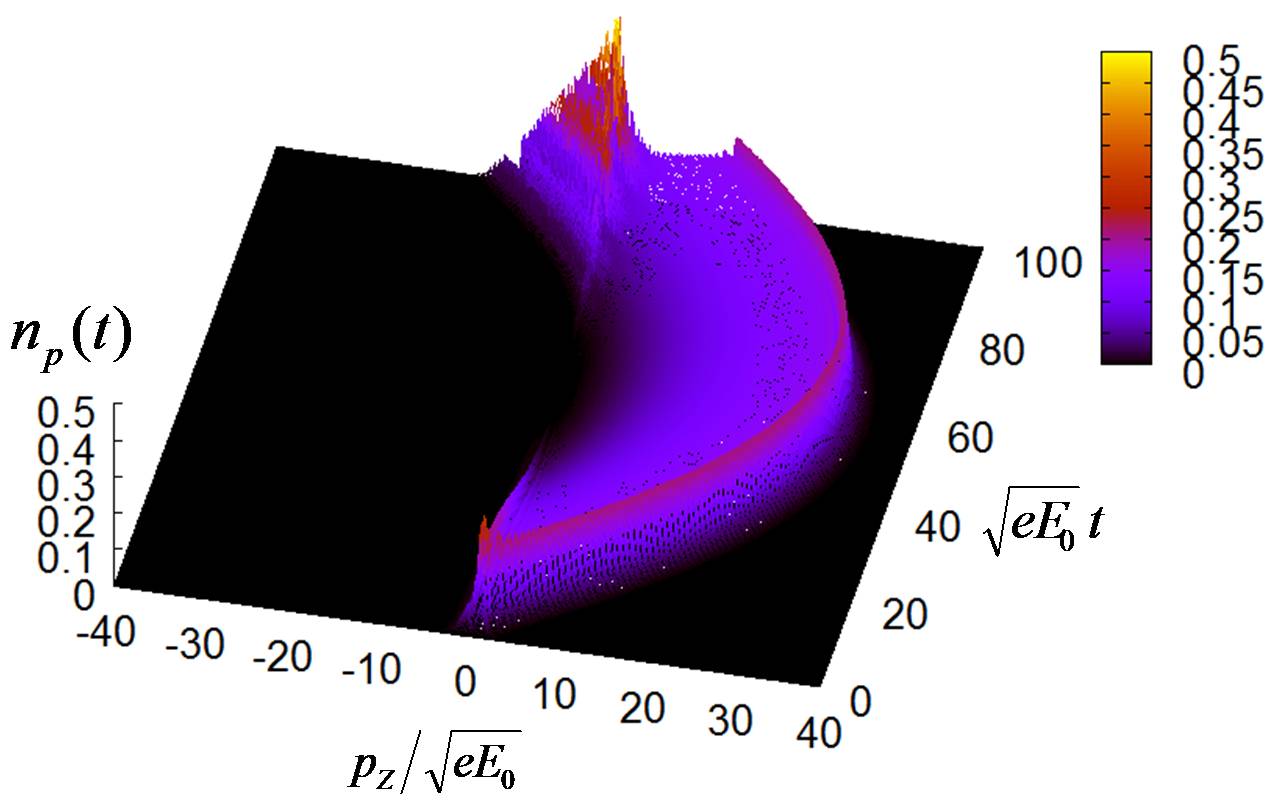} \\
    {\footnotesize (a) boson $a=0.3$ }
   \end{center}
  \end{minipage} &
  \begin{minipage}{0.5\textwidth}
   \begin{center}
    \includegraphics[scale=0.37,bb=0 0 531 400]{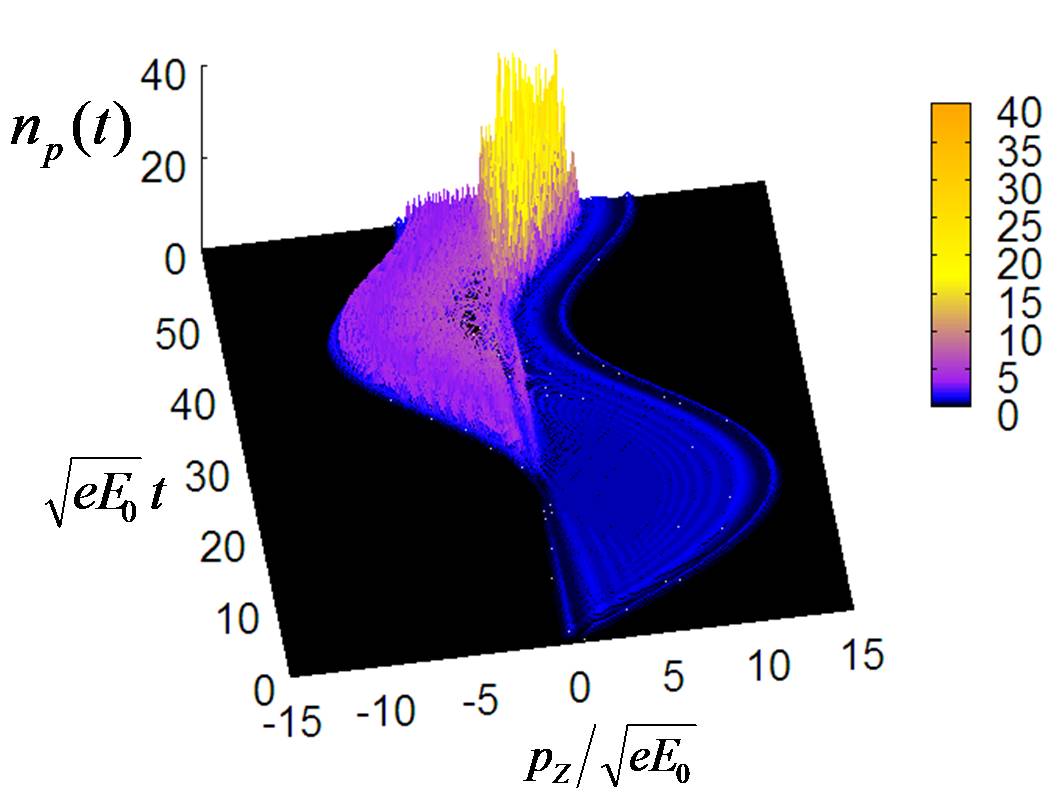} \\
    {\footnotesize (b) boson $a=0.01$ }
   \end{center}
  \end{minipage} \\
  {} & {} \\
  \begin{minipage}{0.5\textwidth}
   \begin{center}
    \includegraphics[scale=0.33,bb=0 0 636 389]{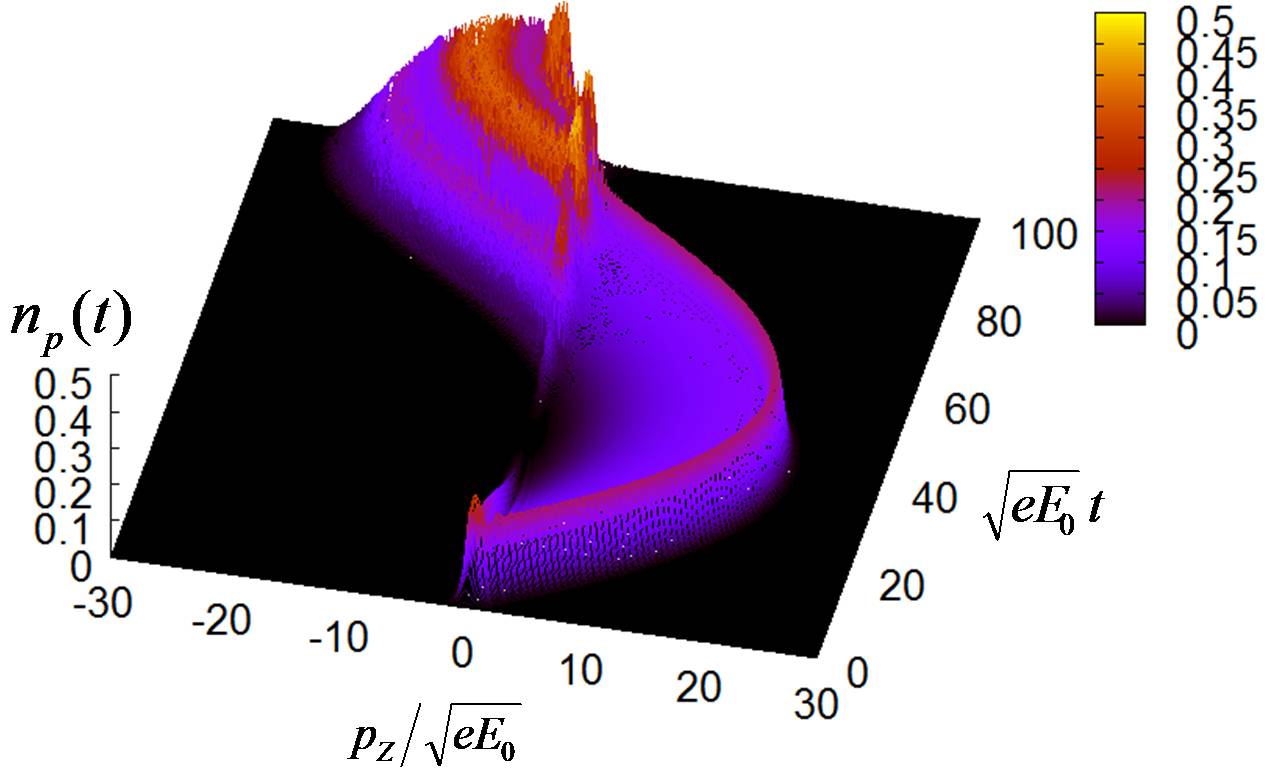} \\
    {\footnotesize (c) fermion $a=0.3$ }
   \end{center}
  \end{minipage} &
  \begin{minipage}{0.5\textwidth}
   \begin{center}
    \includegraphics[scale=0.37,bb=0 0 554 355]{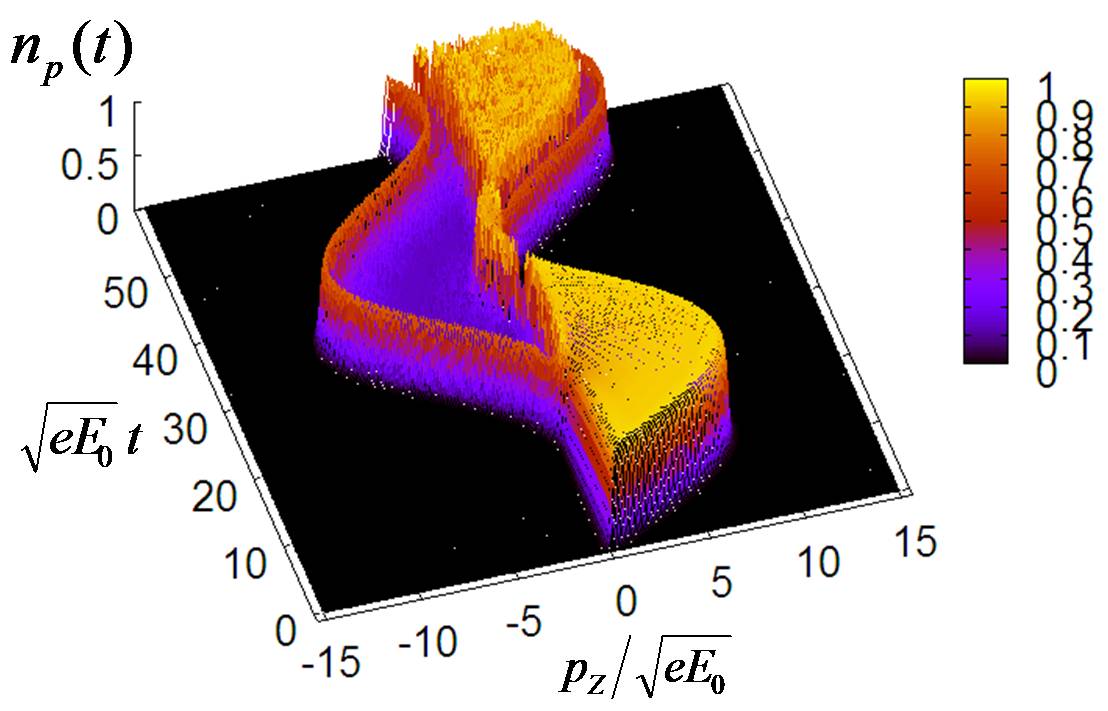} \\
    {\footnotesize (d) fermion $a=0.01$ }
   \end{center}
  \end{minipage} 
 \end{tabular}
 \caption{The time evolution of the longitudinal momentum distributions with back reaction $(e=1)$.}
 \label{fig:BR-distri}
\end{figure}

Now we use the Lagrangian density
\begin{equation}
\mathcal{L} = -\frac{1}{4} F_{\mu \nu } F^{\mu \nu} +(\partial _{\mu} -ieA_{\mu} )\phi ^{\dagger} 
              \! \cdot \! (\partial ^{\mu} +ieA^{\mu} )\phi -m^2 \phi ^{\dagger} \phi ,
\end{equation}
in which $F_{\mu \nu } =\partial _\mu A_\nu -\partial _\nu A_\mu $.
The equations of motion followed from this Lagrangian are the Klein--Gordon equation
\begin{equation}
\left[ (\partial _\mu +ieA_\mu )^2 +m^2 \right] \phi (x) = 0 \label{KG-BR}
\end{equation}
and the Maxwell equations
\begin{equation}
\nabla \cdot \mathbf{E} = \hat{j} _0 , \ \
\frac{d\mathbf{E} }{dt} -\nabla \times \mathbf{B} = -\hat{\mathbf{j} } ,
\end{equation}
in which $\hat{j} _\mu $ is given by Eq.\eqref{current-ope}.
The current operator $\hat{j} _\mu $ is replaced by its expectation value $\langle 0,\mathrm{in} |\hat{j} _\mu |0,\mathrm{in} \rangle $
because we treat the gauge field as a classical field.
A homogeneous initial electric field directing along the $z$-axis reduces 
the Maxwell equations into the single equation
\begin{equation}
\frac{dE_z }{dt} = -\frac{d^2 A^3 }{dt^2 } = -j_z , \label{Maxwell}
\end{equation}
where the $z$-component of the mean current $j_z $ is given by Eq.\eqref{Bcurrent} for bosons or 
Eq.\eqref{Fcurrent} for fermions.

We have solved the Klein--Gordon equation \eqref{KG-BR} [or the Dirac equation] and the Maxwell equation \eqref{Maxwell}
numerically with initial condition of the {\lq \lq}sudden-switch-on": $A^3 =0,\dot{A} ^3 =0$ for $t<0 $
and $A^3 =0,\dot{A} ^3 =-E_0 $ at $t=0$.
We calculate the distribution function and the charge current with the help of Eq.\eqref{gBab} or \eqref{gDab}.

\begin{figure}[t] 
  \begin{tabular}{cc}
  \begin{minipage}{0.5\textwidth}
   \begin{center}
    \includegraphics[scale=0.44,bb=0 0 440 245]{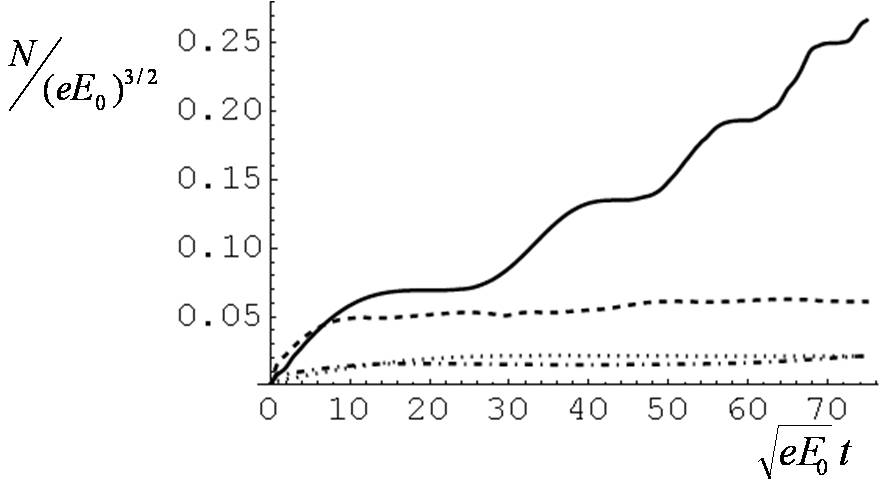} \\
    {\footnotesize (a) the total particle number density }
   \end{center}
  \end{minipage} 
  \begin{minipage}{0.2\textwidth}
   \begin{center}
    \includegraphics[scale=0.52,bb=0 0 202 74]{legend1.jpg} 
   \end{center}
  \end{minipage} \\
  {} & {} \\
  \begin{minipage}{0.5\textwidth}
   \begin{center}
    \includegraphics[scale=0.44,bb=0 0 473 196]{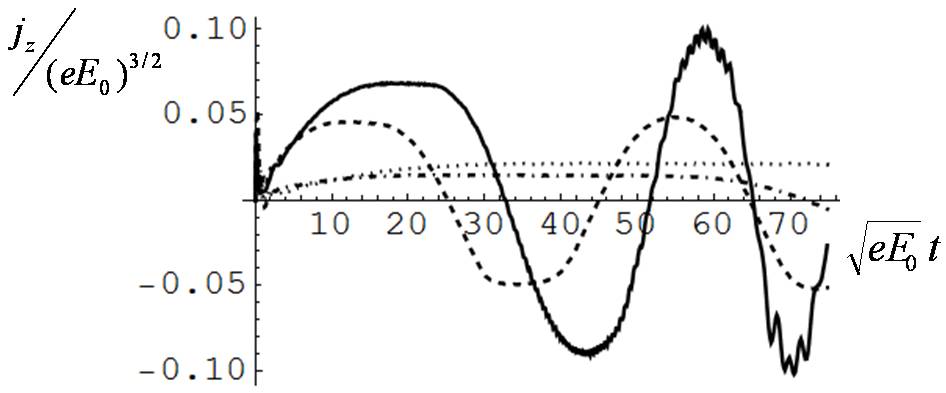} \\
    {\footnotesize (b) the current density }
   \end{center}
  \end{minipage}
  \begin{minipage}{0.5\textwidth}
   \begin{center}
    \includegraphics[scale=0.44,bb=0 0 428 212]{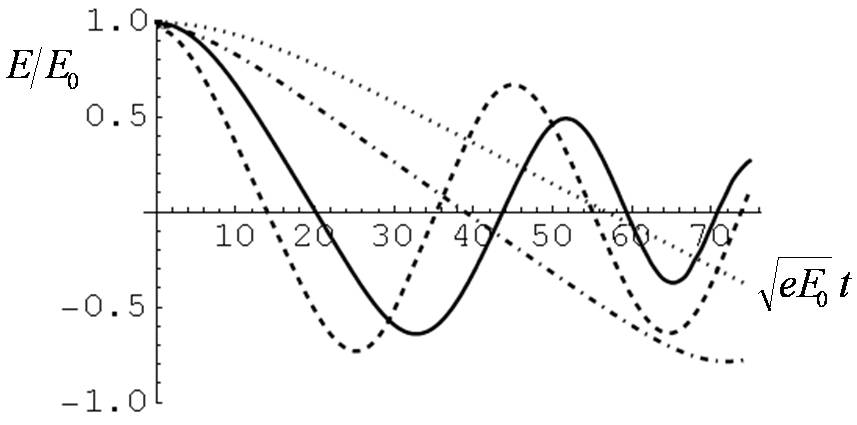} \\
    {\footnotesize (c) the electric field }
   \end{center}
  \end{minipage} 
 \end{tabular}
 \caption{The time evolution of the total particle number density, the current density and the electric field
          $(e=1)$. For fermions, the values averaged over the spin states are shown in (a) and (b). }
 \label{fig:BR-NJE}
\end{figure}

The results of the numerical calculations are shown in Figs.\ref{fig:BR-NJE} and \ref{fig:BR-distri}.
The longitudinal momentum distributions oscillate in momentum space and also
the current and the electric field show oscillation. 
Similar oscillatory behavior have been obtained 
in earlier works \cite{Kluger1991-1993,Kluger1993,Kluger1998,Asakawa-Matsui,Bloch,Vinnik},
although the employed methods are distinct. 

These oscillations can be understood as usual plasma oscillation. 
First, particles [antiparticles] are created with approximately 0 longitudinal momentum. 
They are accelerated to get positive [negative] momentum, and generate positive conduction current.
Then, the electric field is reduced by the current following the Maxwell equation \eqref{Maxwell}. 
After a while its strength reaches 0, at which time acceleration of particles is stopped, however, 
the current is still positive because particles have already gotten positive momentum.
Thus, the electric field continues to decrease through 0 to negative value 
(the direction of the electric field is flipped).
Then, particles start to be decelerated and finally their momentum become 0, 
at which time decrease of the electric field is stopped.
However, particles continue to be decelerated and get negative momentum, which generate negative conduction current, 
because the direction of the electric field is negative at that time.
Then, the negative conduction current increases the electric field and the direction of the electric field
is flipped again. 
These processes are repeated and result in plasma oscillation.

The oscillation is faster for (i) $\bar{a} =0.01 $ than $\bar{a} =0.3$, and (ii) fermions than bosons.
These are because the current is larger for (i) smaller value of $\bar{a} $,
and (ii) fermions than bosons due to spin degrees of freedom.

\begin{figure}[b] 
  \begin{tabular}{cc}
  \begin{minipage}{0.5\textwidth}
   \begin{center}
    \includegraphics[scale=0.44,bb=0 0 463 233]{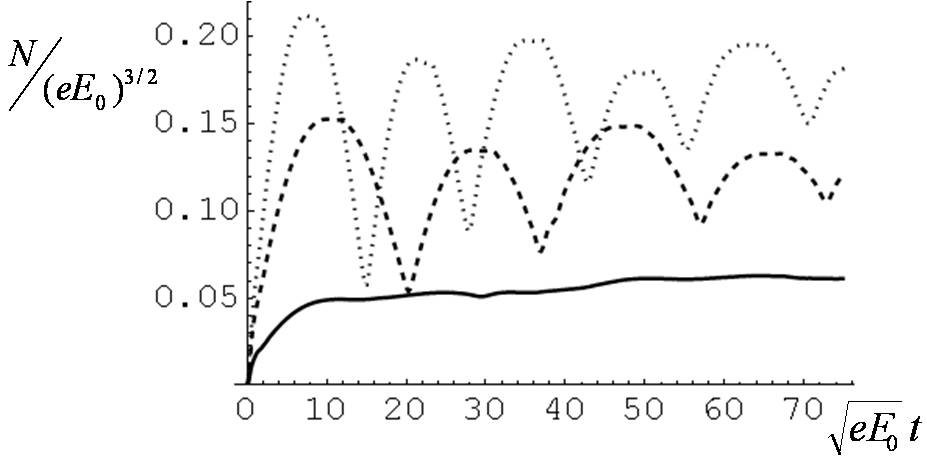} \\
    {\footnotesize (a) the total particle number }
   \end{center}
  \end{minipage} 
  \begin{minipage}{0.2\textwidth}
   \begin{center}
    \includegraphics[scale=0.5,bb=0 0 155 69]{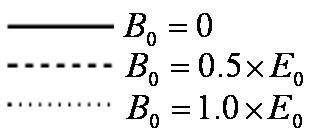} 
   \end{center}
  \end{minipage} \\
  {} & {} \\
  \begin{minipage}{0.5\textwidth}
   \begin{center}
    \includegraphics[scale=0.44,bb=0 0 467 212]{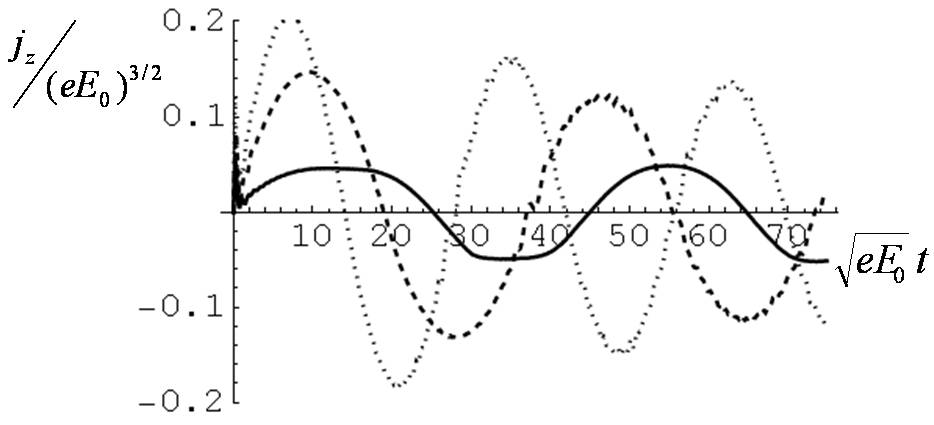} \\
    {\footnotesize (b) the current density }
   \end{center}
  \end{minipage} 
  \begin{minipage}{0.5\textwidth}
   \begin{center}
    \includegraphics[scale=0.44,bb=0 0 428 212]{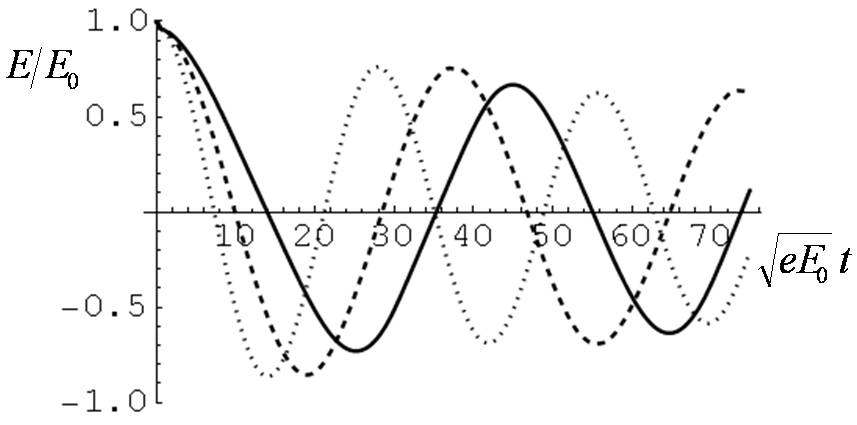} \\
    {\footnotesize (c) the electric field }
   \end{center}
  \end{minipage} 
 \end{tabular}
 \caption{The time evolution of the total particle number density, the current density and the electric field 
          for fermions with $\bar{a} =0.01, e=1$. The values averaged over the spin states are shown in (a) and (b).  }
 \label{fig:BR-NJEinB}
\end{figure}

\vspace{11pt}
Although plasma oscillation itself is a purely classical dynamics,
also quantum effect, namely Bose enhancement or Pauli blocking plays an important role during the time evolution. 
As explained in Sec.\ref{subsec:sudden}, particles of boson [fermion] enhance [suppress] the pair creation 
at the point where they locate in momentum space.  
These effects are more conspicuous for smaller $a$,
because the distribution depends on $a$ roughly as $e^{-2\pi a} $ [see Eq.\eqref{model}] 
and the effects of Bose enhancement and Pauli blocking are stronger for larger value of the distribution
[see Eq.\eqref{ini-n}].

When the distribution crosses the line of $p_z =0 $ in momentum space, the distribution of bosons
with $a=0.01$ shows dramatic increases. 
Because pair creation can occur only in the vicinity of the $p_z =0 $ line,
pair creation is enhanced when old particles come back on the line of $p_z =0$ in momentum space. 
Furthermore, the electric field takes its maximum value when the distribution crosses the $p_z =0$ line,
because the current density is 0 at that time.
This also enhances pair creation. 

On the other hand, the distribution of fermions with $a=0.01$ shows quit different behavior from that of bosons.
It repeats sharp increases and decreases when it crosses the line of $p_z =0$. 
This is because, as seen in Sec.\ref{subsec:sudden}, initial particles with $n_\mathbf{p} >0.5 $ 
cause pair annihilation. 

\begin{figure}[b] 
 \begin{center}
  \begin{tabular}{cc}
  \begin{minipage}{0.5\textwidth}
   \begin{center}
    \includegraphics[scale=0.42,bb=0 0 386 268]{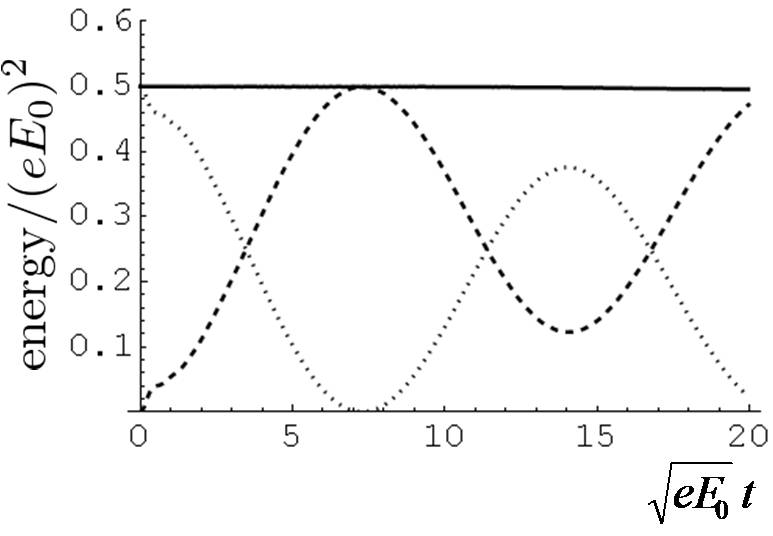}
   \end{center}
  \end{minipage} 
  \begin{minipage}{0.1\textwidth}
    \includegraphics[scale=0.42,bb=0 0 183 74]{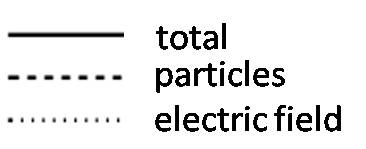} 
  \end{minipage}
  \end{tabular} 
  \caption{Energy balance}
  \label{fig:energy_balance}
 \end{center}
\end{figure}

\vspace{11pt}
Furthermore, we can apply a magnetic field parallel to the electric field.
In this homogeneous system, a magnetic field is unchanged by back reaction.
Thus, the effect of the magnetic field can be taken into account by the replacement \eqref{EtoEB} for bosons
or \eqref{F:EtoEB} for fermions. 
We show the time evolution of the total particle number density, the current density and the electric field 
for fermions in the longitudinal magnetic field in Fig.\ref{fig:BR-NJEinB}. 
In the magnetic field, also the total particle number density shows oscillation. 
This is because of Pauli blocking and suppression of non-zero Landau level: 
(i) the fermion distribution with small $a$ repeats sharp increases and decreases [see Fig.\ref{fig:BR-distri}(d)], 
and (ii) the lowest Landau level, that is the smallest $a$ mode mainly contributes to the total particle number in a 
strong magnetic field, while in the absence of a magnetic filed, larger $a$ modes [e.g. Fig.\ref{fig:BR-distri}(c)] 
also contribute.

The magnetic field enhances the current density of fermions by the mechanism explained in Sec.\ref{subsubsec:fermion-coll}.
With back reaction, enhancement of the current makes the time evolution faster. 
This mechanism may be relevant in the time evolution of the Glasma flux tube. 

\vspace{11pt}
Energy balance between particles and the electric field is shown in Fig.\ref{fig:energy_balance} 
for fermions with $\bar{a} =0.01, e=1$. 
Total energy conservation is automatically guaranteed in our particle picture and associated
regularization method \cite{Kluger1998} (see also \cite{Gatoff}). 
The expectation value of the energy density operator, which is regularized by normal ordering is
$2\int \! \frac{d^3 p}{(2\pi )^3 } \omega _p n_\mathbf{p} (t) $. 
Then, the time-derivative of total energy $\epsilon $ is 
\begin{equation}
\begin{split}
\frac{d\epsilon}{dt} &= \frac{d}{dt} \left[ \frac{1}{2} E^2 
  + 2\int \! \frac{d^3 p}{(2\pi )^3 } \omega _p n_\mathbf{p} (t) \right] \\
 &= E\frac{dE}{dt} +2eE\int \! \frac{d^3 p}{(2\pi )^3 } \frac{p_z }{\omega _p } n_\mathbf{p} (t) 
    +2\int \! \frac{d^3 p}{(2\pi )^3 } \omega _p S(t,\mathbf{p} ) \\
 &= E\left( \frac{dE}{dt} +j^\text{cond} +j^\text{pol} \right) = 0 ,
\end{split}
\end{equation} 
in which $\frac{d}{dt} =\frac{\partial }{\partial t} +eE\frac{\partial }{\partial p_z } $ and
the Maxwell equation \eqref{Maxwell} are used.  

\vspace{11pt}
Before closing this section, let us estimate a time scale of back reaction. 
As a time scale, we define the time when the electric field first reduces its strength to 0 as $t_V $. 
Although we have obtained $t_V $ by the numerical calculation, 
we estimate it analytically to see its dependence on parameter.
As zeroth-order approximation, we employ the modeled expression of the conduction current 
without back reaction \eqref{approxi-jcond}:
\begin{equation}
j_z ^\text{cond} (t) \simeq N_d \frac{2e(eE_0 )^2 }{(2\pi )^3 } e^{-\frac{\pi m^2 }{eE_0 } } t \ \theta (t) , 
 \label{approxi-jcond2}
\end{equation}
in which the electric field is fixed to the initial value and
$N_d$ is introduced as a number of inner degrees of freedom: spin, flavor and color.
Neglecting the polarization current, we substitute Eq.\eqref{approxi-jcond2} into the Maxwell equation \eqref{Maxwell}
and solve it.
Of course, an oscillatory behavior cannot be obtained by this approximation,
however, $t_V $ can:
\begin{equation}
t_V \simeq  \sqrt{ \frac{(2\pi )^3 }{N_d e^3 E_0 } } e^{\frac{\pi m^2 }{2eE_0 } } . \label{tV}
\end{equation} 
Eq.\eqref{tV} is an underestimate because we use the current without back reaction.
Actually, Eq.\eqref{tV} gives a smaller value of $t_V $ 
than the result of numerical calculations.
However, their discrepancy is less than factor 2
and Eq.\eqref{tV} correctly describes the $\bar{a} =\frac{m^2 }{2eE_0 } $ and boson-fermion dependence 
of $t_V $ obtained by numerical calculations. 
Thus, we can safely use Eq.\eqref{tV} for order estimation of $t_V $. 

From Eq.\eqref{tV}, we can notice that the stronger the initial electric field, the shorter time it takes to vanish. 
This is because the conduction current \eqref{approxi-jcond2} increases quadratically with increasing $E_0 $.

In the presence of the magnetic field parallel to the electric field,
$t_V $ for fermions is estimated as
\begin{equation}
t_V \simeq \sqrt{ \frac{(2\pi )^3 }{N_d e^3 E_0 } \frac{\tanh \pi \frac{B_0 }{E_0 } }{\pi \frac{B_0 }{E_0 } } } 
 e^{\frac{\pi m^2 }{2eE_0 } } , \label{tVinB}
\end{equation}
which is a decreasing function of the magnetic field strength $B_0 $.

%%%%%%%%%%%%%%%%%%%%%%%%%%%%%%%%%%%%%%%%%%%%%%%%%%%%%%%%%%%%%%%%%%%%%%%%%%%%%%%%%%%%%%%%%
\section{Summary and discussion} \label{sec:summary}
Our aim has been to make the physics in the Schwinger mechanism clearer.
In particular, we have explored the momentum distributions of created particles and its time evolution. 
It has been accomplished by defining the particle picture at each time using the instantaneous 
positive and negative frequency solutions. 
In the homogeneous and constant electric field, the static distributions have been obtained. 
We have shown that the longitudinal momentum dependence of the distribution arises from the boost-invariant electric
field through the quantization procedure.
Although this situation, namely the homogeneous and constant electric field is artificial,
our consideration on the boost-invariance may have practical importance 
because electric fields formed just after heavy-ion collisions are also (approximately) boost-invariant 
along the longitudinal direction. 

The dynamical view of pair creation has been acquired by dealing with the non-steady electric fields. 
After the electric field is switched on, the distribution rises up around 0-momentum. 
The time the distribution peaks first is earlier for smaller $a=\frac{m^2 +p_\text{T} ^2 }{2eE} $. 
That is, for fixed $m$ and $eE$, particles with $p_\text{T} =0 $ are created first 
and those with larger $p_\text{T} $ are later. 
In particular, the time to create a particle with $a=0$ is 0. 
After sufficiently long time, 
the transverse momentum distribution settles in a Gaussian-like form (Fig.\ref{fig:PT(t)}). 
On the other hand, in the longitudinal direction, particles are accelerated by the electric field 
according to the classical equation of motion $p_z =eEt +\text{const} $
(Fig.\ref{fig:BFn_p(t)}). 
Also the time evolution of the total particle number density and the current density have been computed.
They are suppressed by the longitudinal magnetic field for bosons and enhanced for fermions. 

Taking back reaction into account, we have shown how the coherent initial electric field decays
into particle pairs.
The electric field loses energy owing to creating and accelerating particle pairs. 
We have estimated the time $t_V $ when the electric field first reduces its strength to 0.  
In our calculation, however, the electric field does not vanishes away after $t_V $. 
It continues to exchange energy with particles and shows plasma oscillation. 

This plasma oscillation would be disturbed in reality by collision among particles,
effects of which are not included in our calculation. 
We have treated the gauge field as a classical field.
In this approximation, only particles and antiparticles having approximately 0-momentum
can be created or annihilated by the classical electric field. 
However, in reality, particles and antiparticles having non-zero momentum can collide and 
exchange their momentum through a quantum gauge field. 
These effects are expected to become relevant at $t\gtrsim t_V $, 
because the time evolution at early time $t\lesssim t_V $ is mainly controlled by the strong 
classical electric field. 
Perturbation may serve to treat these effects. 
However, usual perturbation based on asymptotic fields is no use. 
Real time formalism is needed.  
It is hoped to investigate this problem. 

To check whether this decay mechanism of the initial electric field due to back reaction
is relevant to heavy-ion collisions or not, 
let us estimate $t_V $ with parameter motivated by the Glasma; $m=0$ and
$eE_0 \simeq Q_S ^2 \sim 1 \ \text{GeV}^2 $, where $Q_S $ is a saturation scale.
Then, Eq.\eqref{tV} yields
\begin{equation}
t_V \sim \frac{3}{\sqrt{N_d e^2} } \ [\text{fm}]. \label{tVG}
\end{equation}
If one simply assumes that quarks and gluons in a non-Abelian electromagnetic field can be treated as 
fermions (spinor particles) and bosons (scalar particles), respectively 
in an Abelian electromagnetic field\footnote{For quarks this assumption is correct \cite{Gyulassy-Iwazaki}, however, 
for gluons not correct because of intrinsic magnetic moment \cite{Ambjorn-Nielsen-Olesen}. }, 
then $N_d = 2\times N_f \times N_c +2\times [N_c ^2 -1 -(N_c -1)] = 2\times 3\times 3+2\times 6 = 30 $, 
where $N_c -1 $ is the dimension of the Cartan subspace of SU($N_c$). 
Thus, if $e^2 \gtrsim \mathcal{O} (1) $, $t_V $ can be the order or less than 1 fm. 
This result would verify the importance of the Schwinger mechanism at the initial stage of heavy-ion collisions,
although our estimation is rather rough. 
Of course, to discuss thermalization of the system, we must take a quantum gauge field into account.

Furthermore, we have shown that the longitudinal magnetic field speeds up the time evolution
of the fermion system and makes $t_V$ smaller. 
In the case of $eE_0 \simeq eB_0 \simeq Q_S ^2 \sim 1 \ \text{GeV}^2 $, 
$t_V $ gets smaller than Eq.\eqref{tVG} by the factor $\sim \sqrt{\pi } $. 

In this paper, we have dealt with the spatially homogeneous electromagnetic fields. 
However, the flux tube has a finite size in both the transverse and the longitudinal direction,
and expands to the longitudinal direction. 
Thus, for application to particle production in heavy-ion collisions, 
we need to consider finite size effects.
Effects of transverse finite size have been studied in Ref.\cite{Gatoff-Wong,Wong-Wang-Wu,Schonfeld},
however, the flux tube has been treated as a static object. 
In reality, the flux tube evolves dynamically by back reaction. 
In particular, back reaction to a magnetic field can be taken only if one treats a non-uniform field.  
Effects of longitudinal boundaries, especially in the case that boundaries move with the speed of light is also 
open to argument. 
Furthermore, consideration for interaction and correlation between flux tubes may have importance. 
Our present study would give the basis for further investigation of these effects. 

\textit{Note added in proof:}
After the completion of this work, the auther become aware that 
the time scale of back reaction has been studied by S.P.Gavrilov and D.M.Gitman \cite{Gavrilov2008} 
as a consistency restriction on maximal electric field strength,
and the condition corresponding to Eq.\eqref{tV} of us has been obtained. 

%%%%%%%%%%%%%%%%%%%%%%%%%%%%%%%%%%%%%%%%%%%%%%%%%%%%%%%%%%%%%%%%%%%%%%%%%%%%%%%%%%%%%%%%%
\section*{Acknowledgments}
The author would like to thank Professor T.Matsui for helpful discussions, crucial 
comments and careful reading of the manuscript. 
He also acknowledges Professors H.Fujii and K.Itakura for useful discussions, and
Professors T.Hatsuda, O.Morimatsu, T.Hirano, 
S.Sasaki for their interests and comments.
This work is partly supported by JSPS research fellowships for Young Scientists.

%%%%%%%%%%%%%%%%%%%%%%%%%%%%%%%%%%%%%%%%%%%%%%%%%%%%%%%%%%%%%%%%%%%%%%%%%%%%%%%%%%%%%%%%%
%%%%%%%%%%%%%%%%%%%%%%%%%%%%%%%%%%%%%%%%%%%%%%%%%%%%%%%%%%%%%%%%%%%%%%%%%%%%%%%%%%%%%%%%%
%%%%%%%%%%%%%%%%%%%%%%%%%%%%%%%%%%%%%%%%%%%%%%%%%%%%%%%%%%%%%%%%%%%%%%%%%%%%%%%%%%%%%%%%%
\appendix

\section{First-order Klein--Gordon equation} \label{sec:1stKG}
In this appendix we briefly review a method to convert the Klein--Gordon equation
\begin{equation}
\left[ \left( \partial _{\mu } +ieA_{\mu } \right) ^2 +m^2 \right] \phi (x) =0 \label{originalKG}
\end{equation}
into a first-order differential equation with respect to time-derivative \cite{Feshbach}.

We introduce the 2-component expression
\begin{equation}
\Phi (x) \equiv \frac{1}{2} \left( \begin{array}{c}
                            \phi +\frac{i}{m} (\partial _0 +ieA_0 ) \phi \\
                            \phi -\frac{i}{m} (\partial _0 +ieA_0 ) \phi
                          \end{array} \right) . \label{2-component}
\end{equation}
Then, the Klein--Gordon equation \eqref{originalKG} can be rewritten as
\begin{equation}
(\partial _0 +ieA_0 ) \Phi = \left[ \frac{i}{2m} (\nabla -ie\mathbf{A} )^2 (i\sigma _2 +\sigma _3 ) 
                              -im\sigma _3 \right] \Phi , \label{1stKG}
\end{equation}
which is first order with respect to time-derivative ($\sigma _i $ are the Pauli matrices).
The 2-component expression reflects the degrees of freedom of particle and antiparticle.

The inner product
\begin{equation}
(\phi _1 ,\phi _2 ) = i\int \! d^3 \! x \left[ \phi _1 ^{\dagger } \cdot (\partial _0 +ieA_0 )\phi _2
                       -(\partial _0 -ieA_0 ) \phi _1 ^{\dagger } \cdot \phi _2 \right] \label{1-inner}
\end{equation}
is expressed by $\Phi (x) $ as
\begin{equation}
(\phi _1 ,\phi _2 ) = 2m\int \! d^3 \! x \Phi _1 ^{\dagger } \sigma _3 \Phi _2 , \label{2-inner}
\end{equation}
which does not contain time-derivative explicitly unlike Eq.\eqref{1-inner}.

%%%%%%%%%%%%%%%%%%%%%%%%%%%%%%%%%%%%%%%%%%%%%%%%%%%%%%%%%%%%%%%%%
\section{Second-order Dirac equation} \label{sec:2ndD}

We explain a method to convert the Dirac equation
\begin{equation}
\left[ \gamma ^{\mu } (i\partial _{\mu } -eA_{\mu } ) -m\right] \psi (x) = 0 \label{originalDirac}
\end{equation}
into a 2nd order differential equation and solve it 
in the presence of a background electromagnetic field \cite{Nikishov1970a}. 
In the following, we employ the $\gamma$-matrices in the standard Dirac representation:
\begin{equation}
\gamma ^0 = \left( \begin{array}{cc} 1 & 0 \\ 0 & -1 \end{array} \right) , \
\gamma ^i = \left( \begin{array}{cc} 0 & \sigma ^i \\ -\sigma ^i & 0 \end{array} \right) . 
\end{equation}

Inserting the equation
\begin{equation}
\psi (x) = \left[ \gamma ^{\mu } (i\partial _{\mu } -eA_{\mu } ) +m\right] Z(x) \label{ZtoPsi}
\end{equation}
into \eqref{originalDirac} yields the equation for $Z(x) $: 
\begin{equation}
\left[ (\partial _{\mu } +ieA_{\mu } )^2 +\frac{ie}{2} \gamma ^{\mu } \gamma ^{\nu } F_{\mu \nu } +m^2 \right] Z(x) =0 ,
\label{ZDirac}
\end{equation}
where $Z(x) $ is a 4-component spinor and $F_{\mu \nu } $ is the field strength: 
$F_{\mu \nu } =\partial _{\mu } A_{\nu } -\partial _{\nu } A_{\mu } $.
The term $\frac{ie}{2} \gamma ^{\mu } \gamma ^{\nu } F_{\mu \nu } $ represents spin-electromagnetic field interaction.

Because $\frac{i}{2} \gamma ^{\mu } \gamma ^{\nu } F_{\mu \nu } $ has $4\times 4$-matrix structure with respect to 
spinor indices, it has four independent eigenvectors $\Gamma _i (i=1,2,3,4)$.
We can expand $Z(x) $ by $\Gamma _i $ as
\begin{equation}
Z(x) = \sum _i \phi ^i (x) \Gamma _i .
\end{equation}
Then, each $\phi ^i (x) $ obey the Klein--Gordon-type equation 
\begin{equation}
\left[ (\partial _{\mu } +ieA_{\mu } )^2 +e\lambda _i +m^2 \right] \phi ^i (x) =0  \ \ (i=1,2,3,4), \label{2ndDirac}
\end{equation}
where $\lambda _i $ is the eigenvalue of $\frac{i}{2} \gamma ^{\mu } \gamma ^{\nu } F_{\mu \nu } $ corresponding to 
the eigenvector $\Gamma _i $.

In the case of the collinear electromagnetic field $\mathbf{E} = (0,0,E) , \mathbf{B} = (0,0,B) $,
\begin{equation}
\frac{i}{2} \gamma ^{\mu } \gamma ^{\nu } F_{\mu \nu } 
 = iE \left( \begin{array}{cc} 0 & \sigma _3 \\ \sigma _3 & 0 \end{array} \right)
   -B \left( \begin{array}{cc} \sigma _3 & 0 \\ 0 & \sigma _3 \end{array} \right) 
\end{equation}
and we can choose its eigenvectors and corresponding eigenvalues as follows: 
\begin{equation}
\begin{array}{cccc}
 \Gamma _1 = \left( \begin{array}{c} 1 \\ 0 \\ 1 \\ 0 \end{array} \right) &
 \Gamma _2 = \left( \begin{array}{c} 0 \\ 1 \\ 0 \\ -1 \end{array} \right) &
 \Gamma _3 = \left( \begin{array}{c} 1 \\ 0 \\ -1 \\ 0 \end{array} \right) &
 \Gamma _4 = \left( \begin{array}{c} 0 \\ 1 \\ 0 \\ 1 \end{array} \right) \\[+0.8cm]
 \lambda _1 = iE-B & \lambda _2 = iE+B & \lambda _3 = -iE-B & \lambda _4 = -iE+B .
\end{array} \label{Gamma1234}
\end{equation}

%%%%%%%%%%%%%%%%%%%%%%%%%%%%%%%%%%%%%%%%%%%%%%
\subsection{Free field }
First, we apply the method explained above to the free field case ($A^\mu =0$)
and derive solutions of the Dirac equation\footnote{In the free field case, 
it is more general and convenient to derive solutions of the Dirac equation
by boosting solutions in a rest frame. 
However, in the presence of an electromagnetic field, the boosting method
is not convenient because an electromagnetic field is transformed by boost, and free field solutions derived
by the method which is identical to that in electromagnetic fields makes 
the calculations in Sec.\ref{sec:PL},\ref{sec:Nonste},\ref{sec:BR} clear.}.

In this case, we can choose any independent vectors as $\Gamma _i $ 
since $\frac{i}{2} \gamma ^{\mu } \gamma ^{\nu } F_{\mu \nu } = 0 $.
Because the eigenvalue $\lambda _i $ is 0 for any $\Gamma _i $, Eq.\eqref{2ndDirac} becomes
\begin{equation}
\left( \partial _{\mu } ^2 +m^2 \right) \phi ^i (x) =0 \ \ \ (i=1,2,3,4) \label{"KG"}
\end{equation}
and its solutions are
\begin{equation}
{}_+ \phi _\mathbf{p} ^i (x) = \frac{1}{\sqrt{(2\pi )^3 2\omega _\mathbf{p} } } e^{-i\omega _\mathbf{p} t +i\mathbf{p} \cdot \mathbf{x} } , \
{}_- \phi _\mathbf{p} ^i (x) = \frac{1}{\sqrt{(2\pi )^3 2\omega _\mathbf{p} } } e^{i\omega _\mathbf{p} t +i\mathbf{p} \cdot \mathbf{x} } \ \ \
(i=1,2,3,4) . \label{pnkai}
\end{equation}
Define ${}_\pm \! \psi _\mathbf{p} ^i (x) $ as spinors made by Eq.\eqref{ZtoPsi} from the solutions \eqref{pnkai}: 
\begin{equation}
\begin{split}
{}_{\pm } \psi ^i _\mathbf{p} (x) &= \left[ i\slashchar{\partial} +m\right] {}_{\pm } \phi ^i _\mathbf{p} (x) \Gamma _i \\
  &= \left[ \pm \gamma ^0 \omega _\mathbf{p} -\mathbf{\gamma } \cdot \mathbf{p} +m \right] {}_{\pm } \phi ^i _\mathbf{p} (x) \Gamma _i .
  \label{psii}
\end{split}
\end{equation}
Notice that the operator in r.h.s is proportional to the projection operator to positive or negative energy state 
$\Lambda _{\pm } (\mathbf{p} ) $, respectively;
$\left[ \pm \gamma ^0 \omega _\mathbf{p} -\mathbf{\gamma } \cdot \mathbf{p} +m \right] = 2m\Lambda _{\pm } (\pm \mathbf{p} ) $.
For any choice of $\Gamma _i $, the spinor of positive[negative] energy state is gotten from 
the positive[negative] frequency solution of the {\lq \lq}Klein--Gordon" equation \eqref{"KG"}. 
Choice of $\Gamma _i $ affects only a spin state.

To calculate the solutions explicitly, we use $\Gamma _i $ defined by \eqref{Gamma1234}. 
It is sufficient to use only $\Gamma _1 $ and $\Gamma _2 $ because only four spinors are independent among
eight spinors which are derived from four kinds of $\Gamma_i $. 
Spinors made from $\Gamma _3 $ and $\Gamma _4 $ can be expressed by linear combination of 
those from $\Gamma _1 $ and $\Gamma _2 $.

Substituting the explicit form of $\Gamma _1 $ and $\Gamma _2 $ into Eq.\eqref{psii}, and rename
${}_\pm \! \psi _\mathbf{p} ^1 (x) $ as ${}_\pm \! \psi _{\mathbf{p} \uparrow } ^{\mathrm{free} } (x) $ and
${}_\pm \! \psi _\mathbf{p} ^2 (x) $ as ${}_\pm \! \psi _{\mathbf{p} \downarrow } ^{\mathrm{free} } (x) $,
then we obtain
\begin{allowdisplaybreaks}
\begin{gather}
{}_+ \! \psi _{\mathbf{p} \uparrow } ^{\mathrm{free} } (x) = \frac{1}{\sqrt{2} }
  \left[ \frac{m}{\sqrt{\omega _p -p_z } } \Gamma _1 -\frac{p_x +ip_y }{\sqrt{\omega _p -p_z }} \Gamma _2
  +\sqrt{\omega _p -p_z } \Gamma _3 \right]
  \frac{1}{\sqrt{(2\pi )^3 2\omega _p } } e^{-i\omega _p t+i\mathbf{p} \cdot \mathbf{x} } \notag \\
{}_- \! \psi _{\mathbf{p} \uparrow } ^{\mathrm{free} } (x) = \frac{1}{\sqrt{2} }
  \left[ \frac{m}{\sqrt{\omega _p +p_z } } \Gamma _1 -\frac{p_x +ip_y }{\sqrt{\omega _p +p_z }} \Gamma _2
  -\sqrt{\omega _p +p_z } \Gamma _3 \right]
  \frac{1}{\sqrt{(2\pi )^3 2\omega _p } } e^{+i\omega _p t+i\mathbf{p} \cdot \mathbf{x} } \notag \\
{}_+ \! \psi _{\mathbf{p} \downarrow } ^{\mathrm{free} } (x) = \frac{1}{\sqrt{2} }
  \left[ \frac{m}{\sqrt{\omega _p -p_z } } \Gamma _2 +\frac{p_x -ip_y }{\sqrt{\omega _p -p_z }} \Gamma _1
  +\sqrt{\omega _p -p_z } \Gamma _4 \right]
  \frac{1}{\sqrt{(2\pi )^3 2\omega _p } } e^{-i\omega _p t+i\mathbf{p} \cdot \mathbf{x} } \notag \\
{}_- \! \psi _{\mathbf{p} \downarrow } ^{\mathrm{free} } (x) = \frac{1}{\sqrt{2} }
  \left[ \frac{m}{\sqrt{\omega _p +p_z } } \Gamma _2 +\frac{p_x -ip_y }{\sqrt{\omega _p +p_z }} \Gamma _1
  -\sqrt{\omega _p +p_z } \Gamma _4 \right]
  \frac{1}{\sqrt{(2\pi )^3 2\omega _p } } e^{+i\omega _p t+i\mathbf{p} \cdot \mathbf{x} } .
 \label{freeDirac}
\end{gather}
\end{allowdisplaybreaks}
Because these solutions satisfy the orthonormal conditions
\begin{equation}
({}_+ \! \psi _{\mathbf{p} s} ^{\mathrm{free} } ,{}_+ \! \psi _{\mathbf{q} s^\prime } ^{\mathrm{free} } )
 = ({}_- \! \psi _{\mathbf{p} s} ^{\mathrm{free} } ,{}_- \! \psi _{\mathbf{q} s^\prime } ^{\mathrm{free} } )
 = \delta _{s s^\prime } \delta ^3 (\mathbf{p} -\mathbf{q} ) , \
({}_+ \! \psi _{\mathbf{p} s} ^{\mathrm{free} } ,{}_- \! \psi _{\mathbf{q} s^\prime } ^{\mathrm{free} } ) = 0
\ \ (s,s^\prime = \uparrow ,\downarrow ), 
\end{equation}
where the inner product is defined by
\begin{equation}
(\psi _1 ,\psi _2 ) \equiv \int \! d^3 x \psi _1 ^\dagger \psi _2 ,
\end{equation}
the quantized free field operator can be expanded by these solutions: 
\begin{equation}
\psi (x) = \sum _{s=\uparrow ,\downarrow } \int \! d^3 p \left[
 {}_+ \! \psi _{\mathbf{p} s} ^{\mathrm{free} } (x) a_{\mathbf{p} ,s} +
 {}_- \! \psi _{\mathbf{p} s} ^{\mathrm{free} } (x) b_{-\mathbf{p} ,s} ^{\dagger }
 \right] .
\end{equation}

%%%%%%%%%%%%%%%%%%%%%%%%%%%%%%%%%%%%%%%%%%%%%%%%%%%%%%%%
\subsection{In electromagnetic fields}
We solve the Dirac equation in the presence of the collinear electromagnetic field 
$\mathbf{E} = (0,0,E) , \mathbf{B} = (0,0,B) $.
Choice of gauge as $A^\mu = (0,-By,0,-Et) $ converts the {\lq \lq}Klein--Gordon" equation \eqref{2ndDirac} to
\begin{equation}
\left[ \partial _0 ^2 -(\partial _1 +ieBy)^2 -\partial _2 ^2 -(\partial _3 +ieEt)^2 
+m^2 +e\lambda _i \right] \phi ^i (x) =0 .
\end{equation}
This can be solved in the same manner with Sec.\ref{subsubsec:boson-coll}.
A solution is 
\begin{equation}
\phi _\mathbf{p} ^i (t,\mathbf{x} ) = \phi _\mathbf{p} ^i (t) 
 \sqrt{\frac{L}{2\pi } } \left( \frac{eB}{\pi } \right) ^\frac{1}{4} \frac{1}{\sqrt{n!} } D_n (\eta ) 
 \frac{1}{2\pi }e^{i(p_x x+p_z z)} 
\end{equation}
where the subscript $\mathbf{p} $ is the abbreviated expression for $(p_x ,p_z ,n)$ ($n$ is the Landau level) and
variables $\xi $ and $\eta $ are introduced as
\begin{equation}
\xi \equiv \sqrt{\frac{2}{eE} } (eEt +p_z ), \ \eta \equiv \sqrt{\frac{2}{eB} } (eBy +p_x ) .
\end{equation}
The time-dependent part $\phi _\mathbf{p} ^i (t) $ satisfies the equation
\begin{equation}
\left[ \frac{d^2 }{d\xi ^2 } +\frac{\xi ^2 }{4} +\frac{m^2 +e\lambda _i +\zeta _n }{2eE} \right] \phi ^i (t) = 0 ,
\label{"KG"EB}
\end{equation}
where $\zeta _n = (2n+1)eB$. 
We choose two sets of solutions of Eq.\eqref{"KG"EB} with the same motivation as explained in Sec.\ref{subsec:boson}:
\begin{equation}
\begin{matrix}
{}_+ \! \phi _\mathbf{p} ^{\mathrm{in} ,1} (t) = \frac{e^{-\frac{\pi }{4} a_n ^\uparrow } }{(2eE)^{\frac{1}{4} } }
                            D_{-ia_n ^\uparrow -1} ^* (-e^{\frac{\pi }{4} i} \xi ) , &
{}_- \! \phi _\mathbf{p} ^{\mathrm{in} ,1} (t) = \frac{e^{-\frac{\pi }{4} a_n ^\uparrow } }{(2eE)^{\frac{1}{4} } }
                            D_{-ia_n ^\uparrow } (-e^{\frac{\pi }{4} i} \xi ) \\
{}_+ \! \phi _\mathbf{p} ^{\mathrm{in} ,2} (t) = \frac{e^{-\frac{\pi }{4} a_n ^\downarrow } }{(2eE)^{\frac{1}{4} } }
                            D_{-ia_n ^\downarrow -1} ^* (-e^{\frac{\pi }{4} i} \xi ) , &
{}_- \! \phi _\mathbf{p} ^{\mathrm{in} ,2} (t) = \frac{e^{-\frac{\pi }{4} a_n ^\downarrow } }{(2eE)^{\frac{1}{4} } }
                            D_{-ia_n ^\downarrow } (-e^{\frac{\pi }{4} i} \xi )
\end{matrix} \label{outphii}
\end{equation}
and
\begin{equation}
\begin{matrix}
{}_+ \! \phi _\mathbf{p} ^{\mathrm{out} ,1} (t) = \frac{e^{-\frac{\pi }{4} a_n ^\uparrow } }{(2eE)^{\frac{1}{4} } }
                            D_{-ia_n ^\uparrow } (e^{\frac{\pi }{4} i} \xi ) , &
{}_- \! \phi _\mathbf{p} ^{\mathrm{out} ,1} (t) = \frac{e^{-\frac{\pi }{4} a_n ^\uparrow } }{(2eE)^{\frac{1}{4} } }
                            D_{-ia_n ^\uparrow -1} ^* (e^{\frac{\pi }{4} i} \xi ) \\
{}_+ \! \phi _\mathbf{p} ^{\mathrm{out} ,2} (t) = \frac{e^{-\frac{\pi }{4} a_n ^\downarrow } }{(2eE)^{\frac{1}{4} } }
                            D_{-ia_n ^\downarrow } (e^{\frac{\pi }{4} i} \xi ) , &
{}_- \! \phi _\mathbf{p} ^{\mathrm{out} ,2} (t) = \frac{e^{-\frac{\pi }{4} a_n ^\downarrow } }{(2eE)^{\frac{1}{4} } }
                            D_{-ia_n ^\downarrow -1} ^* (e^{\frac{\pi }{4} i} \xi ) 
\end{matrix} \label{inphii}
\end{equation}
where parameter $a_n ^\uparrow $ and $a_n ^\downarrow $ are defined by
\begin{equation}
\begin{matrix}
a_n ^\uparrow \equiv \frac{m^2 +\zeta _n + eB }{2eE} = \frac{m^2 +2neB }{2eE} \\
a_n ^\downarrow \equiv \frac{m^2 +\zeta _n - eB }{2eE} = \frac{m^2 +(2n+2)eB }{2eE}
\end{matrix} \
\ (n=0,1,\cdots ) . \label{an^pm}
\end{equation}

From these solutions, we can construct spinors using Eq.\eqref{ZtoPsi}. 
Define ${}_\pm \! \psi _{\mathbf{p} ,\uparrow} ^{\mathrm{in} /\mathrm{out}} (x) $ as which is derived by substituting
$Z(x)=N {}_\pm \! \phi _\mathbf{p} ^{\mathrm{in} /\mathrm{out} ,1} (t) D_n (\eta ) e^{i(p_x x+p_z z)} \Gamma _1 $
into Eq.\eqref{ZtoPsi} and ${}_\pm \! \psi _{\mathbf{p} ,\downarrow} ^{\mathrm{in} /\mathrm{out}} (x) $ by 
$Z(x)=N {}_\pm \! \phi _\mathbf{p} ^{\mathrm{in} /\mathrm{out} ,2} (t) D_n (\eta ) e^{i(p_x x+p_z z)} \Gamma _2 $
($N$ is a normalization constant).
The indices $\uparrow $ and $\downarrow $ distinguish the spin state: 
$\uparrow $ means spin parallel to the magnetic field
and $\downarrow $ means anti-parallel, which are known from Eq.\eqref{an^pm}.
The resultant spinors are
\begin{allowdisplaybreaks}
\begin{gather}
\begin{split}
{}_+ \! \psi _{\mathbf{p} \uparrow } ^\mathrm{in} (x) 
 &= \frac{e^{-\frac{\pi }{4} a_n ^\uparrow } }{2\sqrt{eE} } 
  \sqrt{\frac{L}{2\pi } } \left( \frac{eB}{\pi } \right) ^\frac{1}{4} \frac{1}{\sqrt{n!} }
  \left[ mD_{-ia_n ^\uparrow -1} ^* (-e^{\frac{\pi }{4} i} \xi ) D_n (\eta ) \Gamma _1 \right. \\[-8pt]
 & \left.
  -\sqrt{2eB} D_{-ia_n ^\uparrow -1} ^* (-e^{\frac{\pi }{4} i} \xi ) nD_{n-1} (\eta ) \Gamma _2
  +\sqrt{2eE} e^{\frac{\pi }{4} i} D_{-ia_n ^\uparrow } ^* (-e^{\frac{\pi }{4} i} \xi ) D_n (\eta ) \Gamma _3 \right]
  \frac{e^{i(p_x x+p_z z)} }{2\pi} 
\end{split} \notag \\
\begin{split}
{}_- \! \psi _{\mathbf{p} \uparrow } ^\mathrm{in} (x) 
 &= \frac{e^{-\frac{\pi }{4} a_n ^\uparrow } }{2\sqrt{eE} } 
  \sqrt{\frac{L}{2\pi } } \left( \frac{eB}{\pi } \right) ^\frac{1}{4} \frac{1}{\sqrt{n!} }
  \left[ \frac{m}{\sqrt{a_n ^\uparrow } } D_{-ia_n ^\uparrow } (-e^{\frac{\pi }{4} i} \xi ) D_n (\eta ) \Gamma _1 \right. \\[-8pt]
 & \left.
  -\sqrt{\frac{2eB}{a_n ^\uparrow } } D_{-ia_n ^\uparrow } (-e^{\frac{\pi }{4} i} \xi ) nD_{n-1} (\eta ) \Gamma _2
  -\sqrt{2eEa_n ^\uparrow } e^{\frac{\pi }{4} i} D_{-ia_n ^\uparrow -1 } (-e^{\frac{\pi }{4} i} \xi ) 
  D_n (\eta ) \Gamma _3 \right] \frac{e^{i(p_x x+p_z z)} }{2\pi}
\end{split} \notag \\
\begin{split}
{}_+ \! \psi _{\mathbf{p} \downarrow } ^\mathrm{in} (x) 
 &= \frac{e^{-\frac{\pi }{4} a_n ^\downarrow } }{2\sqrt{eE} } 
  \sqrt{\frac{L}{2\pi } } \left( \frac{eB}{\pi } \right) ^\frac{1}{4} \frac{1}{\sqrt{n!} }
  \left[ mD_{-ia_n ^\downarrow -1} ^* (-e^{\frac{\pi }{4} i} \xi ) D_n (\eta ) \Gamma _2 \right. \\[-8pt]
 & \left.
  +\sqrt{2eB} D_{-ia_n ^\downarrow -1} ^* (-e^{\frac{\pi }{4} i} \xi ) D_{n+1} (\eta ) \Gamma _1
  +\sqrt{2eE} e^{\frac{\pi }{4} i} D_{-ia_n ^\downarrow } ^* (-e^{\frac{\pi }{4} i} \xi ) D_n (\eta ) \Gamma _4 \right]
  \frac{e^{i(p_x x+p_z z)} }{2\pi} 
\end{split} \notag \\
\begin{split}
{}_- \! \psi _{\mathbf{p} \downarrow } ^\mathrm{in} (x) 
 &= \frac{e^{-\frac{\pi }{4} a_n ^\downarrow } }{2\sqrt{eE} } 
  \sqrt{\frac{L}{2\pi } } \left( \frac{eB}{\pi } \right) ^\frac{1}{4} \frac{1}{\sqrt{n!} }
  \left[ \frac{m}{\sqrt{a_n ^\downarrow } } D_{-ia_n ^\downarrow } (-e^{\frac{\pi }{4} i} \xi ) D_n (\eta ) \Gamma _2 \right. \\[-8pt]
 & \left.
  +\sqrt{\frac{2eB}{a_n ^\downarrow } } D_{-ia_n ^\downarrow } (-e^{\frac{\pi }{4} i} \xi ) D_{n+1} (\eta ) \Gamma _1
  -\sqrt{2eEa_n ^\downarrow } e^{\frac{\pi }{4} i} D_{-ia_n ^\downarrow -1 } (-e^{\frac{\pi }{4} i} \xi ) D_n (\eta ) \Gamma _4 \right]
  \frac{e^{i(p_x x+p_z z)} }{2\pi} 
\end{split} \label{psi^in_EM}
\end{gather}
\end{allowdisplaybreaks}
and
\begin{allowdisplaybreaks}
\begin{gather}
\begin{split}
{}_+ \! \psi _{\mathbf{p} \uparrow } ^\mathrm{out} (x) 
 &= \frac{e^{-\frac{\pi }{4} a_n ^\uparrow } }{2\sqrt{eE} } 
  \sqrt{\frac{L}{2\pi } } \left( \frac{eB}{\pi } \right) ^\frac{1}{4} \frac{1}{\sqrt{n!} }
  \left[ \frac{m}{\sqrt{a_n ^\uparrow } } D_{-ia_n ^\uparrow } (e^{\frac{\pi }{4} i} \xi ) 
  D_n (\eta ) \Gamma _1 \right. \\[-8pt]
 & \left.
  -\sqrt{\frac{2eB}{a_n ^\uparrow } } D_{-ia_n ^\uparrow } (e^{\frac{\pi }{4} i} \xi ) nD_{n-1} (\eta ) \Gamma _2
  +\sqrt{2eEa_n ^\uparrow } e^{\frac{\pi }{4} i} D_{-ia_n ^\uparrow -1 } (e^{\frac{\pi }{4} i} \xi ) D_n (\eta ) \Gamma _3 \right]
  \frac{e^{i(p_x x+p_z z)} }{2\pi} 
\end{split} \notag \\
\begin{split}
{}_- \! \psi _{\mathbf{p} \uparrow } ^\mathrm{out} (x) 
 &= \frac{e^{-\frac{\pi }{4} a_n ^\uparrow } }{2\sqrt{eE} } 
  \sqrt{\frac{L}{2\pi } } \left( \frac{eB}{\pi } \right) ^\frac{1}{4} \frac{1}{\sqrt{n!} }
  \left[ mD_{-ia_n ^\uparrow -1} ^* (e^{\frac{\pi }{4} i} \xi ) D_n (\eta ) \Gamma _1 \right. \\[-8pt]
 & \left.
  -\sqrt{2eB} D_{-ia_n ^\uparrow -1} ^* (e^{\frac{\pi }{4} i} \xi ) nD_{n-1} (\eta ) \Gamma _2
  -\sqrt{2eE} e^{\frac{\pi }{4} i} D_{-ia_n ^\uparrow } ^* (e^{\frac{\pi }{4} i} \xi ) D_n (\eta ) \Gamma _3 \right]
  \frac{e^{i(p_x x+p_z z)} }{2\pi} 
\end{split} \notag \\
\begin{split}
{}_+ \! \psi _{\mathbf{p} \downarrow } ^\mathrm{out} (x) 
 &= \frac{e^{-\frac{\pi }{4} a_n ^\downarrow } }{2\sqrt{eE} } 
  \sqrt{\frac{L}{2\pi } } \left( \frac{eB}{\pi } \right) ^\frac{1}{4} \frac{1}{\sqrt{n!} }
  \left[ \frac{m}{\sqrt{a_n ^\downarrow } } D_{-ia_n ^\downarrow } (e^{\frac{\pi }{4} i} \xi ) D_n (\eta ) \Gamma _2 \right. \\[-8pt]
 & \left.
  +\sqrt{\frac{2eB}{a_n ^\downarrow } } D_{-ia_n ^\downarrow } (e^{\frac{\pi }{4} i} \xi ) D_{n+1} (\eta ) \Gamma _1
  +\sqrt{2eEa_n ^\downarrow } e^{\frac{\pi }{4} i} D_{-ia_n ^\downarrow -1 } (e^{\frac{\pi }{4} i} \xi ) D_n (\eta ) \Gamma _4 \right]
  \frac{e^{i(p_x x+p_z z)} }{2\pi} 
\end{split} \notag \\
\begin{split}
{}_- \! \psi _{\mathbf{p} \downarrow } ^\mathrm{out} (x) 
 &= \frac{e^{-\frac{\pi }{4} a_n ^\downarrow } }{2\sqrt{eE} } 
  \sqrt{\frac{L}{2\pi } } \left( \frac{eB}{\pi } \right) ^\frac{1}{4} \frac{1}{\sqrt{n!} }
  \left[ mD_{-ia_n ^\downarrow -1} ^* (e^{\frac{\pi }{4} i} \xi ) D_n (\eta ) \Gamma _2 \right. \\[-8pt]
 & \left.
  +\sqrt{2eB} D_{-ia_n ^\downarrow -1} ^* (e^{\frac{\pi }{4} i} \xi ) D_{n+1} (\eta ) \Gamma _1
  -\sqrt{2eE} e^{\frac{\pi }{4} i} D_{-ia_n ^\downarrow } ^* (e^{\frac{\pi }{4} i} \xi ) D_n (\eta ) \Gamma _4 \right]
  \frac{e^{i(p_x x+p_z z)} }{2\pi} . 
\end{split} \label{psi^out_EM}
\end{gather}
\end{allowdisplaybreaks}
These solutions satisfy the orthonormal condition
\begin{equation}
\begin{matrix}
({}_+ \! \psi _{\mathbf{p} s} ^{\mathrm{as} } ,{}_+ \! \psi _{\mathbf{q} s^{\prime } } ^{\mathrm{as} } ) 
  = ({}_- \! \psi _{\mathbf{p} s} ^{\mathrm{as} } ,{}_- \! \psi _{\mathbf{q} s^{\prime } } ^{\mathrm{as} } ) 
  = \delta (p_x -q_x )\delta (p_z -q_z ) \frac{L}{2\pi } \delta _{n,m} \delta _{s s^{\prime } } \\[+7pt]
({}_+ \! \psi _{\mathbf{p} s} ^{\mathrm{as} } ,{}_- \! \psi _{\mathbf{q} s^{\prime } } ^{\mathrm{as} } ) =0
\end{matrix} \ \
\left( \begin{matrix}
\mathrm{as} =\mathrm{in} ,\mathrm{out} \\
s,s^{\prime } = \uparrow ,\downarrow 
\end{matrix} \right) .
\end{equation}
The quantized field operator in the presence of the collinear electromagnetic field 
can be expanded in two ways: 
\begin{equation}
\begin{split}
\psi (x) &= \sum _{s=\uparrow ,\downarrow } \frac{2\pi }{L} \sum _n \int \! dp_x \int \! dp_z
            \left[ {}_+ \! \psi _{\mathbf{p} s} ^\mathrm{in} (x) a_{\mathbf{p} s} ^\mathrm{in}
            + {}_- \! \psi _{\mathbf{p} s} ^\mathrm{in} (x) b_{-\mathbf{p} s} ^{\mathrm{in} \dagger } \right] \\
         &= \sum _{s=\uparrow ,\downarrow } \frac{2\pi }{L} \sum _n \int \! dp_x \int \! dp_z
            \left[ {}_+ \! \psi _{\mathbf{p} s} ^\mathrm{out} (x) a_{\mathbf{p} s} ^\mathrm{out}
            + {}_- \! \psi _{\mathbf{p} s} ^\mathrm{out} (x) b_{-\mathbf{p} s} ^{\mathrm{out} \dagger } \right] .
\end{split}
\end{equation}

\vspace{11pt}
We can obtain solutions in the pure electric field, $A^\mu =(0,0,0,-Et) $,
from Eq.\eqref{psi^in_EM} and \eqref{psi^out_EM} by the replacements
\begin{equation}
\begin{matrix}
\sqrt{\frac{L}{2\pi } } \left( \frac{eB}{\pi } \right) ^\frac{1}{4} \frac{1}{\sqrt{n!} } D_n (\eta ) 
 \rightarrow \frac{e^{ip_y y} }{\sqrt{2\pi } } \\
\sqrt{\frac{L}{2\pi } } \left( \frac{eB}{\pi } \right) ^\frac{1}{4} \frac{1}{\sqrt{n!} }
 \sqrt{2eB} D_{n+1} (\eta ) 
 \rightarrow (p_x -ip_y ) \frac{e^{ip_y y} }{\sqrt{2\pi } } \\
\sqrt{\frac{L}{2\pi } } \left( \frac{eB}{\pi } \right) ^\frac{1}{4} \frac{1}{\sqrt{n!} }
 \sqrt{2eB} nD_{n-1} (\eta ) 
 \rightarrow (p_x +ip_y ) \frac{e^{ip_y y} }{\sqrt{2\pi } } \\
a_n ^{\uparrow /\downarrow } \rightarrow a=\frac{m_\mathrm{T} ^2 }{2eE} .
\end{matrix}
\end{equation}
The results are
\begin{allowdisplaybreaks}
\begin{gather}
\begin{split}
{}_+ \! \psi _{\mathbf{p} \uparrow } ^\mathrm{in} (x) &= \frac{e^{-\frac{\pi }{4} a}}{2\sqrt{eE} } \left[ 
  m D_{-ia-1} ^* (-e^{\frac{\pi }{4} i} \xi ) \Gamma _1
  -(p_x +ip_y ) D_{-ia-1} ^* (-e^{\frac{\pi }{4} i} \xi ) \Gamma _2 \right. \\[-8pt]
 & \hspace{5cm} \left.
  +\sqrt{2eE} e^{\frac{\pi }{4} i } D_{-ia} ^* (-e^{\frac{\pi }{4} i} \xi )\Gamma _3 \right]
  \frac{e^{i\mathbf{p} \cdot \mathbf{x} } }{\sqrt{(2\pi )^3 } } 
\end{split} \notag \\[-5pt]
\begin{split}
{}_- \! \psi _{\mathbf{p} \uparrow } ^\mathrm{in} (x) &= \frac{e^{-\frac{\pi }{4} a}}{2\sqrt{eE} } \left[ 
  a^{-\frac{1}{2} } m D_{-ia} (-e^{\frac{\pi }{4} i} \xi ) \Gamma _1
  -a^{-\frac{1}{2} } (p_x +ip_y ) D_{-ia} (-e^{\frac{\pi }{4} i} \xi ) \Gamma _2 \right. \\[-8pt]
 & \hspace{5cm} \left.
  -m_{\mathrm{T} } e^{\frac{\pi }{4} i } D_{-ia-1} (-e^{\frac{\pi }{4} i} \xi )\Gamma _3 \right]
  \frac{e^{i\mathbf{p} \cdot \mathbf{x} } }{\sqrt{(2\pi )^3 } } 
\end{split} \notag \\[-5pt]
\begin{split}
{}_+ \! \psi _{\mathbf{p} \downarrow } ^\mathrm{in} (x) &= \frac{e^{-\frac{\pi }{4} a}}{2\sqrt{eE} } \left[ 
  mD_{-ia-1} ^* (-e^{\frac{\pi }{4} i} \xi )\Gamma _2
  +(p_x -ip_y ) D_{-ia-1} ^* (-e^{\frac{\pi }{4} i} \xi )\Gamma _1  \right. \\[-8pt]
 & \hspace{5cm} \left.
  +\sqrt{2eE} e^{\frac{\pi }{4} i } D_{-ia} ^* (-e^{\frac{\pi }{4} i} \xi )\Gamma _4 \right]
  \frac{e^{i\mathbf{p} \cdot \mathbf{x} } }{\sqrt{(2\pi )^3 } } 
\end{split} \notag \\[-5pt]
\begin{split}
{}_- \! \psi _{\mathbf{p} \downarrow } ^\mathrm{in} (x) &= \frac{e^{-\frac{\pi }{4} a}}{2\sqrt{eE} } \left[ 
  a^{-\frac{1}{2} } m D_{-ia} (-e^{\frac{\pi }{4} i} \xi )\Gamma _2
  +a^{-\frac{1}{2} } (p_x -ip_y ) D_{-ia} (-e^{\frac{\pi }{4} i} \xi )\Gamma _1 \right. \\[-8pt]
 & \hspace{5cm} \left.
  -m_{\mathrm{T} } e^{\frac{\pi }{4} i } D_{-ia-1} (-e^{\frac{\pi }{4} i} \xi )\Gamma _4 \right]
  \frac{e^{i\mathbf{p} \cdot \mathbf{x} } }{\sqrt{(2\pi )^3 } }
\end{split} \label{inpsi}
\end{gather}
\end{allowdisplaybreaks}
and
\begin{allowdisplaybreaks}
\begin{gather}
\begin{split}
{}_+ \! \psi _{\mathbf{p} \uparrow } ^\mathrm{out} (x) &= \frac{e^{-\frac{\pi }{4} a}}{2\sqrt{eE} } \left[ 
  a^{-\frac{1}{2} } m D_{-ia} (e^{\frac{\pi }{4} i} \xi ) \Gamma _1
  -a^{-\frac{1}{2} } (p_x +ip_y ) D_{-ia} (e^{\frac{\pi }{4} i} \xi ) \Gamma _2 \right. \\[-8pt]
 & \hspace{5cm} \left.
  +m_{\mathrm{T} } e^{\frac{\pi }{4} i } D_{-ia-1} (e^{\frac{\pi }{4} i} \xi )\Gamma _3 \right]
  \frac{e^{i\mathbf{p} \cdot \mathbf{x} } }{\sqrt{(2\pi )^3 } }
\end{split} \notag \\[-5pt]
\begin{split}
{}_- \! \psi _{\mathbf{p} \uparrow } ^\mathrm{out} (x) &= \frac{e^{-\frac{\pi }{4} a}}{2\sqrt{eE} } \left[ 
  m D_{-ia-1} ^* (e^{\frac{\pi }{4} i} \xi ) \Gamma _1
  -(p_x +ip_y ) D_{-ia-1} ^* (e^{\frac{\pi }{4} i} \xi ) \Gamma _2 \right. \\[-8pt]
 & \hspace{5cm} \left.
  -\sqrt{2eE} e^{\frac{\pi }{4} i } D_{-ia} ^* (e^{\frac{\pi }{4} i} \xi )\Gamma _3 \right]
  \frac{e^{i\mathbf{p} \cdot \mathbf{x} } }{\sqrt{(2\pi )^3 } }
\end{split} \notag \\[-5pt]
\begin{split}
{}_+ \! \psi _{\mathbf{p} \downarrow } ^\mathrm{out} (x) &= \frac{e^{-\frac{\pi }{4} a}}{2\sqrt{eE} } \left[ 
  a^{-\frac{1}{2} } m D_{-ia} (e^{\frac{\pi }{4} i} \xi )\Gamma _2
  +a^{-\frac{1}{2} } (p_x -ip_y ) D_{-ia} (e^{\frac{\pi }{4} i} \xi )\Gamma _1 \right. \\[-8pt]
 & \hspace{5cm} \left.
  +m_{\mathrm{T} } e^{\frac{\pi }{4} i } D_{-ia-1} (e^{\frac{\pi }{4} i} \xi )\Gamma _4 \right]
  \frac{e^{i\mathbf{p} \cdot \mathbf{x} } }{\sqrt{(2\pi )^3 } }
\end{split} \notag \\[-5pt]
\begin{split}
{}_- \! \psi _{\mathbf{p} \downarrow } ^\mathrm{out} (x) &= \frac{e^{-\frac{\pi }{4} a}}{2\sqrt{eE} } \left[ 
  mD_{-ia-1} ^* (e^{\frac{\pi }{4} i} \xi )\Gamma _2
  +(p_x -ip_y ) D_{-ia-1} ^* (e^{\frac{\pi }{4} i} \xi )\Gamma _1 \right. \\[-8pt]
 & \hspace{5cm} \left.
  -\sqrt{2eE} e^{\frac{\pi }{4} i } D_{-ia} ^* (e^{\frac{\pi }{4} i} \xi )\Gamma _4 \right]
  \frac{e^{i\mathbf{p} \cdot \mathbf{x} } }{\sqrt{(2\pi )^3 } } . 
\end{split} \label{outpsi}
\end{gather}
\end{allowdisplaybreaks}

\vspace{11pt}
Solutions in the pure magnetic field, $A^\mu = (0,-By,0,0) $, can be obtained 
from the free field solutions \eqref{freeDirac} by the replacements
\begin{equation}
\begin{matrix}
\frac{e^{ip_y y} }{\sqrt{2\pi } } \rightarrow 
\sqrt{\frac{L}{2\pi } } \left( \frac{eB}{\pi } \right) ^\frac{1}{4} \frac{1}{\sqrt{n!} } D_n (\eta ) \\
(p_x -ip_y ) \frac{e^{ip_y y} }{\sqrt{2\pi } } \rightarrow 
\sqrt{\frac{L}{2\pi } } \left( \frac{eB}{\pi } \right) ^\frac{1}{4} \frac{1}{\sqrt{n!} } 
\sqrt{2eB} D_{n+1} (\eta ) \\
(p_x +ip_y ) \frac{e^{ip_y y} }{\sqrt{2\pi } } \rightarrow 
\sqrt{\frac{L}{2\pi } } \left( \frac{eB}{\pi } \right) ^\frac{1}{4} \frac{1}{\sqrt{n!} }
\sqrt{2eB} nD_{n-1} (\eta ) \\
\omega _p \rightarrow \left \{ \begin{matrix} 
  \omega _p ^\uparrow \equiv \sqrt{m^2 +p_z ^2 +2neB} \\
  \omega _p ^\downarrow \equiv \sqrt{m^2 +p_z ^2 +(2n+2)eB}
  \end{matrix} \right. .
\end{matrix}
\end{equation}
The results are
\begin{allowdisplaybreaks}
\begin{gather}
\begin{split}
{}_+ \! \psi _{\mathbf{p} \uparrow } ^B (x)
 &= \frac{1}{\sqrt{2} }
  \sqrt{\frac{L}{2\pi } }  \left( \frac{eB}{\pi } \right) ^\frac{1}{4} \frac{1}{\sqrt{n!} }
  \left[ \frac{m}{\sqrt{\omega _p ^\uparrow -p_z } } D_n (\eta ) \Gamma _1
  -\frac{\sqrt{2eB} }{\sqrt{\omega _p ^\uparrow -p_z }} nD_{n-1} (\eta ) \Gamma _2 \right. \\[-3pt]
 & \hspace{3cm} \left.
  +\sqrt{\omega _p ^\uparrow -p_z } D_n (\eta ) \Gamma _3 \right]
  \frac{1}{\sqrt{(2\pi )^2 2\omega _p ^\uparrow } } e^{-i\omega _p ^\uparrow t+i(p_x x+p_z z) } 
\end{split} \notag \\[-5pt]
\begin{split}
{}_- \! \psi _{\mathbf{p} \uparrow } ^B (x)
 &= \frac{1}{\sqrt{2} }
  \sqrt{\frac{L}{2\pi } } \left( \frac{eB}{\pi } \right) ^\frac{1}{4} \frac{1}{\sqrt{n!} }
  \left[ \frac{m}{\sqrt{\omega _p ^\uparrow +p_z } } D_n (\eta ) \Gamma _1 
  -\frac{\sqrt{2eB} }{\sqrt{\omega _p ^\uparrow +p_z }} nD_{n-1} (\eta ) \Gamma _2 \right. \\[-3pt]
 & \hspace{3cm} \left.
  -\sqrt{\omega _p ^\uparrow +p_z } D_n (\eta ) \Gamma _3 \right]
  \frac{1}{\sqrt{(2\pi )^2 2\omega _p ^\uparrow } } e^{+i\omega _p ^\uparrow t+i(p_x x+p_z z) } 
\end{split} \notag \\[-5pt]
\begin{split}
{}_+ \! \psi _{\mathbf{p} \downarrow } ^B (x)
 &= \frac{1}{\sqrt{2} } 
  \sqrt{\frac{L}{2\pi } } \left( \frac{eB}{\pi } \right) ^\frac{1}{4} \frac{1}{\sqrt{n!} }
  \left[ \frac{m}{\sqrt{\omega _p ^\downarrow -p_z } } D_n (\eta ) \Gamma _2
  +\frac{\sqrt{2eB} }{\sqrt{\omega _p ^\downarrow -p_z }} D_{n+1} (\eta ) \Gamma _1 \right. \\[-3pt]
 & \hspace{3cm} \left.
  +\sqrt{\omega _p ^\downarrow -p_z } D_n (\eta ) \Gamma _4 \right]
  \frac{1}{\sqrt{(2\pi )^2 2\omega _p ^\downarrow } } e^{-i\omega _p ^\downarrow t+i(p_x x+p_z z) }  
\end{split} \notag \\[-5pt]
\begin{split}
{}_- \! \psi _{\mathbf{p} \downarrow } ^B (x)
 &= \frac{1}{\sqrt{2} }
  \sqrt{\frac{L}{2\pi } } \left( \frac{eB}{\pi } \right) ^\frac{1}{4} \frac{1}{\sqrt{n!} }
  \left[ \frac{m}{\sqrt{\omega _p ^\downarrow +p_z } } D_n (\eta ) \Gamma _2 
  +\frac{\sqrt{2eB} }{\sqrt{\omega _p ^\downarrow +p_z }} D_{n+1} (\eta ) \Gamma _1 \right. \\[-3pt]
 & \hspace{3cm} \left.
  -\sqrt{\omega _p ^\downarrow +p_z } D_n (\eta ) \Gamma _4 \right]
  \frac{1}{\sqrt{(2\pi )^2 2\omega _p ^\downarrow } } e^{+i\omega _p ^\downarrow t+i(p_x x+p_z z) } . 
\end{split} \label{psi_pM}
\end{gather}
\end{allowdisplaybreaks}
Notice that, in this case, distinction between in and out solutions is unnecessary
since the magnetic field affects only transverse part of the solution and
time-dependent part $e^{\mp i\omega _p t} $ is unchanged, which is separable
into positive and negative frequency in all time.
In other words, a pure magnetic field cannot cause pair creation.

%%%%%%%%%%%%%%%%%%%%%%%%%%%%%%%%%%%%%%%%%%%%%%%%%%%%%%%%%%%%%%%%%%%%%%%%%%%%%%%%%%%%%%%%%%%%%%%%%%%%%%%%%%%%%%%

\end{document}